\documentclass[prb,twocolumn,showpacs,amsmath,amssymb,superscriptaddress,floatfix]{revtex4-1}
\usepackage{amsmath}
\usepackage{amssymb}
\usepackage{amsthm}
\usepackage{amsfonts}
\usepackage{dsfont}
\usepackage{listings}
\usepackage{tabularx}
\usepackage{enumerate}
\usepackage{latexsym}
\usepackage{mathdots}
\usepackage{bigints}

\usepackage{psfrag}

\usepackage{bm}
\usepackage{graphicx}
\usepackage{subfigure}
\usepackage{xcolor}

\newtheorem{theorem}{Theorem}[section]

\theoremstyle{definition}
\newtheorem{definition}[theorem]{Definition}

\newcommand{\spa}{\text{ }}

\newcommand{\beq}{\begin{equation}}
\newcommand{\eneq}{\end{equation}}

\newcommand{\bra}[1]{\left\langle#1\right|}
\newcommand{\ket}[1]{\left|#1\right\rangle}

\newcommand{\mH}[1]{\mathcal{H}^{(#1)}}
\newcommand{\mK}[1]{\mathcal{K}^{(#1)}}
\newcommand{\mQ}[1]{\mathcal{Q}^{(#1)}}
\newcommand{\mG}[1]{\mathcal{G}^{(#1)}}
\newcommand{\mC}[1]{\mathcal{C}^{(#1)}}

\newcommand{\ml}{\mathcal{L}}

\newcommand{\mt}{\mathcal{T}}
\newcommand{\mKdef}[2]{\mathcal{K}\left(#1, #2\right)}
\newcommand{\Kp}{\mathcal{K}^{(p)}}
\newcommand{\Hp}{\mathcal{H}^{(p)}}
\newcommand{\Qp}{\mathcal{Q}^{(p)}}
\newcommand{\Gp}{\mathcal{G}^{(p)}}
\newcommand{\Cp}{\mathcal{C}^{(p)}}

\newcommand{\mI}{\mathcal{I}}
\newcommand{\mP}{\mathcal{P}}
\newcommand{\quant}{X}

\newcommand{\cd}{c^\dagger}

\newcommand{\ztwo}[1]{\ket{\mathbb{Z}^{(#1)}_2}}
\newcommand{\fsa}[1]{F^{(#1)}}
\newcommand{\norm}[1]{c^{(#1)}}
\newcommand{\tridiag}[1]{\beta^{(#1)}}
\newcommand{\ztwop}[1]{\ket{\mathbb{Z}'^{(#1)}_2}}
\newcommand{\init}[1]{\ket{R^{(#1)}}}

\newcommand{\that}{\hat{t}}

\newcommand{\Np}{N_{\Phi}}

\newcommand{\prodal}[2]{\underset{#1}{\overset{#2}{\prod}}}
\newcommand{\sumal}[2]{\underset{#1}{\overset{#2}{\sum}}}

\newcommand{\oneket}[1]{\ket{\ \fbox{$#1$}\ }}
\newcommand{\twoket}[2]{\ket{\ \fbox{$#1$}\ \fbox{$#2$}\ }}

\newcommand{\zero}{\textrm{o}}
\newcommand{\plus}{$+$}
\newcommand{\minus}{$-$}
\newcommand{\bzero}{\fbox{\parbox[c][0.5em]{0.5em}{\zero}}}
\newcommand{\bplus}{\fbox{\parbox[c][0.5em]{0.7em}{\plus}}}
\newcommand{\bminus}{\fbox{\parbox[c][0.5em]{0.7em}{\minus}}}
\newcommand{\bxd}[1]{\fbox{\parbox[c][0.5em]{0.5em}{#1}}}

\input{epsf}

\newcommand{\nn}{\nonumber}

\newcommand{\twopartdef}[4]
{
	\left\{
		\begin{array}{ll}
			#1 & \mbox{if } #2 \\
			#3 & \mbox{if } #4
		\end{array}
	\right.
}

\newcommand{\twopartdefoth}[3]
{
	\left\{
		\begin{array}{ll}
			#1 & \mbox{if } #2 \\
			#3 & \mbox{otherwise}
		\end{array}
	\right.
}

\newcommand{\threepartdef}[6]
{
	\left\{
		\begin{array}{ll}
			#1 & \mbox{\textrm{if} } #2 \\
			#3 & \mbox{\textrm{if} } #4 \\
			#5 & \mbox{\textrm{if} } #6
		\end{array}
	\right.
}

\newcommand{\fourpartdef}[8]
{
	\left\{
		\begin{array}{ll}
			#1 & \mbox{\textrm{if} } #2 \\
			#3 & \mbox{\textrm{if} } #4 \\
			#5 & \mbox{\textrm{if} } #6 \\
			#7 & \mbox{\textrm{if} } #8
		\end{array}
	\right.
}

\begin{document}
\tolerance 10000
\newcommand{\vk}{\boldsymbol{k}}   
\newcommand{\br}{\boldsymbol{r}}
\newcommand{\bq}{\boldsymbol{q}}
\title{Quantum Many-body Scars in a Landau Level on a Thin Torus}
\author{Sanjay Moudgalya}
\affiliation{Department of Physics, Princeton University, Princeton, NJ 08544, USA}
\author{B. Andrei Bernevig}
\affiliation{Department of Physics, Princeton University, Princeton, NJ 08544, USA}
\author{Nicolas Regnault}
\affiliation{Laboratoire de Physique de l'Ecole normale sup\'{e}rieure, ENS, Universit\'{e} PSL, CNRS, Sorbonne Universit\'{e}, Universit\'{e} Paris-Diderot, Sorbonne Paris Cit\'{e}, Paris, France}
\date{\today}
\begin{abstract}
We study a kinetically constrained pair hopping model that arises within a Landau level in the quantum Hall effect. 
At filling $\nu = 1/3$, the model exactly maps onto the so-called ``PXP model", a constrained model for the Rydberg atom chain that is numerically known to exhibit ETH-violating states in the middle of the spectrum or quantum many-body scars. 
Indeed, particular charge density wave configurations exhibit the same revivals seen in the PXP model. 
We generalize the mapping to fillings factors $\nu = p/(2p+1)$, and show that the model is equivalent to non-integrable spin-chains within particular constrained Krylov Hilbert spaces. %
These lead to new examples of quantum many-body scars which manifest as revivals and slow thermalization of particular charge density wave states.
Finally, we investigate the stability of the quantum scars under certain Hamiltonian perturbations motivated by the fractional quantum Hall physics.
\end{abstract}
\maketitle
\section{Introduction}
The breakdown of thermalization in non-integrable isolated quantum systems has been a subject of recent interest. 
This is believed to be equivalent to the failure of the Eigenstate Thermalization Hypothesis (ETH), \cite{deutsch1991quantum, srednicki1994chaos, rigol2008thermalization, d2016quantum} which is expected to hold for generic non-integrable models. 
A well-known instance of such a failure occurs in Many-body Localization (MBL)\cite{gornyi2005interacting, basko2006metal, pal2010many, nandkishore2015many} which happens in the presence of disorder,\cite{pal2010many,  kjall2014many} quasiperiodicity,\cite{iyer2013many, khemani2017two, lee2017many} or strong electric fields. \cite{schulz2018stark, van2018bloch} 
The failure of the Eigenstate Thermalization Hypothesis (ETH) in MBL systems can be explained by the emergence of quasilocal integrability, also reflected in the absence of level repulsion in the MBL phase. 
A different kind of breakdown of thermalization occurs when some ETH violating eigenstates in the middle of the spectrum coexist with otherwise ETH satisfying eigenstates.
Such a scenario was rigorously shown for the first time in the (non-integrable) spin-$S$ Affleck-Kennedy-Tasaki-Lieb (AKLT) model,\cite{affleck1988valence} where a quasiparticle tower of states in the middle of the spectrum was obtained in Ref.~[\onlinecite{moudgalya2018exact}] and [\onlinecite{moudgalya2018entanglement}].
These states, which so far remain the only tower of states to be analytically tractable without having a conservation law, were shown to have a sub-thermal (logarithmic) scaling of entanglement entropy,\cite{moudgalya2018entanglement} thus violating the strong ETH prediction. 
For a set of analytically tractable states of low entanglement stemming out of a conservation law, see Refs.~[\onlinecite{vafek2017entanglement, sala2019ergodicity, khemani2019local, pai2019dynamical}].
In addition, Ref.~[\onlinecite{shiraishi2017systematic}] showed that ground states of certain models can be embedded in the middle of the spectra of deformed models, thus violating ETH. A similar construction was worked out for topological models in Ref.~[\onlinecite{ok2019topological}].
These constitute a violation of strong ETH,\cite{kim2014testing, garrison2018does} which states that \emph{all} eigenstates in the middle of the spectrum obey ETH. 
Violations of strong ETH have also been found (independently and from a different perspective) numerically in certain systems with constrained Hilbert spaces.
Ref.~[\onlinecite{bernien2017probing}] observed non-thermal oscillations after a quench in cold atom experiments with Rydberg atoms, where the Hamiltonian imposes a penalty on neighboring atoms can both be excited,\cite{urban2009observation, lesanovsky2011many} forming an effective low-energy constrained Hilbert space.
To explain the oscillations from a N\'{e}el-like state, Refs.~[\onlinecite{turner2017quantum}] and [\onlinecite{turner2018quantum}] studied the so-called PXP model, a toy model for Rydberg atoms,\cite{lesanovsky2012interacting} and reported the presence of strong-ETH violating eigenstates that numerically are found to have a sub-thermal (logarithmic) growth of entanglement entropy. 
This phenomenon was called many-body scarring,\cite{turner2017quantum} analogous to the well-known phenomenon of scarring due to unstable periodic orbits in phase space when a classical system is quantized.\cite{heller1984bound} 
The many-body analogue of classical phase space was conjectured to be the Time-Dependent Variation Principle (TDVP) manifold of states,\cite{haegeman2011time} a correspondence that was illustrated using Matrix Product States (MPS) for the PXP model.\cite{ho2018periodic} 
Subsequently, four exact strong ETH violating states were analytically obtained for the PXP model.\cite{lin2018exact}
However, several questions about quantum scars in constrained models remain open, including their origin,\cite{khemani2018signatures, james2019nonthermal, robinson2018signatures, choi2018emergent, bull2019systematic, sala2019ergodicity, khemani2019local} the nature of the scars,\cite{lin2018exact, surace2019lattice, iadecola2019quantum} and constructions of scars in fermionic systems.  
It is thus important to search for other constrained systems that exhibit similar features. 
A natural system to look for constrained dynamics is the quantum Hall effect. 
It is well known that projection onto a Landau level imposes kinetic constraints on the electrons. 
These constraints are particularly apparent in the one-dimensional mapping of a Landau level on a cylinder.\cite{bergholtz2006one} 
There momentum conservation in the transverse direction leads to center of mass conservation of hopping terms along the cylinder.  
While such models generally involve long range hopping terms, longer range terms are exponentially suppressed when the circumference of the cylinder is small.
Models that arise in these so-called ``thin-torus" limits of quantum Hall systems have been extensively studied in the quantum Hall literature.
Notable among these are the development of the one-dimensional theory at half-filling,\cite{bergholtz2005half, bergholtz2006one, bergholtz2008quantum} mapping on to effective spin-chains for several filling factors,\cite{wang2012spin}  and exact solutions at half-filling\cite{bergholtz2005half, bergholtz2008quantum, nakamura2011beyond} and one-third filling.\cite{nakamura2012exactly} 
However, most of the existing literature has focused only on the ground state and not on excited states or dynamical properties, with a few recent exceptions.\cite{fremling2018dynamics, liu2018geometric}
In this work, we explore interesting features in the rest of the Hilbert space in these models, and show the appearance of constrained Hilbert spaces and quantum many-body scars. 
This paper is organized as follows.
In Sec.~\ref{sec:1Dmapping}, we review the one-dimensional mapping of a single Landau level in a two-dimensional quantum Hall system, and we discuss the thin-torus limit and introduce the truncated ``pair-hopping" model.
In Sec.~\ref{sec:arbitfilling}, we show the emergence of constrained Krylov subspaces at filling factors $\nu = p/(2p + 1)$. 
There we discuss the mapping of a particular constrained subspace to the PXP model at filling $\nu = 1/3$, and the effective spin-chains models that arise at filling $\nu = p/(2p + 1)$ in Sec.~\ref{sec:arbitfilling}. 
We also discuss the properties of the constrained Hilbert spaces that arise out of this models.
In Sec.~\ref{sec:properties}, we discuss various properties of the constrained Hamiltonians such as symmetries and zero-modes. 
In Sec.~\ref{sec:scars}, we discuss the Forward Scattering Approximation for the models at filling $\nu = p/(2p + 1)$ and show the existence of many-body scars and slow thermalizing product states for $\nu = 2/5$ and $\nu = 3/7$. These initial states are charge density waves in the quantum Hall language. 
In Sec.~\ref{sec:stability}, we briefly discuss the stability of scars to electrostatic terms that arise in the quantum Hall setup. 
There we numerically present some evidence that these many-body scars survive for small strengths of electrostatic terms.
Finally we conclude with open problems in Sec.~\ref{sec:conclusions}.
The appendices are reserved for technical details on results presented in the main text. 
\section{One-dimensional Mapping of a Landau level}\label{sec:1Dmapping}
We briefly review the origin of 1D pair-hopping models via a mapping of a single Landau level of a 2D quantum Hall system on a cylinder/torus to a 1D chain.\cite{haldane1985many, haldane1985periodic, bergholtz2005half, bergholtz2006one, bergholtz2008quantum} 
For the sake of illustration, we consider electrons either on an infinite cylinder for open boundary conditions (OBC) or on a torus for periodic boundary conditions (PBC) of length $L_x$ and circumference $L_y$ with $\Np = L_x L_y/(2 \pi)$ flux quanta.
We consider the Landau gauge $\vec{A} = B x \hat{y}$, where $B$ is the transverse magnetic field and $\hat{y}$ is the direction along the circumference of the cylinder. 
Setting the magnetic length to $\sqrt{\frac{\hbar}{eB}} = 1$, within a Landau level, we define the single-particle magnetic translation operators as\cite{fradkin2013field, bergholtz2008quantum}
\begin{equation}
    \that_x = \exp\left(\frac{L_x}{\Np} \left(\frac{\partial}{\partial x} + i y\right)\right),\;\;\; \that_y = \exp\left(\frac{L_y}{\Np}\frac{\partial}{\partial y}\right),
\label{eq:spmagtrans}
\end{equation}
which obey the commutation relation\cite{haldane1985many}
\begin{equation}
    \that_x \that_y = \exp\left(-i\frac{2 \pi}{\Np}\right)\that_y\that_x.
\label{eq:spmagcomm}
\end{equation}
A complete orthornormal basis of single-particle orbitals $\{\psi_{l,j}(\br)\}$, $1\leq j \leq \Np$ in the $l$-th Landau level can be constructed using eigenstates of $\that_y$.
These eigenstates satisfy
\begin{equation}
    \that_y \psi_{l, j}\left(\br\right) =  \exp\left(i\frac{2 \pi j}{\Np}\right) \psi_{l,j}\left(\br\right),\;\;\; \that_x \psi_{l,j}\left(\br\right) = \psi_{l, j+1}\left(\br\right).
\label{eq:speigproperty}
\end{equation}
For example, these wavefunctions in the lowest Landau level ($l = 0$) on a torus read\cite{haldane1985periodic, bergholtz2008quantum} 
\begin{eqnarray}
    &\psi_{0, j}\left(\br\right) = \frac{1}{\sqrt{\sqrt{\pi} L_y}}\sumal{m = -\infty}{\infty}{\left[\exp\left(i y L_x \left(m + \frac{j}{\Np}\right)\right)\right.}\nn \\
    &\left.\exp\left(- \frac{1}{2}\left(x + \left(m + \frac{j}{\Np}\right)L_x\right)^2\right)\right].
\label{eq:LLLorbitals}
\end{eqnarray}
For spinless electrons within a Landau level, the electron-electron interaction term can be written in the second-quantized form\cite{bergholtz2008quantum}
\begin{equation}
    H_l = \sumal{j_1, j_2, j_3, j_4 = 0}{\Np - 1}{V^{(l)}_{j_1,j_2,j_3,j_4}\cd_{j_1}\cd_{j_2}c_{j_3}c_{j_4}},
\label{eq:eehamilform}
\end{equation}
where $\cd_j$ and $c_j$ are the fermionic creation and annihilation operators for the single-particle orbital $\psi_{l, j}(\br)$, and 
\begin{eqnarray}
    &V^{(l)}_{j_1,j_2,j_3,j_4} \equiv \frac{1}{2}\iint_{\mathbb{T}^2}{\mathrm{d}^2 \br_1\ \mathrm{d}^2 \br_2\ \left(\psi^\ast_{l, j_1}\left(\br_1\right)\psi^\ast_{l, j_2}\left(\br_2\right)\right.} \nn \\
    &\left. \times V\left(\br_1 - \br_2\right)\psi_{l, j_3}\left(\br_2\right)\psi_{l, j_4}\left(\br_1\right)\right).
\label{eq:matrixel}
\end{eqnarray}
Since $j_1$, $j_2$, $j_3$ and $j_4$ are the $\hat{y}$ momentum eigenvalues (see Eq.~(\ref{eq:speigproperty})), for any interaction $V$ that is translation invariant in the $y$ direction we obtain
\begin{equation}
    j_1 + j_2 = j_3 + j_4 \;\; \textrm{mod}\;\Np.
\label{eq:momentumconservation}
\end{equation}
Further using translation invariance in the $\hat{x}$ direction, the Hamiltonian $H_l$ of Eq.~(\ref{eq:eehamilform}) can be reparametrized as\cite{bergholtz2005half, bergholtz2006one, bergholtz2008quantum}
\begin{eqnarray}
    &H_l = \sumal{m = 0}{\frac{\Np}{2}-1}{\sumal{k = m + 1}{\frac{\Np}{2}}{\left[\frac{V^{(l)}_{k,m}}{\left(1 + \delta_{m,0}\right)\left(1 + \delta_{k, \Np/2}\right)}\right.}} \nn \\
    &\times \left. \sumal{j = 0}{\Np - 1}{\left(\cd_{j}\cd_{j+k+m} c_{j+k} c_{j+m} + h.c.\right)}\right] \nn \\
    &\equiv \sumal{m = 0}{\frac{\Np}{2}-1}{\sumal{k = m + 1}{\frac{\Np}{2}}{V^{(l)}_{k,m} C_{k,m}}},
\label{eq:qhhamil}
\end{eqnarray}
where 
\begin{eqnarray}
    &V^{(l)}_{k,m} \equiv V^{(l)}_{j+m, j+k, j+k+m, j} - V^{(l)}_{j+m, j+k, j, j+k+m}\nn \\
    &+ V^{(l)}_{j+k, j+m, j, j+k+m} - V^{(l)}_{j+k, j+m, j+k+m, j}.
\label{eq:Vmn}
\end{eqnarray}
A procedure for obtaining $V^{(l)}_{k,m}$ given the potential $V(\br)$ is outlined in App.~\ref{app:Vkmgeneral}. We also refer readers to Ref.~[\onlinecite{lee2015geometric}] for a more general analysis.
For example, consider the short-range Haldane-Trugman-Kivelson potential,\cite{haldane1983fractional, trugman1985exact} 
\begin{equation}
    V\left(\br_1 - \br_2\right) \propto \nabla^2 \delta\left(\br_1 - \br_2\right).
\label{eq:tkpotential}
\end{equation}
The matrix elements $V^{(0)}_{k,m}$ for this potential in the lowest Landau level follow (when $L_x, \Np \rightarrow \infty$) (see Eq.~(\ref{eq:TKlist}))
\begin{equation}
    V^{(0)}_{k,m} \propto \left(k^2 - m^2\right) e^{-2\pi^2\frac{k^2 + m^2}{L_y^2}}.
\label{eq:Vkmtrugmankivelson}
\end{equation}
Note that whenever $m = 0$, $C_{k, 0}$ of Eq.~(\ref{eq:qhhamil}) reads
\begin{equation}
    C_{k,0} = \sumal{j = 0}{\Np-1}{\cd_j c_j \cd_{j+k} c_{j+k}} \equiv \sumal{j = 0}{\Np-1}{\hat{n}_j \hat{n}_{j+k}},
\end{equation}
where $\hat{n}_j \equiv \cd_j c_j$. Thus $C_{k,0}$ is a pure electrostatic term.
In the thin-torus limit ($L_y \rightarrow 0$), the strength of the terms $C_{k,m}$ decreases exponentially with increasing $(k^2 + m^2)$ (see Eqs.~(\ref{eq:Vkmtrugmankivelson}) and (\ref{eq:TKlist})). 
Thus, taking into account the terms up to the largest non-electrostatic term for a short-range potential in the lowest Landau level (remembering $k \geq m + 1$), we obtain an effective Hamiltonian $H_{\textrm{LLL}}$
\begin{eqnarray}
    &H_{\textrm{LLL}} = \sumal{j = 0}{\Np -1 }{\left(V^{(0)}_{1,0} \hat{n}_j  \hat{n}_{j+1} + V^{(0)}_{2,0} \hat{n}_j \hat{n}_{j+2}\right.} \nn \\
    &\left.+ V^{(0)}_{2,1} \left(\cd_j \cd_{j+3} c_{j+2} c_{j+1} + h.c.\right)\right).
\label{eq:uptopairhop}
\end{eqnarray}
We refer to the term $C_{2,1}$ as the ``pair-hopping term".
Note that these Hamiltonians can be generalized to three-body hopping terms, but we do not consider them in this work.
In this work, we focus on the study of the model Eq.~(\ref{eq:uptopairhop}).
These models preserve the center-of-mass position of the electrons in addition to their center-of-mass momentum, a property that is not respected by one-body electron hopping terms.
Thus, we expect these Hamiltonians to lead to novel dynamical phenomena.\cite{seidel2005incompressible}
In Secs.~\ref{sec:arbitfilling}-\ref{sec:scars}, we set $V^{(0)}_{1,0} = V^{(0)}_{2,0} = 0$, and $V^{(0)}_{2,1} = 1$, i.e. considering only the pair-hopping term.
We will discuss the effect of electrostatic terms in Sec.~\ref{sec:stability}.
We thus obtain a one-dimensional chain with $L \equiv \Np$ sites with the Hamiltonian that reads
\begin{equation}
    H \equiv \sumal{j = 1}{L_b}{H_j} =  \sumal{j = 1}{L_b}{\left(\cd_j \cd_{j+3} c_{j + 2} c_{j + 1} + h.c.\right)}
\label{eq:pairhopping}
\end{equation}
where $j + m$, $m = 1, 2, 3$ are defined modulo $L$, and $L_b = L$ (resp. $L_b = L - 3$) for PBC (resp. OBC).
Note that $H_j$ is non-vanishing only on the following configurations of sites $j$ to $j + 3$:
\begin{eqnarray}
    H_j \overset{\;\;j\;\;\;\;\;\;\;j+3}{\ket{0\spa 1 \spa 1 \spa 0}} &=& \overset{\;\;j\;\;\;\;\;\;\;j+3}{\ket{1 \spa 0 \spa 0 \spa 1}} \nn \\
    H_j \overset{\;\;j\;\;\;\;\;\;\;j+3}{\ket{1 \spa 0 \spa 0 \spa 1}} &=& \overset{\;\;j\;\;\;\;\;\;\;j+3}{\ket{0 \spa 1 \spa 1 \spa 0}}.
\label{eq:pairhoppingrules}
\end{eqnarray}
The rules of Eq.~(\ref{eq:pairhoppingrules}) correspond to so-called ``squeezing" and ``antisqueezing" processes in the Fractional Quantum Hall physics.\cite{bernevig2008model}
\section{Effective Spin-Chains and Constrained Hilbert spaces}\label{sec:arbitfilling}
We now review the mapping of the Hamiltonian of Eq.~(\ref{eq:pairhopping}) on to spin-1 chain models, first discussed in Ref.~[\onlinecite{wang2012spin}].
In this work, we restrict ourselves to filling factors of the form $\nu = p/(2p + 1)$ and system sizes of the form $L = (2p + 1) N$, $N \in \mathbb{N}$. 
We numerically observe that the ground state of $H$ is at half filling for even $L$, and filling $\nu = (L \pm 1)/(2L)$ for odd $L$.
Thus the sectors we study are in the middle of the full spectrum of $H$.  
\subsection{Mapping onto spin-1 chains}
Here we provide a summary of the mapping, which we illustrate with the details in Secs.~\ref{sec:1b3} and \ref{sec:pb2p+1}.
A crucial property of the pair-hopping Hamiltonian of Eq.~(\ref{eq:pairhopping}) that we rely on is the the existence of Krylov subspaces that, crucially, are \emph{smaller} than the full Hilbert space and are closed under the action of the Hamiltonian $H$. 
Given a state $\ket{R}$ and an operator $O$, a Krylov subspace $\mKdef{\ket{R}}{O}$ is defined as
\begin{equation}
    \mKdef{\ket{R}}{O} \equiv \mathrm{Span}\{\ket{R}, O\ket{R}, O^2\ket{R}, \cdots \}.
\label{eq:krylovdefn}
\end{equation}
For the pair-hopping model at filling $\nu = p/(2p + 1)$, we focus on the Krylov subspace $\Kp$, defined as
\begin{equation}
    \Kp \equiv \mKdef{{\init{p}}}{H},
\label{eq:Kpdefn}
\end{equation}
where ${\init{p}}$ is the state
\begin{equation}
    {\init{p}} = \bigotimes_{j = 1}^{N} \ket{\ \fbox{$0(10)^p$}\ },
\label{eq:puniformstate}
\end{equation}
where $(01)^p$ denotes the repetition of $(01)$ $p$ times.
Each of the $N$ boxed units in Eq.~(\ref{eq:puniformstate}) is referred to as a \emph{unit cell} (the system size $L = (2p + 1) N$).
For example, for $p = 1$ ($\nu = 1/3$) and $p = 2$ ($\nu = 2/5$), the states ${\init{1}}$ and ${\init{2}}$ read
\begin{eqnarray}
    {\init{1}} &=& \ket{\ \fbox{010}\ \fbox{010}\ \cdots\ \fbox{010}\ },\nn \\
    {\init{2}} &=& \ket{\ \fbox{01010}\ \fbox{01010}\ \cdots\ \fbox{01010}\ }.
\label{eq:12uniformexample}
\end{eqnarray}
An important property of the Krylov subspaces $\Kp$ is that it is \emph{not} the full Hilbert space of the pair-hopping system at that particular filling, a property discussed in Refs.~[\onlinecite{bergholtz2006one}] and [\onlinecite{wang2012spin}] and which will be illustrated in Secs.~\ref{sec:1b3} and \ref{sec:pb2p+1}.
Thus, to study the dynamics of states in $\Kp$ under the Hamiltonian $H$, it is sufficient to study the eigenstates of $\Hp$, the restriction of $H$ to $\Kp$. 
Since the Hamiltonian $H$ is a four-site Hamiltonian, in terms of unit cells, the $\Hp$ is a two unit cell Hamiltonian that has both intra- and inter- unit cell terms. That is, it is of the form
\begin{equation}
    \Hp = \sumal{j = 1}{N}{\Hp_j} + \sumal{j = 1}{N_b}{\Hp_{j,j+1}},
\label{eq:Hpgeneral}
\end{equation}
where $\Hp_j$ and $\Hp_{j,j+1}$ are one unit cell and two unit cell terms respectively, and 
\begin{equation}
    N_b = \twopartdef{N-1}{\textrm{OBC}}{N}{\textrm{PBC}}.
\label{eq:Nbdefn}
\end{equation}
We will obtain the explicit expressions for $\Hp_j$ and $\Hp_{j,j+1}$ for various $p$ in Secs.~\ref{sec:1b3} and \ref{sec:pb2p+1}.
As we will show there, $\Hp$ can be mapped onto a spin-1 Hamiltonian with each unit cell (as defined in $\init{p}$ in Eq.~(\ref{eq:puniformstate})) replaced by $p$ spin-1's.
\footnote{For PBC, the $(2p + 1)$ different ways of grouping the sites into unit cells results in an effective Hamiltonian with the same spectrum since $H$ is a center-of-mass preserving Hamiltonian that has a $(2p + 1)$-fold degenerate spectrum.\cite{seidel2005incompressible}} 
This mapping proceeds by \emph{appropriately} inserting $(p - 1)$ fictitious $0$'s (pseudozeroes) into each unit cell, grouping the resulting $3p$ sites into $p$ blocks of 3, and identifying each block with one of the following spin-1 configurations\cite{wang2012spin}
\begin{eqnarray}
    \ket{\zero} &\equiv& \ket{0\spa 1\spa 0} \nn \\
    \ket{+} &\equiv& \ket{0 \spa 0 \spa 1} \nn \\
    \ket{-} &\equiv& \ket{1 \spa 0 \spa 0}.
\label{eq:spin1dof}
\end{eqnarray}
As we will see in Secs.~\ref{sec:1b3} and \ref{sec:pb2p+1}, these three spin-1 configurations are sufficient to obtain a faithful mapping.  
In the rest of the paper, we denote configurations of $N$ unit cells as $\ket{\vec{\sigma}} = \ket{\sigma_1\sigma_2\cdots\sigma_N}$, where $\sigma_j$ is the configuration of spin-1's in the $j$-th unit cell (i.e. $\sigma_j$ encodes the configuration of $p$ spin-1's). 
For example, when $N = 2$, $p = 3$ ($\nu = 3/7$), consider the configuration $\ket{\psi} = \ket{\ \fbox{0011010}\ \fbox{0110010}\ }$, consisting of 14 orbitals.
As we will explain in Sec.~\ref{sec:pb2p+1}, this configuration can be uniquely mapped onto the spin-1 configuration $\ket{\vec{\sigma}} = \ket{\ \fbox{+\zero\zero}\ \fbox{\zero \minus\zero}\ }$, which consists of $N = 2$ unit cells and $p = 3$ spin-1's in each unit cell.
Then, we denote the configuration of each unit cell by the three spin configuration $\sigma_1 = \fbox{+\zero\zero}$ and $\sigma_2 = \fbox{\zero \minus \zero}$. 
We denote the spin-1 operators acting on the $n$-th spin on the $j$-th unit cell as $S^\alpha_{j,n}$, where $\alpha = x, y, z, +, -$. 
Further, we define operators $T_{j,n}$ and $U_{j,n}$ that will be used frequently throughout our analysis:
\begin{equation}
    T_{j,n} \equiv \frac{S^z_{j,n} S^-_{j,n}}{\sqrt{2}},\;\;\; U_{j,n} \equiv \frac{S^-_{j,n} S^z_{j,n}}{\sqrt{2}}.
\label{eq:TUdefn}
\end{equation}
The properties of these operators are provided in App.~\ref{app:TUproperties} (see Eq.~(\ref{eq:TUactions})). The only non-vanishing actions of the operators $T_{j,n}$ and $U_{j,n}$ are given by
\begin{equation}
	T_{j,n} \ket{\zero} = -\ket{-},\;\;\; U_{j,n}\ket{+} = \ket{\zero}, 
\label{eq:TUactionsmain}
\end{equation}  
where $\ket{\ast}_{j,n}$, with $\ast = +, \zero, -$ is the configuration of the spin-1 on the $n$-th site in the $j$-th unit cell.
Further, given a configuration $\ket{\vec{\sigma}} = \ket{\sigma_1\sigma_2\cdots\sigma_N}$, we define the following quantities that we use later in the paper. 
\begin{eqnarray}
    &P_{\sigma_j} = \sumal{l = 1}{p}{\delta_{\sigma_{jl}, +}}, \;\;\; M_{\sigma_j} = \sumal{l = 1}{p}{\delta_{\sigma_{jl}, -}} \nn \\
	&X^{(P)}_{\sigma_j} = \sumal{l = 1}{p}{(p + 1 - l)\  \delta_{\sigma_{jl}, +}} \nn \\
	&X^{(M)}_{\sigma_j} = \sumal{l = 1}{p}{l\  \delta_{\sigma_{jl}, -}},
\label{eq:Xdefn}
\end{eqnarray}
where $\sigma_{jl}$ is the configuration of the $l$-th spin in the $j$-th unit cell.
$P_{\sigma_j}$ (resp. $M_{\sigma_j}$) is the number of $+$'s (resp. $-$'s) in the $j$-th unit cell, $X^{(P)}_{\sigma_j}$ (resp. $X^{(M)}_{\sigma_j}$) is the sum of positions of $+$'s (resp. $-$'s) within the $j$-th unit cell counted from right (resp. left). 
For example, if $\sigma_j = \overset{1\ 2\ 3\ 4\ 5\ 6\ 7\ }{\fbox{\zero\minus\minus\plus\zero\plus\plus}}$, we have $P_{\sigma_j} = 3$, $M_{\sigma_j} = 2$, $X^{(P)}_{\sigma_j} = 7$, and $X^{(M)}_{\sigma_j} = 5$. 
\subsection{Filling $\boldsymbol{\nu = 1/3}$}\label{sec:1b3}
\subsubsection{Effective Hamiltonian}\label{sec:effhamil}
We now derive the effective Hamiltonian $\mH{1}$ at filling $\nu = 1/3$, i.e. $p = 1$, that acts on the Krylov subspace $\mK{1}$.
Consider the pair-hopping Hamiltonian on a system of size $L = 3N$. 
Using the spin-1 mapping of Eq.~(\ref{eq:spin1dof}), the state ${\init{1}}$ of Eq.~(\ref{eq:12uniformexample}) in the spin-1 language with the same choice of unit cells reads
\begin{equation}
    {\init{1}} = \ket{\ \bzero\ \bzero\ \cdots\ \bzero\ }.
\end{equation}
Using Eqs.~(\ref{eq:pairhopping}) and (\ref{eq:pairhoppingrules}), acting on $\init{1}$ with $H$ once results in a sum of  configurations that have one pair of $\ket{\ \bplus\ \bminus\ }$ on neighboring unit cells in a ``vacuum" of unit cells in the configuration $\ket{\ \bzero\ }$, i.e. configurations of the form
\begin{equation}
    \ket{\ \bzero\ \bzero\ \cdots\ \bzero\ \bplus\ \bminus\ \bzero\ \cdots\ \bzero\ }. 
\label{eq:1actstate}
\end{equation}
Using Eqs.~(\ref{eq:pairhoppingrules}) and (\ref{eq:spin1dof}), we immediately deduce that $\mH{1}_j = 0$ and that the actions of a single term $\mH{1}_{j, j+1}$ on the spin-1 configurations  read 
\begin{eqnarray}
    &&\mH{1}_{j, j+1} \overset{j\;\;j+1}{\ket{\cdots {\bzero} \spa {\bzero} \cdots}} = \overset{j\;\;j+1}{\ket{\cdots\spa {\bplus} \spa {\bminus} \cdots}} \nn \\
    &&\mH{1}_{j, j+1}\overset{j\;\;j+1}{\ket{\cdots \bplus \spa \bminus\cdots}} = \overset{j\;\;j+1}{\ket{\cdots\bzero \spa \bzero\cdots}} \nn \\
    &&\mH{1}_{j, j+1}\overset{j\;\;j+1}{\ket{\cdots \bzero \spa \bplus \cdots}} = 0 \nn \\
    &&\mH{1}_{j, j+1}\overset{j\;\;j+1}{\ket{\cdots \bminus \spa \bplus \cdots}} = 0\nn \\
    &&\mH{1}_{j, j+1}\overset{j\;\;j+1}{\ket{\cdots \bminus \spa \bzero \cdots}} = 0.
\label{eq:1b3scattering}
\end{eqnarray}
Using Eq.~(\ref{eq:1b3scattering}), further actions of $\mH{1}$ on the state of Eq.~(\ref{eq:1actstate}) either (i) destroy the nearest neighbor configuration $\bplus\ \bminus$, or (ii) create the nearest neighbor state $\bplus\ \bminus$ from the configuration $\bzero\ \bzero$.
Thus, one never obtains the nearest neighbor configurations $\bplus\ \bplus$, $\bplus\ \bzero$, $\bzero\ \bminus$,  or $\bminus\ \bminus$, and it is sufficient to consider the rules of Eq.~(\ref{eq:1b3scattering}).
Note that (i) and (ii) are the \emph{only} possible actions of $\mH{1}$ on subsequent configurations as well. In the language of the original orbitals, these processes correspond to squeezing and antisqueezing of close configurations respectively.
Thus any state $\ket{\vec{\sigma}} = \ket{\sigma_1\cdots\sigma_N}$ in the Krylov subspace $\mK{1}$ obeys the following constraints:
\begin{enumerate}
    \item[(c1)] The only allowed configuration of nearest neighbor unit cells are 
    \begin{equation}
        \bplus\ \bminus, \bzero\ \bzero, \bzero\ \bplus, \bminus\ \bzero, \bminus\ \bplus.
    \label{eq:nnallowed}
    \end{equation}
    This constraint can be compactly stated as $P_{\sigma_j} = M_{\sigma_{j+1}}$ $\forall j$, $1 \leq j \leq N - 1$. 
    \item[(c2)] With OBC, within the Krylov subspace, the leftmost (resp. rightmost) unit cell cannot have the configuration $\bminus$ (resp. $\bplus$), i.e. in the Krylov subspace $M_{\sigma_1} = 0$ (resp. $P_{\sigma_N} = 0$). This follows from Eq.~(\ref{eq:1b3scattering}) where $+$ and $-$ are created together only on nearest neighboring unit cells with $+$ in the left unit cell and $-$ in the right unit cell.  
\end{enumerate}
This is an example of a constrained Hilbert space.
Thus, using Eq.~(\ref{eq:1b3scattering}), $\mH{1}_{j,j+1}$ reads\cite{wang2012spin} 
\begin{eqnarray}
    \mH{1}_{j,j+1}  &=& \overset{\; j\;\;\;j+1}{\ket{\ {\bplus}\ {\bminus}\ }}\overset{\;\; j\;\;\;j+1}{\bra{\ {\bzero}\ {\bzero}\ }} + h.c. \nn \\
    &=& \left(\frac{S^z_{j,1} S^+_{j,1}}{\sqrt{2}}\right) \left(-\frac{S^z_{j+1,1} S^-_{j+1,1}}{\sqrt{2}}\right) + h.c. 
\end{eqnarray}
Thus, using Eq.~(\ref{eq:TUdefn}), we obtain the following expression for the Hamiltonian $\mH{1}$
\begin{equation}
\mH{1} = -\sumal{j = 1}{N_b}{\left(U^\dagger_{j,1} T_{j+1, 1} + \textrm{h.c.}\right)}.
\label{eq:effectivehamil1b3}
\end{equation}
Note that although the subscript ``$\cdots,1$" in the spin operators is redundant for this case because the unit cell contains a single spin-1, we continue to use it in order to smoothly transition to arbitrary values of $p$.  
\subsubsection{Mapping on to the PXP model}
The constrained Hilbert space $\mK{1}$ can be alternately specified by moving to the dual lattice of the spin-1 lattice, i.e. the sites $\{j + \frac{1}{2}\}$ defined on the bonds $\{(j, j+1)\}$. 
Thanks to the highly constrained Hilbert space, configurations of $N$ unit cells in $\mK{1}$ can be written in terms of spin-1/2 degrees of freedom on the dual lattice of $N - 1$ (resp. $N$) sites for OBC (resp. PBC) using the mapping
\begin{eqnarray}
    &&\ket{\ \bplus\ \bminus\ }_{j, j + 1} \rightarrow \ket{\ \uparrow\ }_{j+\frac{1}{2}} \nn \\
    &&\ket{\ \bzero\ \bplus\ }_{j, j+1} \rightarrow \ket{\ \downarrow\ }_{j+\frac{1}{2}} \nn \\
    &&\ket{\ \bminus\ \bzero\ }_{j, j+1} \rightarrow \ket{\ \downarrow\ }_{j+\frac{1}{2}} \nn \\
    &&\ket{\ \bzero\ \bzero\ }_{j, j+1} \rightarrow \ket{\ \downarrow\ }_{j+\frac{1}{2}}, \nn \\
    &&\ket{\ \bminus\ \bplus\ }_{j, j+1} \rightarrow \ket{\ \downarrow\ }_{j+\frac{1}{2}},
\label{eq:duallatticemapping}
\end{eqnarray}
where $\ket{\ \bxd{$\ast$}\ \bxd{$\ast$}\ }_{j,j+1}$, $\ast = +, \zero, -$ is the configuration of the $j$-th and $(j + 1)$-th unit cells on the spin-1 lattice and $\ket{\ast}_{j + \frac{1}{2}}$, $\ast = \uparrow,\downarrow$ is the configuration of the site $j + \frac{1}{2}$ on the dual lattice. The subscripts are taken to be modulo $N$ for PBC.  In other words, the nearest neighbor configuration $\bplus\ \bminus$ maps onto $\uparrow$ whereas all other nearest neighbor configurations in the Krylov subspace map onto $\downarrow$.
While the mapping appears to be many to one, we will shortly show that it is in fact invertible for both PBC and OBC as a result of the constraints of $\mK{1}$.
For example, the configuration 
\begin{equation}
    \ket{\psi_1} = \overset{1\;\;\;\;\;2\;\;\;\;\;\;3\;\;\;\;\;\;4\;\;\;\;\;\;5\;\;\;\;\;6}{\ket{\ \bzero\ \bplus\ \bminus\ \bplus\ \bminus\ \bzero\ }}
\label{eq:psi1config}
\end{equation}
maps on to the configurations $\ket{\psi^{(\textrm{PBC})}}$ and $\ket{\psi^{(\textrm{OBC})}}$ for PBC and OBC respectively, where
\begin{eqnarray}
    \ket{\psi^{(\textrm{PBC})}} = \overset{\frac{3}{2}\;\frac{5}{2}\;\frac{7}{2}\;\frac{9}{2}\;\frac{11}{2}\;\frac{13}{2}}{\ket{\ \downarrow\ \uparrow\ \downarrow\ \uparrow\ \downarrow\ \downarrow\ }} \label{eq:psidconfigpbc} \\
    \ket{\psi^{(\textrm{OBC})}} = \overset{\frac{3}{2}\;\frac{5}{2}\;\frac{7}{2}\;\frac{9}{2}\;\frac{11}{2}}{\ket{\ \downarrow\ \uparrow\ \downarrow\ \uparrow\ \downarrow\ }}.
\label{eq:psidconfigobc}
\end{eqnarray}
Note that the mapping of Eq.~(\ref{eq:duallatticemapping}) does not allow the dual lattice configuration $\ket{\ \uparrow\ \uparrow\ }$ even though it includes all possible nearest neighbor configurations allowed in $\mK{1}$ (Eq.~(\ref{eq:nnallowed})).
Thus, the constraint (c1) on $\mK{1}$ defined in Sec.~\ref{sec:effhamil} translates to the constraint that no nearest neighbor spins can be $\uparrow$ in the dual lattice (a hallmark of the PXP model\cite{turner2017quantum}). 
The mapping from the dual lattice back to the spin-1 lattice reads
\begin{eqnarray}
    &&\ket{\ \downarrow\ \uparrow\ }_{j-\frac{1}{2}, j+\frac{1}{2}} \rightarrow \ket{\ \bplus\ }_{j} \nn \\
    &&\ket{\ \uparrow\ \downarrow\ }_{j-\frac{1}{2}, j+\frac{1}{2}} \rightarrow \ket{\ \bminus\ }_{j} \nn \\
    &&\ket{\ \downarrow\ \downarrow\ }_{j-\frac{1}{2}, j+\frac{1}{2}} \rightarrow \ket{\ \bzero\ }_{j}, \nn \\
\label{eq:dualreversemapping}
\end{eqnarray}
where $\ket{\ \bxd{$\ast$}\ \bxd{$\ast$}\ }_{j-\frac{1}{2},j+\frac{1}{2}}$, $\ast = \uparrow, \downarrow$ is the configuration of the $\left(j-1/2\right)$-th and $\left(j+1/2\right)$-th sites on the dual lattice and $\ket{\ast}_{j}$, $\ast = +, \zero, -$ is the configuration of the $j$-th unit cell on the spin-1 lattice. The subscripts are taken to be modulo $N$ for PBC. 
Note that Eqs.~(\ref{eq:duallatticemapping}) and (\ref{eq:dualreversemapping}) ensure that the mapping is one to one for PBC.
For example, the configuration $\ket{\psi^{(PBC)}}$ of Eq.~(\ref{eq:psidconfigpbc}) maps onto $\ket{\psi_1}$ of Eq.~(\ref{eq:psi1config}) under Eq.~(\ref{eq:dualreversemapping}).\footnote{Note that since the leftmost and rightmost sites in the spin-1 language are labelled by $j = 1$ and $j = N$ respectively, the leftmost and rightmost sites on the dual lattice are labelled by $j = \frac{3}{2}$ and $j = N + \frac{1}{2}$ for PBC ($j = N - \frac{1}{2}$ for OBC).}
With OBC, Eq.~(\ref{eq:dualreversemapping}) can be applied to obtain the configuration of the unit cells $j$, $2 \leq j \leq N -2$.
The configurations of the leftmost ($j = 1$) and rightmost ($j = N$) unit cells can then be uniquely obtained using the constraint (c2) defined in Sec.~\ref{sec:effhamil}. 
For example, using the rules of Eq.~(\ref{eq:dualreversemapping}), the configuration $\ket{\psi^{(\textrm{OBC})}}$ of Eq.~(\ref{eq:psidconfigobc}) maps onto $\ket{\ \fbox{$\ast$}\ \bplus\ \bminus\ \bplus\ \bminus\ \fbox{$\ast$}\ }$, and the $\fbox{$\ast$}$ on the leftmost and rightmost unit cells are $\fbox{\zero}$, since that is the only allowed configuration allowed by the constraints (c1) and (c2).
Thus $\ket{\psi^{\textrm{OBC}}}$ of Eq.~(\ref{eq:psidconfigobc}) maps onto $\ket{\psi_1}$ of Eq.~(\ref{eq:psi1config}).
Since the Hamiltonian $\mH{1}$ consists of two unit cell terms, using the mapping of Eq.~(\ref{eq:duallatticemapping}), the corresponding Hamiltonian $H^{(d)}$ in the dual lattice consists of three site terms $H^{(d)}_{j-\frac{1}{2},j+\frac{1}{2},j+\frac{3}{2}}$ in the bulk and two site terms on the boundaries $H^{(d)}_{j-\frac{1}{2},j+\frac{1}{2}}$.
For example, the non-vanishing actions of $\mH{1}_{j,j+1}$ in the bulk of the chain translate to
\begin{eqnarray}
    &\mH{1}_{j,j+1}\overset{\;\;j\;\;\;j+1}{\ket{\ \fbox{$\ast$}\ \bplus\ \bminus\ \fbox{$\ast$}\ }} = \overset{\;\;j\;\;\;j+1}{\ket{\ \fbox{$\ast$}\ \bzero\ \bzero\ \fbox{$\ast$}\ }} \nn \\
    &\implies H^{(d)}_{j-\frac{1}{2},j+\frac{1}{2},j+\frac{3}{2}}\overset{j+\frac{1}{2}}{\ket{\ \downarrow\ \uparrow\ \downarrow\ }} = \overset{j+\frac{1}{2}}{\ket{\ \downarrow\ \downarrow\ \downarrow\ }}, \nn \\
    &\mH{1}_{j,j+1}\overset{\;\;j\;\;\;j+1}{\ket{\ \fbox{$\ast$}\ \bzero\ \bzero\ \fbox{$\ast$}\ }} = \overset{\;\;j\;\;\;j+1}{\ket{\ \fbox{$\ast$}\ \bplus\ \bminus\ \fbox{$\ast$}\ }} \nn \\
    &\implies H^{(d)}_{j-\frac{1}{2},j+\frac{1}{2},j+\frac{3}{2}}\overset{j+\frac{1}{2}}{\ket{\ \downarrow\ \downarrow\ \downarrow\ }} = \overset{j+\frac{1}{2}}{\ket{\ \downarrow\ \uparrow\ \downarrow\ }}, \nn \\
\label{eq:effectivespin1b2}
\end{eqnarray}
where $\ast$ corresponds to any allowed configuration of the unit cell, and subscripts are taken modulo $N$ for PBC. 
However with OBC, the actions of $\mH{1}_{j,j+1}$ on the left and right boundaries read (using the constraint (c2))
\begin{eqnarray}
    &\mH{1}_{1,2}\overset{\hspace{-6mm}1\;\;\;\;\; 2}{\ket{\ \bplus\ \bminus\ \fbox{$\ast$}\ }} = \overset{\hspace{-6mm}1\;\;\;\;\; 2}{\ket{\ \bzero\ \bzero\ \fbox{$\ast$}\ }} \nn \\
    &\implies H^{(d)}_{\frac{3}{2},\frac{5}{2}}\overset{\frac{3}{2}\;\;\frac{5}{2}}{\ket{\ \uparrow\ \downarrow\ }} = \overset{\frac{3}{2}\;\;\frac{5}{2}}{\ket{\ \downarrow\ \downarrow\ }}, \nn \\
    &\mH{1}_{1,2}\overset{\hspace{-6mm}1\;\;\;\;\; 2}{\ket{\ \bzero\ \bzero\ \fbox{$\ast$}\ }} = \overset{\hspace{-6mm}1\;\;\;\;\; 2}{\ket{\ \bplus\ \bminus\ \fbox{$\ast$}\ }} \nn \\
    &\implies H^{(d)}_{\frac{3}{2},\frac{5}{2}}\overset{\frac{3}{2}\;\;\frac{5}{2}}{\ket{\ \downarrow\ \downarrow\ }} = \overset{\frac{3}{2}\;\;\frac{5}{2}}{\ket{\ \uparrow\ \downarrow\ }}, \nn \\
    &\mH{1}_{N-1,N}\overset{\;\;\;\;N - 1\;\;\;N}{\ket{\ \fbox{$\ast$}\ \bplus\ \bminus\ }} = \overset{\;\;\;\;N - 1\;\;\;N}{\ket{\  \fbox{$\ast$}\ \bzero\ \bzero\ }} \nn \\
    &\implies H^{(d)}_{N-\frac{3}{2},N-\frac{1}{2}}\overset{N - \frac{3}{2}\;\;N - \frac{1}{2}}{\ket{\ \downarrow\ \uparrow\ }} = \overset{N - \frac{3}{2}\;\;N - \frac{1}{2}}{\ket{\ \downarrow\ \downarrow\ }}, \nn \\
    &\mH{1}_{N-1,N}\overset{\;\;\;\;N - 1\;\;\;N}{\ket{\ \fbox{$\ast$}\ \bzero\ \bzero\ }} = \overset{\;\;\;\;N - 1\;\;\;N}{\ket{\ \fbox{$\ast$}\ \bplus\ \bminus\ }} \nn \\
    &\implies H^{(d)}_{N-\frac{3}{2},N-\frac{1}{2}}\overset{N - \frac{3}{2}\;\;N - \frac{1}{2}}{\ket{\ \downarrow\ \downarrow\ }} = \overset{N - \frac{3}{2}\;\;N - \frac{1}{2}}{\ket{\ \downarrow\ \uparrow\ }}, \nn \\
\label{eq:effectivespin1b2obc}
\end{eqnarray}
The terms of the dual lattice Hamiltonian in Eqs.~(\ref{eq:effectivespin1b2}) and (\ref{eq:effectivespin1b2obc}) are terms the PXP model, studied in several contexts in the literature\cite{moessner2001ising, fendley2004competing, turner2017quantum} i.e.
\begin{equation}
    H^{(d)} = \twopartdef{\sumal{l = \frac{3}{2}}{N + \frac{1}{2}}{\mathcal{P}_{l-1} \sigma^x_l \mathcal{P}_{l+1}}}{\textrm{PBC}}{\sumal{l = \frac{5}{2}}{N-\frac{3}{2}}{\mathcal{P}_{l-1} \sigma^x_l \mathcal{P}_{l+1}} + \sigma^x_{\frac{3}{2}} \mathcal{P}_{\frac{5}{2}} + \mathcal{P}_{N - \frac{3}{2}} \sigma^x_{N-\frac{1}{2}}}{\textrm{OBC}} 
\label{eq:bosoniceffective}
\end{equation}
where $\sigma^x_l$ is a Pauli matrix on site $l$, and $\mathcal{P}_l$ is a projector on site $l$ on to $\ket{\downarrow}$, i.e.
\begin{equation}
    \mathcal{P}_l \equiv \frac{(1 - \sigma^z_l)}{2}.
\label{eq:njdefn}
\end{equation}
Thus, the pair-hopping Hamiltonian $H$ restricted to the Krylov subspace $\mK{1}$ is exactly the PXP Hamiltonian.
In Sec.~\ref{sec:scars} we will rely on this mapping and show the existence of quantum many-body scars\cite{turner2017quantum, turner2018quantum} in the pair-hopping Hamiltonian $H$.
To easily generalize to other filling factors, even when $p = 1$ we stick to the original spin-1 degrees of freedom language instead of the spin-1/2 degrees of freedom of the PXP model.
The expression of the Hilbert space dimension for OBC is derived in App.~\ref{app:dimension} and the  dimension of the Krylov subspace $\mK{1}$ is shown to be $D^{(1)}_N = F_{N+1}$, where $F_n$ is the $n$-th Fibonacci number.
Thus $\mK{1}$ is isomorphic to the Hilbert space of the PXP model,\cite{turner2017quantum} and $D^{(1)}_N$ thus scales as $\varphi^N$, where the \emph{quantum dimension} $\varphi$ is the Golden ratio $\varphi = \frac{1 + \sqrt{5}}{2}$ (see Eq.~(\ref{eq:lambdap1})). 
\subsection{Filling \boldsymbol{$\nu = p/(2p + 1)$}}\label{sec:pb2p+1}
We now move to the effective Hamiltonians at filling factors $\nu = p/(2p + 1)$ of the original Hamiltonian $H$.
Here we focus on the Krylov subspace $\mK{p} = \mKdef{{\init{p}}}{H}$ defined in Eq.~(\ref{eq:Kpdefn}), where ${\init{p}}$ is defined in Eq.~(\ref{eq:puniformstate}).
To understand the structure of $\mK{p}$, we invoke an important property shown in Ref.~[\onlinecite{wang2012spin}] (see Eq.~(15) and Appendix B therein).
First note that $\init{p}$ of Eq.~(\ref{eq:puniformstate}) is of the form
\begin{equation}
    {\init{p}} = \ket{\cdots \overset{i}{0}(10)^q \cdots}
\label{eq:Rppattern}
\end{equation}
for some $i$, $q$, where $i$, $1 \leq i \leq L$ for PBC (or $1 \leq i \leq L - 2$ for OBC) denotes the position of the orbital, and $q$, $1 \leq q \leq p$ denotes the number of times the pattern ``10" appears consecutively. 
As a consequence of the squeezing property of Eq.~(\ref{eq:pairhoppingrules}), this pattern implies that\cite{wang2012spin}
\begin{equation}
    \sumal{k = i}{i + 2q}{\bra{\psi}\cd_k c_k\ket{\psi}} \geq q\;\; \textrm{for any} \ket{\psi} \in \mK{p}. 
\label{eq:impproperty}
\end{equation}
This property constrains the allowed unit cell configurations in $\mK{p}$. 
For example, for $\nu = 2/5$ ($p = 2$), $\init{2}$ is defined in Eq.~(\ref{eq:12uniformexample}), and all the unit cells configurations read $\fbox{01010}$. 
According to Eq.~(\ref{eq:impproperty}), any unit cell for some $\ket{\psi} \in \mK{2}$ of the form $\fbox{$n_1 n_2 n_3 n_4 n_5$}$ should satisfy $n_1 + n_2 + n_3 \geq 1$ (resp. $n_3 + n_4 + n_5 \geq 1$), which is obtained by choosing $q = 1$ and $i$ in Eq.~(\ref{eq:impproperty}) to be the first (resp. third) site within any unit cell of $\init{2}$ of Eq.~(\ref{eq:puniformstate}). 
From these inequalities, we deduce that the unit cell configurations $\fbox{00011}$ (resp. $\fbox{11000}$) violate Eq.~(\ref{eq:impproperty}), and are thus not allowed for any configuration in $\mK{2}$.
To summarize, we obtain eight allowed unit cell configurations for $p = 2$:
\begin{eqnarray}
    &\fbox{00110}\ \fbox{01100}\ \fbox{10100}\ \fbox{00101}\nn \\
    &\fbox{01010}\ \fbox{10010}\ \fbox{01001}\ \fbox{10001}.
\label{eq:nu25config}
\end{eqnarray}
The unit cell configurations of Eq.~(\ref{eq:nu25config}) can be uniquely mapped onto configurations of two spin-1's by adding one fictitious pseudozero in between any two consecutive (although not necessarily adjacent) 1's in Eq.~(\ref{eq:nu25config}):\cite{wang2012spin}
\begin{equation}
    \begin{tabular}{cc}
        $\fbox{00110} = \fbox{001[0]10} \equiv \fbox{\plus\zero}$ & $\fbox{01100} = \fbox{01[0]100} \equiv \fbox{\zero \minus}$ \\
        $\fbox{10100} = \fbox{10[0]100} \equiv \fbox{\minus\minus}$ & $\fbox{00101} = \fbox{001[0]01} \equiv \fbox{\plus\plus}$ \\
        $\fbox{01010} = \fbox{01[0]010} \equiv \fbox{\zero\zero}$ & $\fbox{10001} = \fbox{100[0]01} \equiv \fbox{\minus\plus}$ \\
        $\fbox{01001} = \fbox{010[0]01} \equiv \fbox{\zero \plus}$ & $\fbox{10010} = \fbox{100[0]10} \equiv \fbox{\minus\zero}$ \\
    \end{tabular}
\label{eq:p2mapping}
\end{equation}
where $+$, $-$ and $\zero$ are spin-1 configurations defined in Eq.~(\ref{eq:spin1dof}) and $[0]$ is the pseudozero.
The addition of pseudozeroes and mapping on to spin-1's can be reversed by deleting a $0$ between the $1$'s within a unit cell after inverting the spin-1 mapping using Eq.~(\ref{eq:spin1dof}).
For example, the configuration $\fbox{\zero+}$ maps onto $\fbox{010001}$ using Eq.~(\ref{eq:spin1dof}), which corresponds to $\fbox{01001}$ since we know one of the $0$'s between the $1$'s is a pseudozero.
Similarly, for general $p$, $z = p - 1$ pseudozeroes are added between two consecutive 1's so that the size of the configuration in each unit cell is $3p$, which can then be mapped on to a configuration of $p$ spin-1's using Eq.~(\ref{eq:spin1dof}).
Such a mapping is one to one as a consequence of Eqs.~(\ref{eq:Rppattern}) and (\ref{eq:impproperty}), and we refer readers to Ref.~[\onlinecite{wang2012spin}] for a complete discussion of this property.
The action of the Hamiltonian $H$ in Eq.~(\ref{eq:pairhopping}) can be written in terms of spin-1 variables using the mapping of Eq.~(\ref{eq:spin1dof}).
Here we show this action separately when the term $H_j$ in Eq.~(\ref{eq:pairhopping}) acts between neighboring unit cells, and when it acts within a unit cell.
When $H_j$ acts between neighboring unit cells,
\begin{eqnarray}
    H_j\overset{j\;\;\;\;\;\;\;\;\;j+3}{\ket{\ \fbox{$\cdots\spa 1\spa0$}\ \fbox{$0\spa1\spa\cdots$}\ }} &=& \overset{j\;\;\;\;\;\;\;\;\;j+3}{\ket{\ \fbox{$\cdots\spa 0\spa 1$}\ \fbox{$1\spa 0\spa \cdots$}\ }} \nn \\
    H_j\overset{j\;\;\;\;\;\;\;\;\;j+3}{\ket{\ \fbox{$\cdots\spa 0\spa 1$}\ \fbox{$1\spa 0\spa \cdots$}\ }} &=& \overset{j\;\;\;\;\;\;\;\;\;j+3}{\ket{\ \fbox{$\cdots\spa 1\spa 0$}\ \fbox{$0\spa 1\spa \cdots$}\ }}.
\label{eq:Hjbetweenaction}
\end{eqnarray}
Using the spin-1 mapping of Eq.~(\ref{eq:spin1dof}), and noting that configurations within a unit cell do not have adjacent 1's (``11") after the addition of pseudozeroes (hence $\fbox{$\cdots 10$}$ and $\fbox{$01\cdots$}$ in Eq.~(\ref{eq:Hjbetweenaction}) respectively read  $\fbox{$\cdots 010$}$ and  $\fbox{$010\cdots$}$ \emph{or} $\fbox{$\cdots[0]10$}$ and $\fbox{$01[0]\cdots$}$ after the addition of pseudozeroes), the action of the effective Hamiltonian reads
\begin{eqnarray}
    \mH{p}_{j,j+1}\overset{j\;\;\;j+1}{\twoket{\cdots \zero}{\zero \cdots}} &=& \overset{j\;\;\;j+1}{\twoket{\cdots +}{- \cdots}} \nn \\
    \mH{p}_{j,j+1}\overset{j\;\;\;j+1}{\twoket{\cdots +}{- \cdots}} &=& \overset{j\;\;\;j+1}{\twoket{\cdots \zero}{\zero \cdots}},
\label{eq:borderunitaction}
\end{eqnarray}
where $j$ is the unit cell index. 
Similarly, when $H_j$ acts within a unit cell, 
\begin{eqnarray}
    H_j \overset{j\;\;\;\;\;\;\;\;\;\;\;\;j+3}{\ket{\ \fbox{$\cdots\spa 0\spa 1\spa [0] \spa 1 \spa 0\spa \cdots$}\ }} &=& \overset{j\;\;\;\;\;\;\;\;\;\;\;\;j+3}{\ket{\ \fbox{$\cdots \spa 1 \spa 0 \spa [0] \spa 0 \spa 1 \spa \cdots$}\ }} \nn \\
    H_j \overset{j\;\;\;\;\;\;\;\;\;\;\;\;j+3}{\ket{\ \fbox{$\cdots \spa 1 \spa 0 \spa [0] \spa 0 \spa 1\spa \cdots $}\ }} &=& \overset{j\;\;\;\;\;\;\;\;\;\;\;\;j+3}{\ket{\ \fbox{$\cdots\spa 0 \spa 1 \spa [0] \spa 1 \spa 0 \spa \cdots$}\ }} \nn \\
\label{eq:pairhoppingrulespseudo}
\end{eqnarray}
Using Eqs.~(\ref{eq:pairhoppingrulespseudo}) and (\ref{eq:spin1dof}), relying once again on the absence of adjacent 1's (``11") after the addition of pseudozeroes, the action in the spin-1 language thus reads
\begin{eqnarray}
    &&\left(\mH{p}_j\right)_{n,n+1}\overset{n\;\;n+1}{\oneket{\cdots \zero + \cdots}} = \overset{n\;\;n+1}{\oneket{\cdots + \zero \cdots}},\nn \\
    &&\left(\mH{p}_j\right)_{n,n+1}\overset{n\;\;n+1}{\oneket{\cdots + \zero \cdots}} = \overset{n\;\;n+1}{\oneket{\cdots \zero + \cdots}}, \nn \\
    &&\left(\mH{p}_j\right)_{n,n+1}\overset{n\;\;n+1}{\oneket{\cdots \zero - \cdots}} = \overset{n\;\;n+1}{\oneket{\cdots - \zero \cdots}},\nn \\
    &&\left(\mH{p}_j\right)_{n,n+1}\overset{n\;\;n+1}{\oneket{\cdots - \zero \cdots}} = \overset{n\;\;n+1}{\oneket{\cdots \zero - \cdots}},
\label{eq:withinunitaction}
\end{eqnarray}
where $j$ is the unit cell index. 
The effective Hamiltonian within $\mK{p}$ can thus be written as\cite{wang2012spin}
\begin{eqnarray}
    &\mH{p} = \sumal{j = 1}{N}{\left[\sumal{n = 1}{p-1}{\left(T^\dagger_{j,n} T_{j, n+1} + U^\dagger_{j,n} U_{j,n+1} + h.c.\right)}\right.} \nn \\
    &\left.- \left(U^\dagger_{j,p} T_{j+1,1} + h.c.\right)\right]
\label{eq:genericfillinghamil}
\end{eqnarray}
where $T_{j,n}$ and $U_{j,n}$ are spin-1 operators defined in Eq.~(\ref{eq:TUdefn}), and their actions are shown in App.~\ref{app:TUproperties}.
Note that Eq.~(\ref{eq:genericfillinghamil}) reduces to Eq.~(\ref{eq:effectivehamil1b3}) when $p = 1$. 
We now describe the Krylov subspace $\Kp$. 
After adding $(p-1)$ pseudozeroes in between two $1$'s within a unit cell (and not between $1$'s in different unit cells), $\init{p}$ of Eq.~(\ref{eq:puniformstate}) reads
\begin{equation}
    \init{p} = \bigotimes_{j = 1}^{N}{\ket{\ \fbox{$(01[0])^{p-1}$010}\ }},
\label{eq:initp}
\end{equation}
where $(01[0])^{p-1}$ indicates that $(01[0])$ is repeated $p - 1$ times. 
Thus, using Eq.~(\ref{eq:spin1dof}), in terms of spin-1 variables ${\init{p}}$ of Eq.~(\ref{eq:initp}) reads
\begin{equation}
    {\init{p}} = \ket{\ \fbox{$\zero\cdots\zero$}\ \cdots\ \fbox{$\zero\cdots\zero$}\ }.
\label{eq:rootstategen}
\end{equation}
As a consequence of the Eqs.~ (\ref{eq:borderunitaction}), (\ref{eq:withinunitaction}) and (\ref{eq:rootstategen}), any configuration of $N$ unit cells $\ket{\vec{\sigma}} = \ket{\sigma_1\sigma_2\cdots\sigma_N}$ in $\mK{p}$ has the following constraints:
\begin{enumerate}
    \item[(c1)] $P_{\sigma_j} = M_{\sigma_{j+1}}$, where $P_{\sigma_j}$ (resp. $M_{\sigma_{j+1}}$) is the number of $+$ (resp. $-$) in the $j$-th (resp. $(j + 1)$-th unit cell): This follows from Eq.~(\ref{eq:borderunitaction}), where starting from ${\init{p}}$ of Eq.~(\ref{eq:rootstategen}), $+$ and $-$ are created together on neighboring unit cells. 
    \item[(c2)] Within each unit cell $-$ appears to the left of $+$: This follows from Eq.~(\ref{eq:borderunitaction}) because starting from $\init{p}$, $+$ and $-$ are always created to the left and right of the unit cells respectively, and they cannot cross each other due to Eq.~(\ref{eq:withinunitaction}). That is, there is no term of the Hamiltonian that allows the process $+- \leftrightarrow -+$ within a unit cell.
    \item[(c3)] With OBC, the leftmost (resp. rightmost) unit cell $\sigma_1$ (resp. $\sigma_N$) cannot have a $-$ (resp. $+$), i.e. $M_{\sigma_1} = 0$ (resp. $P_{\sigma_N} = 0$): This follows from Eq.~(\ref{eq:borderunitaction}), where starting $+$ and $-$ are created together on neighboring unit cells with $+$ in the left unit cell and $-$ in the right unit cell, and thus the leftmost (resp. rightmost) unit cell cannot have a $-$ (resp. $+$). 
\end{enumerate}
The expression for the Hilbert space dimension $D^{(p)}$ of $\Kp$ is obtained for OBC in App.~\ref{app:dimension}.
For general $p$ and large $N$,  $D^{(p)}$ grows with $N$ as $D^{(p)} \sim (\lambda^{(p)})^N$ 
where the \emph{quantum dimension} $\lambda^{(p)}$ is given by  (see Eq.~(\ref{eq:qdimroughscaling}))
\begin{equation}
    \lambda^{(p)} \sim 2^{p-1} \varphi,
\end{equation}
where $\varphi$ is the Golden ratio.
\section{Properties of the Effective Hamiltonians}\label{sec:properties}
\begin{figure}
    \centering
    \includegraphics[scale = 0.43]{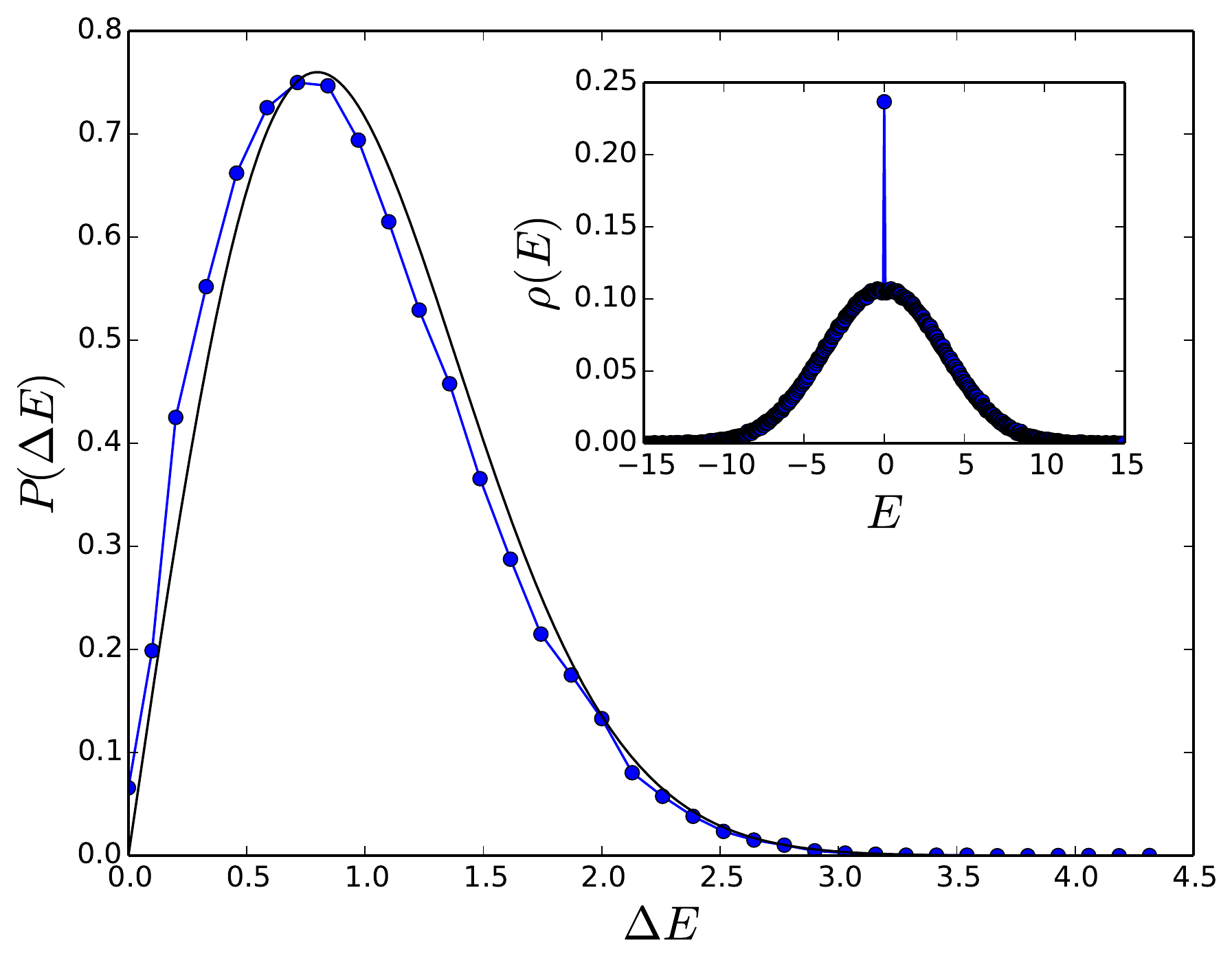}
\caption{Level statistics of the eigenstates of the Hamiltonian $\mH{2}$. The black line shows the expected GOE curve for non-integrable models. The standard parameter $\langle r\rangle \approx 0.5284$, close to the GOE value of $\langle r \rangle \approx 0.5295$.\cite{atas2013distribution} (Inset) Density of states for the eigenstates of the Hamiltonian $\mH{2}$. The peak at $E = 0$ indicates the presence of a large number of zero-modes in an otherwise non-integrable model with a Gaussian density of states. Data is shown for a system with $p = 2$ and $N = 12$ in the quantum number sector $(k, \quant) = (0, +1)$, where $\quant$ is the quantum number corresponding to the symmetry $\mI\mP$ (see Sec.~\ref{sec:symnon}).}
\label{fig:hamilprop}
\end{figure}
We now review some important properties of the Hamiltonians $\mH{p}$ of Eq.~(\ref{eq:genericfillinghamil}).
\subsection{Symmetries and Non-integrability}\label{sec:symnon}
\begin{figure*}[ht]
    \centering
    \includegraphics[scale=0.45]{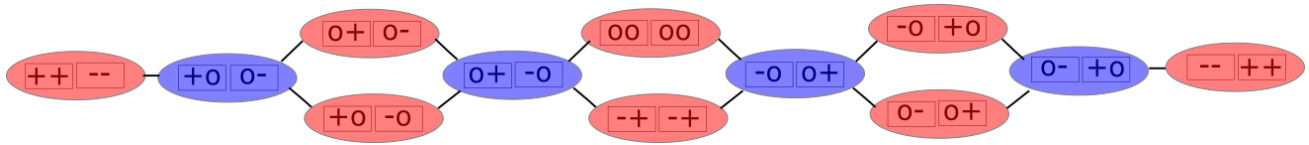}
    \caption{(Color online) The graph $\mG{2}$ for $N = 2$ with PBC. Each node (ellipsoid) corresponds to a many-body basis state of the Krylov subspace $\mK{2}$. A link is drawn between two nodes if there exists a non-zero matrix element in the Hamiltonian that couples the corresponding many-body states. Nearest neighbor nodes have charges $\mQ{2}$ that differ by 1. The red and blue nodes represent the ones with $\mC{2}_{\sigma} = +1$ and $\mC{2}_{\sigma} = -1$ respectively (see Eqs.~(\ref{eq:Z2index}) and (\ref{eq:chargegenp})).}
    \label{fig:graph}
\end{figure*}
We first discuss the symmetries of the Hamiltonian $\mH{p}$.
Consider the transformation of the Hamiltonian $\mH{p}$ under spin flips of spin-1's in the chain, given by the unitary operator
\begin{equation}
    \mP = \prodal{j = 1}{N}{\prodal{n = 1}{p}{\exp\left(i \pi S^x_{j,n}\right)}}.
\label{eq:spinflip}
\end{equation}
Since $\mP S^{\pm}_{j,n} \mP^\dagger =  S^{\mp}_{j,n}$ and $\mP S^z_{j,n} \mP^\dagger = -S^z_{j,n}$, the operators $T_{j,n}$ and $U_{j,n}$ of Eq.~(\ref{eq:TUdefn}) transform as
\begin{equation}
    \mP T_{j,n} \mP^\dagger = - U^\dagger_{j,n},\;\;\; \mP U_{j,n} \mP^\dagger = - T^\dagger_{j,n}.
\label{eq:spinflipaction}
\end{equation}
Thus, under spin-flips the Hamiltonian $\mH{p}$ transforms to 
\begin{eqnarray}
    &\mP \mH{p} \mP^\dagger = \sumal{j = 1}{N}{\left[\sumal{n = 1}{p-1}{\left(U_{j,n} U^\dagger_{j, n+1} + T_{j,n} T^\dagger_{j,n+1} + h.c.\right)}\right.} \nn \\
    &\left.- \left(T_{j,p} U^\dagger_{j+1,1} + h.c.\right)\right]
\label{eq:spinfliphamiltransform}
\end{eqnarray}
Further, we consider inversion symmetry $\mI$, which transforms operators $T_{j,n}$ and $U_{j,n}$ as
\begin{equation}
    \mI T_{j,n} \mI^\dagger = T_{N + 1 - j, p + 1 - n},\;\;\; \mI U_{j,n} \mI^\dagger = U_{N + 1 - j, p + 1 - n}.
\label{eq:inversionaction}
\end{equation}
Acting $\mI$ and $\mI^\dagger$ on the left and right of Eq.~(\ref{eq:spinfliphamiltransform}), after rearranging the sum we obtain
\begin{eqnarray}
    &\mI \mP \mH{p} \mP^\dagger\mI^\dagger =\sumal{j = 1}{N}{\left[\sumal{n = 1}{p-1}{\left(T^\dagger_{j,n} T_{j, n+1} + U^\dagger_{j,n} U_{j,n+1} + h.c.\right)}\right.} \nn \\
    &\left.- \left(U^\dagger_{j,p} T_{j+1,1} + h.c.\right)\right] \nn \\
    &= \mH{p}
\label{eq:Hpsymmetry}
\end{eqnarray}
Thus, the Hamiltonian $\mH{p}$ has a $\mathbb{Z}_2$ symmetry generated by the unitary $\mI\mP$.
We denote its quantum numbers by $\quant = \pm 1$.
The $\mI\mP$ symmetry of the Hamiltonians $\Hp$ is the same as inversion symmetry of the pair-hopping Hamiltonian $H$. 
In the original language, this symmetry is the inversion symmetry of the pair hopping Hamiltonian $H$ of Eq.~(\ref{eq:pairhopping}). 
We illustrate this with a simple example with $N = 2$ and $p = 2$. 
Under the operator $\mI\mP$, the configuration $\twoket{\zero +}{\zero -}$ transforms to $\twoket{+\zero}{-\zero}$. 
Using the mapping of Eq.~(\ref{eq:p2mapping}), in the original language these states read $\twoket{010[0]01}{01[0]100}$ (resp. $\twoket{01001}{01100}$) and $\twoket{001[0]10}{10[0]010}$ (resp. $\twoket{00110}{10010}$) with (resp. without) the pseudozeroes, which are related by the action of inversion. 
With PBC, $\mH{p}$ is also trivially invariant under a translation by one unit cell.
Thus it can be block-diagonalized into $N$ blocks labelled by momenta $\{k = 2\pi j/N\}, 0 \leq j \leq N - 1$. 
However, the translation operator $\mathcal{T}$ and $\mathcal{I}$ (hence $\mI\mathcal{P}$) do not commute unless $k = 0$ or $k = \pi$.
The Hamiltonians $\mH{p}$ are non-integrable. 
The characteristic property of non-integrable models is the appearance of Wigner-Dyson energy level statistics within a given quantum number sector. 
When $p = 1$, $\mH{1}$ can be exactly mapped on to the PXP model as discussed in Sec.~\ref{sec:1b3}, and the Wigner-Dyson level statistics of the PXP model was observed in Ref.~[\onlinecite{turner2017quantum}].
We find similar level statistics for all the quantum number sectors of the Hamiltonian $\mH{p}$ for $p \leq 3$ up to the system sizes we are able to study numerically. 
In Fig.~\ref{fig:hamilprop}, we show the level statistics in the $(k,\quant) = (0, +1)$ sector of the Hamiltonian $\mH{2}$ for a system with $N = 12$ unit cells, where $\quant$ is the quantum number corresponding to the $\mI\mP$ symmetry.
\subsection{Charge Operators}
To unravel the properties relevant for many-body scars, we map $\mH{p}$ onto a single particle hopping on a graph $\Gp$, where each node of the graph represents a product configuration $\ket{\vec{\sigma}} \in \mK{p}$.
This idea was employed to study the PXP model in Refs.~[\onlinecite{turner2017quantum}] and [\onlinecite{turner2018quantum}].
The links of the graph indicate the non-vanishing matrix elements of the Hamiltonian between the corresponding node configurations. 
For example, the graph $\mG{2}$ for $N = 2$ with PBC is shown in Fig.~\ref{fig:graph}.
To better understand the structure of the graph $\Gp$, it is useful to define a charge $\mQ{p}_{\vec{\sigma}}$ associated with each node (each configuration $\ket{\vec{\sigma}}$.
We start with $p = 1$ as an example. Here, we define a charge $\mQ{1}_{\vec{\sigma}}$ as   
\begin{equation}
    \mQ{1}_{\vec{\sigma}} \equiv \sumal{j = 1}{N_b}{(-1)^{j}\left(\frac{P_{\sigma_{j}} + M_{\sigma_{j+1}}}{2}\right)}.
\label{eq:chargep1}
\end{equation}
Note that $P_{\sigma_j} = M_{\sigma_{j+1}}$ according to the constraint (c1) in Sec.~\ref{sec:effhamil}.
From Eq.~(\ref{eq:1b3scattering}), we deduce that the action of each term in the Hamiltonian changes the number of $+$ spins in the $j$-th unit cell and the number of $-$ spins in the $(j+1)$-th unit cell by 1. 
That is, for product states $\ket{\vec{\sigma}}$ and $\ket{\vec{\tau}}$, the following holds for a single value of $j^\ast$, $1 \leq j^\ast \leq N_b$ (since $p = 1$, $P_{\sigma_j}, M_{\sigma_j} \in \{0, 1\}\;\forall \;\sigma_j$):
\begin{equation}
    \mH{1}_{j^\ast, j^\ast+1} \ket{\vec{\sigma}} = \ket{\vec{\tau}} \implies P_{\tau_{j^\ast}} = 1 - P_{\sigma_{j^\ast}}, M_{\tau_{j^\ast+1}} = 1 - M_{\sigma_{j^\ast+1}}.
\label{eq:hamilp1config}
\end{equation}
As shown in App.~\ref{app:symmetries} using Eqs.~(\ref{eq:chargep1}) and (\ref{eq:hamilp1config}), the Hamiltonian $\mH{1}$ can be split into two parts as
\begin{equation}
    \mH{1} = \mH{1}_+ + \mH{1}_-,
\label{eq:hamilsplittext}
\end{equation}
where all the basis states $\ket{\vec{\tau}}$ that appear in $\mH{1}_+\ket{\vec{\sigma}}$ (resp. $\mH{1}_-\ket{\vec{\sigma}}$) satisfy $\mQ{1}_{\vec{\tau}} = \mQ{1}_{\vec{\sigma}} + 1$ (resp. $\mQ{1}_{\vec{\tau}} = \mQ{1}_{\vec{\sigma}} - 1$).
Thus, $\mH{1}$ can be written as a sum of charge-raising and charge-lowering operators $\mH{1}_+$ and $\mH{1}_-$ respectively. 
This splitting will be useful when we discuss quantum many-body scars in Sec.~\ref{sec:scars}.
All of the structure of $\mH{1}$ described in the previous paragraph generalizes to any $p$. 
We now define charges $\mQ{p}_{\vec{\sigma}}$ as  
\begin{equation}
    \Qp_{\vec{\sigma}} \equiv \sumal{j = 1}{N_b}{(-1)^{j+1} \left(\frac{P_{\sigma_j} + M_{\sigma_{j+1}}}{2} - \left(X^{(P)}_{\sigma_j} + X^{(M)}_{\sigma_{j+1}}\right)\right)},
\label{eq:chargegenp}
\end{equation}
where $P_{\sigma_j}$, $M_{\sigma_j}$, $X^{(P)}_{\sigma_j}$, and $X^{(M)}_{\sigma_j}$ are defined in Eq.~(\ref{eq:Xdefn}).
Note that $P_{\sigma_j} = M_{\sigma_{j+1}}$ according to the constraint (c1) in Sec.~\ref{sec:pb2p+1}.
For example, when $p = 3$ and $N = 2$ consider the configuration $\twoket{\zero + +}{- \zero -}$ with PBC.
Here, $\sigma_1 = \fbox{$\zero + +$}$ and $\sigma_2 = \fbox{$- \zero -$}$.
Using Eq.~(\ref{eq:Xdefn}), we obtain $P_{\sigma_1} =  2$, $M_{\sigma_1} =  0$, $X^{(P)}_{\sigma_1} = 3$, $X^{(M)}_{\sigma_1} = 0$, and $P_{\sigma_2} =  0$, $M_{\sigma_2} =  2$, $X^{(P)}_{\sigma_2} = 0$, $X^{(M)}_{\sigma_2} = 4$.
Thus the charge of this configuration is $\Qp_{\sigma} = 5$. 
Note that when $p = 1$, $X^{(P)}_{\sigma_j} = P_{\sigma_j}$ and $X^{(M)}_{\sigma_j} = M_{\sigma_j}$, and thus Eq.~(\ref{eq:chargegenp}) reduces to Eq.~(\ref{eq:chargep1}).
When the charge is defined as in Eq.~(\ref{eq:chargegenp}), as shown in App.~\ref{app:symmetries} we can in fact write the Hamiltonian $\mH{p}$ as
\begin{equation}
    \mH{p} = \mH{p}_+ + \mH{p}_-,
\label{eq:Hpsplit}
\end{equation}
where for product configurations $\ket{\vec{\sigma}}$ and $\ket{\vec{\tau}}$,
\begin{equation}
    \mH{p}_{\pm} \ket{\vec{\sigma}} = \ket{\vec{\tau}} + \cdots \implies \Qp_{\vec{\tau}} = \Qp_{\vec{\sigma}} \pm 1. 
\label{eq:Hpsplitprop}
\end{equation}
\subsection{Zero-modes}\label{sec:zeromodes}
The PXP model is known to exhibit exponentially many zero energy eigenstates (i.e. $E = 0$) at the center of its spectrum.\cite{turner2017quantum, schecter2018many, turner2018quantum}
We show that the Hamiltonians $\mH{p}$ share the same feature.
For example, the inset of Fig.~\ref{fig:hamilprop} shows the density of states of the Hamiltonian $\mH{2}$, which clearly exhibits a sharp peak at $E = 0$. 
Their origin can be traced to the structure of the graphs $\mG{p}$ introduced in the previous section. 
From Eqs.~(\ref{eq:Hpsplit}) and (\ref{eq:Hpsplitprop}), adjacent nodes of the graph $\Gp$, corresponding to product configurations $\ket{\vec{\sigma}}$ and $\ket{\vec{\tau}}$, have charges that differ by 1.
Thus, we define a $\mathbb{Z}_2$ index for each node of the $\Gp$ as
\begin{equation}
    \Cp_{\vec{\sigma}} \equiv (-1)^{\Qp_{\vec{\sigma}}},
\label{eq:Z2index}
\end{equation}
such that neighboring nodes have different indices.
Hence the graph $\Gp$ is bipartite. We henceforth refer to the sublattices with $\Cp = +1$ and $\Cp = -1$ as \emph{even} and \emph{odd} sublattices respectively.   
Single-particle zero energy eigenstates of fermions hopping on a bipartite lattice have been known for long.\cite{sutherland1986localization}
If the number of nodes on the even and odd sublattices are $N_{e}$ and $N_{o}$ respectively, a lower-bound for the number of zero-energy eigenstates $Z$ for any hopping Hamiltonian on a bipartite graph $\Gp$ is given by\cite{sutherland1986localization, inui1994unusual} 
\begin{equation}
	Z \geq |N_{e} - N_{o}|.
\label{eq:zerobound}
\end{equation}
However, for the Hamiltonians $\Hp$, applying the bound of Eq.~(\ref{eq:zerobound}) on $\Gp$ alone does not provide the best lower bound on the number of zero modes, even in the case of the PXP model.\cite{schecter2018many, turner2018quantum} 
The symmetry $\mI \mP$ of Eq.~(\ref{eq:Hpsymmetry}) can be used to construct graphs $\Gp_+$ and $\Gp_-$ for the quantum number sectors $\quant = +1$ and $\quant = -1$. 
The nodes of these graphs $\Gp_+$ and $\Gp_-$ are no longer product states, but they are respectively $\{\ket{\vec{\sigma}_+}\}$ and $\{\ket{\vec{\sigma}_-}\}$, symmetric and antisymmetric superpositions of $\ket{\vec{\sigma}}$ and $\mI\mP\ket{\vec{\sigma}}$.
That is,
\begin{equation}
    \ket{\vec{\sigma}_+} = \frac{\ket{\vec{\sigma}} + \mI\mP\ket{\vec{\sigma}}}{\mathcal{N}},\;\;\; \ket{\vec{\sigma}_-} = \frac{\ket{\vec{\sigma}} - \mI\mP\ket{\vec{\sigma}}}{\sqrt{2}},
\label{eq:symantisymsup}
\end{equation}
where $\mathcal{N} = \sqrt{2}$ if $\ket{\vec{\sigma}} \neq \mI\mP\ket{\vec{\sigma}}$, and $\mathcal{N} = 2$ when $\ket{\vec{\sigma}} = \mI\mP\ket{\vec{\sigma}}$ (i.e. when $\ket{\vec{\sigma_-}} = 0$). 
Since the Hamiltonian $\Hp$ is $\mI\mP$-symmetric, if the nodes corresponding to $\ket{\vec{\sigma}}$ and $\ket{\vec{\tau}}$ are connected in graph $\Gp$, the nodes corresponding to $\ket{\vec{\sigma}_+}$ and $\ket{\vec{\tau}_+}$ are connected in the graph $\Gp_+$.
Similarly, the nodes corresponding to $\ket{\vec{\sigma}_-}$ and $\ket{\vec{\tau}_-}$ are connected in $\Gp_-$ unless either of them vanish, which happens when $\ket{\vec{\sigma}} = \mI\mP\ket{\vec{\sigma}}$. 

\begin{figure*}
\centering
\begin{tabular}{ccc}
 \includegraphics[scale = 0.31]{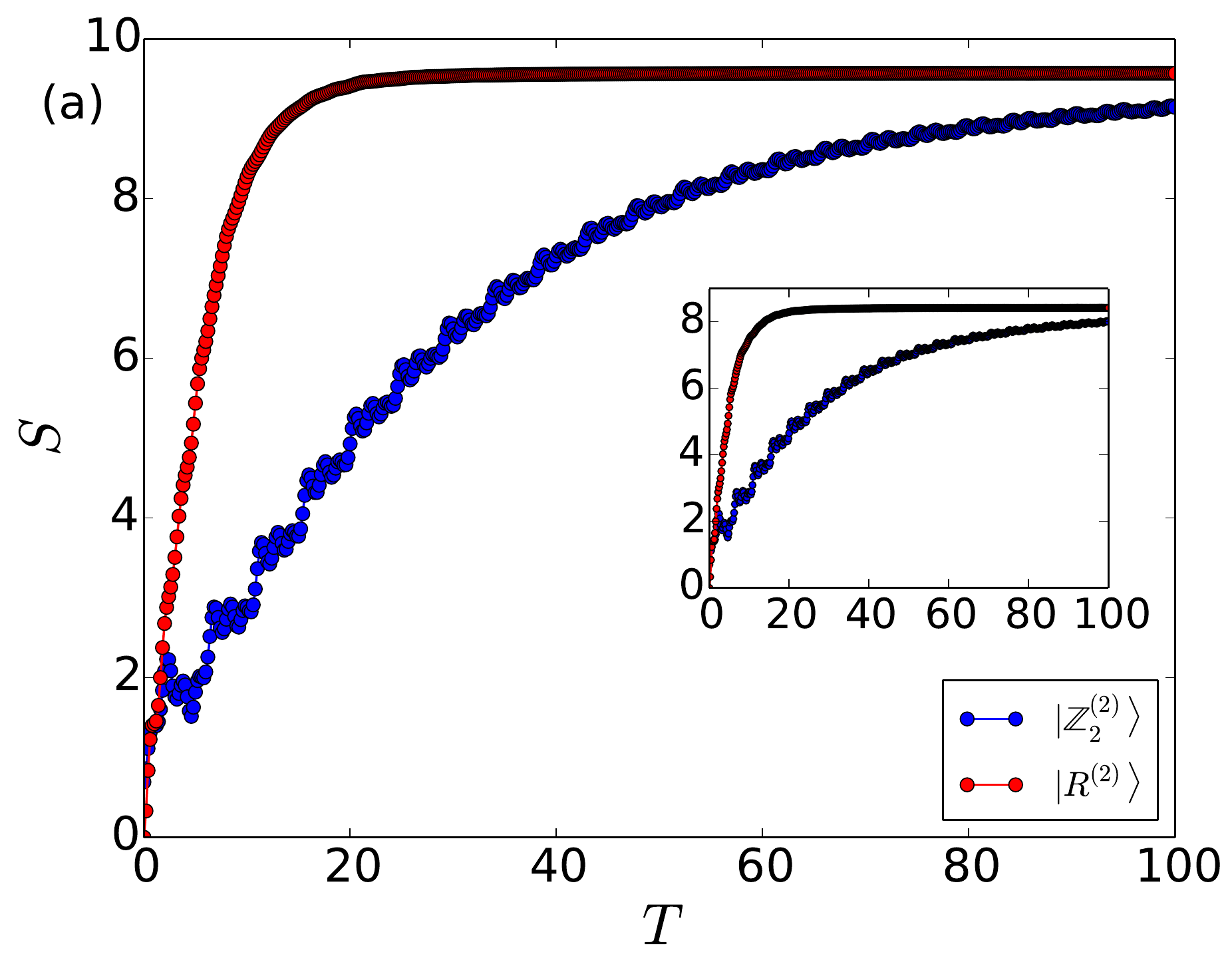}&\includegraphics[scale=0.31]{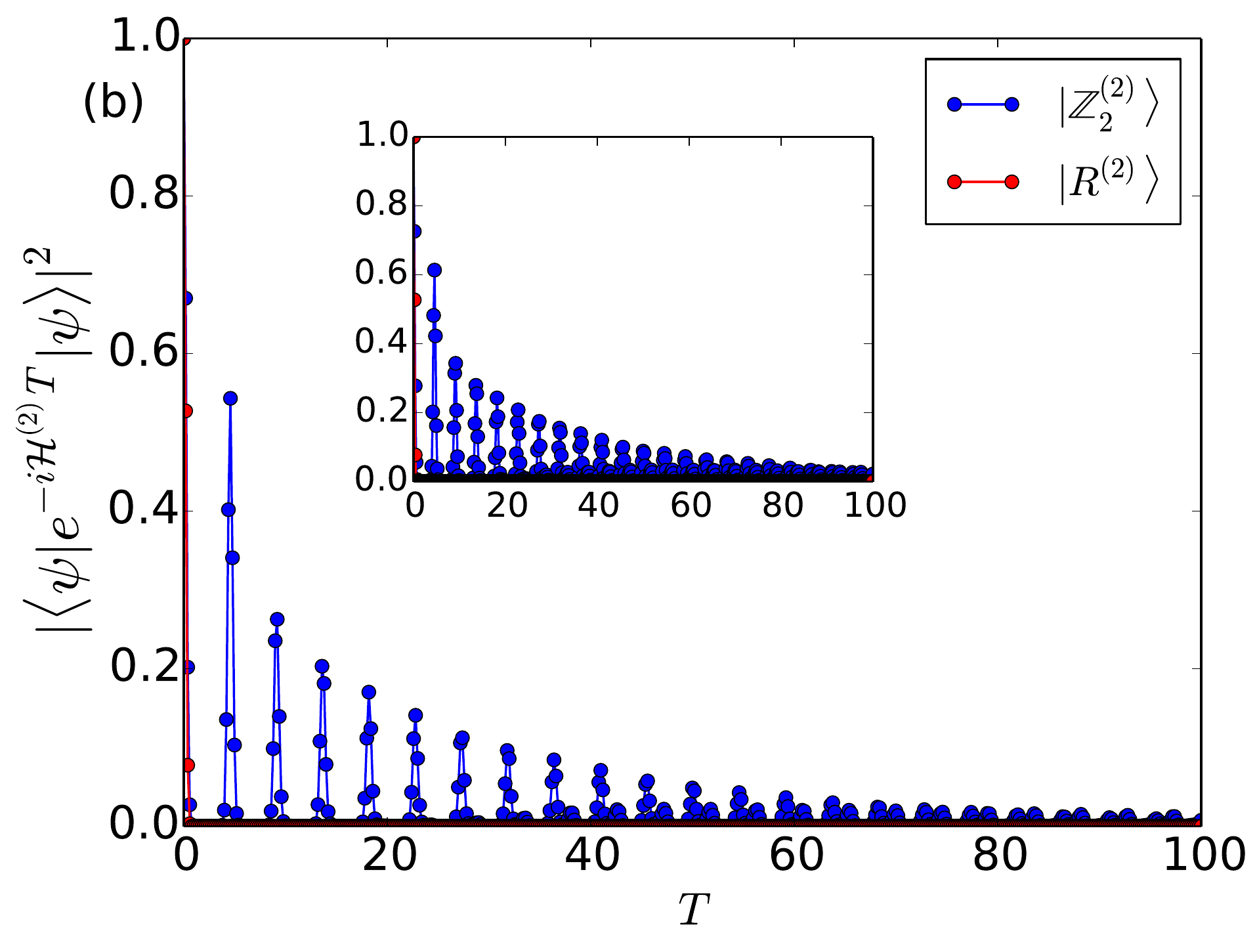}&\includegraphics[scale=0.31]{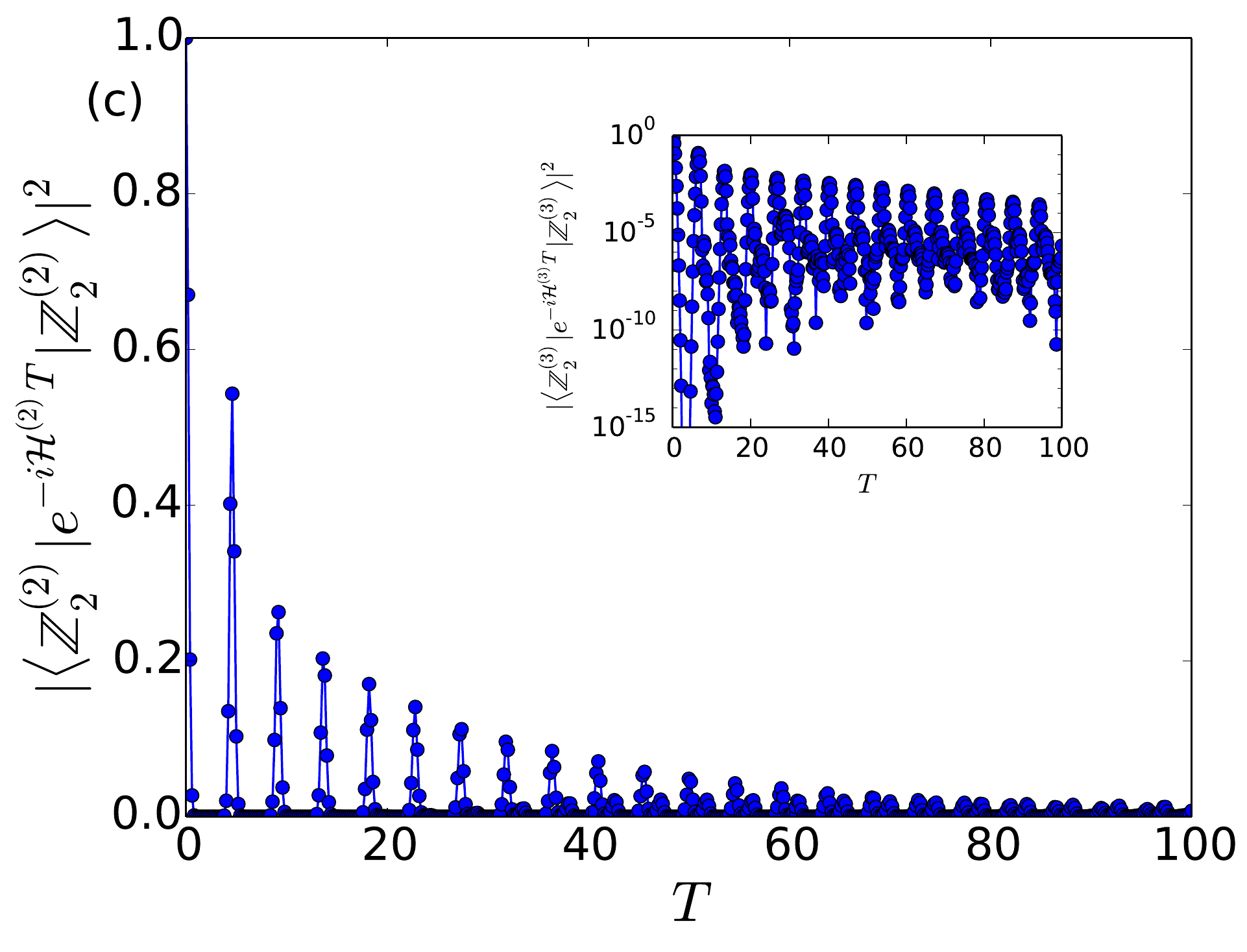}
\end{tabular}
\caption{(Color online) Time evolution of (a) entanglement entropies, and (b) fidelities of the $\ztwo{2}$ and $\init{2}$ states  (with $k = 0$) under the Hamiltonian $\mH{2}$. Data is shown for PBC in the quantum number sector $(k,\quant) = (0, +1)$ for $N = 16$ (insets display the same quantities for $N = 14$). (c) The fidelity for the evolution of $\ztwo{2}$ state (with $k = 0$) for $N = 20$. The inset shows the fidelity for $p = 3$ of the $\ket{\mathbb{Z}^{(3)}_2}$ state (with $k = 0$) for $N = 12$ (for clarity, we have used a log scale for the fidelity). While the entanglement entropies are not accessible, we observe that this larger size still exhibits clear revivals.}
\label{fig:timevolscars}
\end{figure*}

In App.~\ref{sec:chargesym} we discuss the transformation of the $\Qp_{\vec{\sigma}}$ under the symmetry $\mI\mP$ and in Eq.~(\ref{eq:Z2symtransform}) we show that the $\mathbb{Z}_2$ index of Eq.~(\ref{eq:Z2index}) is invariant under the $\mI\mP$ symmetry.
Thus, the configuration $\mI\mP \ket{\vec{\sigma}}$ has the same $\mathbb{Z}_2$ index $\Cp_{\vec{\sigma}}$ as the configuration $\ket{\vec{\sigma}}$.
Thus the $\mathbb{Z}_2$ indices for the symmetric (resp. antisymmetric) superposition $\ket{\vec{\sigma}_+}$ (resp. $\ket{\vec{\sigma}_-}$) of Eq.~(\ref{eq:symantisymsup}) on graph $\Gp_+$ (resp. $\Gp_-$) can be defined to be $\Cp_{\vec{\sigma}}$. 
Eq.~(\ref{eq:zerobound}) can then be applied separately to $\Gp_+$ and $\Gp_-$ and then summed over both sectors to obtain a stronger lower-bound on the total number of zero modes:
\begin{equation}
	Z  \geq \sumal{c \in \{+,-\}}{}{|N^c_{e} - N^c_{o}|},
\label{eq:zeroboundcomp} 
\end{equation}
where $N^+_e$ (resp. $N^-_e$) and $N^+_o$ (resp. $N^-_o$) are the number of nodes of the graph $\Gp_+$ (resp. $\Gp_-$) on the even and odd sublattices respectively. 
To obtain a computable lower-bound on the zero-modes, we use an important observation.
The structure of $\Gp_+$ and $\Gp_-$ are identical except for symmetric product states that satisfy
\begin{equation}
    \ket{\vec{\sigma}} = \mI\mP\ket{\vec{\sigma}}.
\label{eq:productinvariant}
\end{equation}
Indeed, such product states always form the nodes of $\Gp_+$ whereas they do not appear in $\Gp_-$. 
If we denote the number of symmetric product states (satisfying Eq.~(\ref{eq:productinvariant})) on the even (resp. odd) sublattices of $\Gp_+$ as $N^{p}_e$ (resp. $N^p_o$), the bound of Eq.~(\ref{eq:zeroboundcomp}) can be written as
\begin{eqnarray}
    Z &\geq& |N^+_e - N^+_o| + |N^-_e - N^-_o| \nn \\
    & = & |N^-_e + N^p_e - N^-_o - N^p_o| +  |N^-_e - N^-_o| \nn \\
    &\geq& |N^p_e - N^p_o|. 
\label{eq:zerobounduseful}
\end{eqnarray}
Thus, to obtain a lower-bound on the number of zero-modes, it is sufficient to study the product states that satisfy Eq.~(\ref{eq:productinvariant}) (i.e. symmetric product states).
Such states can be uniquely determined by the configuration of half of the chain, and as we show in App.~\ref{app:zeromodes}, we expect the number of zero modes of $\Hp$ to scale as $\sqrt{D^{(p)}_N}$, where $D^{(p)}_N$ is the Hilbert space dimension of $\Kp$.
While we believe a detailed counting of zero modes in $\Kp$ can be done using the machinery introduced in App.~\ref{app:dimension} or using the methods of Ref.~[\onlinecite{schecter2018many}], do not pursue this calculation in this work.
\section{Quantum Many-body Scars}\label{sec:scars}
We now discuss the fate of quantum many-body scars in the Hamiltonians $\Hp$.
We first discuss the case of $\mH{1}$, which maps on to the PXP model. 
In the PXP model, the anomalous dynamics of the N\'eel state was studied,\cite{turner2017quantum, turner2018quantum} which reads (for PBC and even system size)
\begin{equation}
    \ztwo{\textrm{PXP}} = \ket{\ \uparrow\ \downarrow\ \uparrow\ \downarrow\ \cdots\ \uparrow\ \downarrow\ }. 
\label{eq:Z2pxp}
\end{equation}
In particular, the entanglement growth of the $\ket{\mathbb{Z}^{\textrm{PXP}}_2}$ state for the PXP model shows oscillations about a sub-thermal value in spite of the Wigner-Dyson level statistics and thus the non-integrability of the PXP model.\cite{turner2017quantum}
This anomalous behavior was explained by the existence of eigenstates in the PXP Hamiltonian that have a subthermal entanglement entropy and a anomalously large overlap with the $\ztwo{\textrm{PXP}}$ state for any finite system size.
Such states were then approximated using the so-called Forward Scattering Approximation (FSA),\cite{turner2017quantum} which we elaborate below for the Hamiltonians $\Hp$. 
Before we move on to general $p$, we translate the scar physics of the PXP model in terms of the Hamiltonian $\mH{1}$ and the Krylov subspace $\mK{1}$, allowing a direct generalization to arbitrary $p$. 
Using the mapping of Eq.~(\ref{eq:dualreversemapping}), we map the $\ztwo{\textrm{PXP}}$ state of Eq.~(\ref{eq:Z2pxp}) on to the $\ztwo{1}$ state of the constrained Hilbert space $\mK{1}$ defined in Eq.~(\ref{eq:Kpdefn}), which thus reads (for PBC and even $N$)
\begin{eqnarray}
    &&\ztwo{1} = \ket{\ \bplus\ \bminus\ \bplus\ \cdots\ \bminus\ \bplus\ \bminus\ }.
\label{eq:Z2p1}
\end{eqnarray}
In the orbital occupation basis of $H$, using the mapping of Eq.~(\ref{eq:spin1dof}), they are density-wave configurations of the form
\begin{eqnarray}
    &&\ztwo{1} = \ket{\ \fbox{001}\ \fbox{100}\ \fbox{001}\ \cdots\ \fbox{100}\ \fbox{001}\ \fbox{100}\ },\nn \\
\end{eqnarray}
which is the ``maximally squeezed state"\cite{bernevig2008model, bernevig2009anatomy} at $\nu = 1/3$ in the quantum Hall language, i.e. the configuration that cannot be ``squeezed" further but it can be ``antisqueezed" (see Eq.~(\ref{eq:pairhoppingrules})). (Note that in the quantum Hall case, the presence of longer range squeezing terms leads to a different maximally squeezed configuration).   
As shown in App.~\ref{app:charge} (see Eqs.~(\ref{eq:chargez2pbc}) and (\ref{eq:chargez2obc})) the $\ztwo{1}$ state of Eq.~(\ref{eq:Z2p1}) is the state in $\mK{1}$, with the lowest charge $\mQ{1}_{\mathbb{Z}_2} = -N/2$. 
Since the PXP Hamiltonian maps on to $\mH{1}$, the anomalous dynamics of the $\ztwo{\textrm{PXP}}$ state of Eq.~(\ref{eq:Z2pxp}) for the PXP model thus maps on to the dynamics of the $\ztwo{1}$ state of Eq.~(\ref{eq:Z2p1}) for the Hamiltonian $\mH{1}$ in Eq.~(\ref{eq:effectivehamil1b3}).
Thus, generalizing to $p > 1$, we conjecture that the Hamiltonian $\Hp$ shows anomalous dynamics for the lowest charge states in $\Kp$, which are the $\ztwo{p}$ states that read (see App.~\ref{app:chargegenp} for their derivation)
\begin{equation}
    \ztwo{p} = \ket{\ \fbox{$+\cdots+$}\ \fbox{$- \cdots-$}\ \cdots\ \fbox{$+\cdots+$}\ \fbox{$- \cdots-$}}.
\label{eq:Z2genp}
\end{equation}
When written in the orbital basis without the pseudozeroes ($[0]$'s), these are density wave configurations are the ``maximally squeezed states" at $\nu = p/(2p + 1)$, and they read 
\begin{eqnarray}
    &\ztwo{p} = \nn \\
    &\ket{\ \fbox{$00101\cdots01$}\ \fbox{$10\cdots10100$}\ \cdots\ \fbox{$00101\cdots01$}\ \fbox{$10\cdots10100$}}\nn \\
\end{eqnarray}

\begin{figure*}
\centering
\begin{tabular}{cc}
    \includegraphics[scale = 0.43]{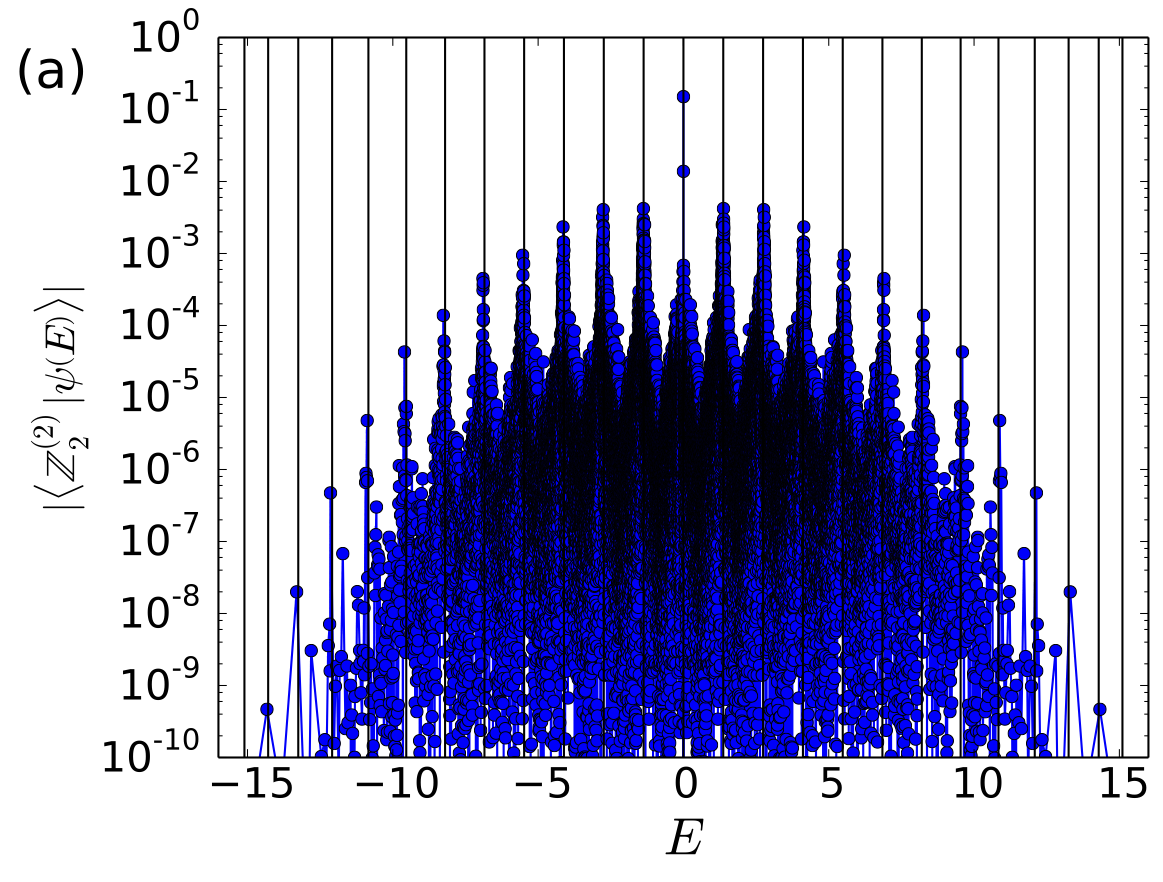}&\includegraphics[scale=0.43]{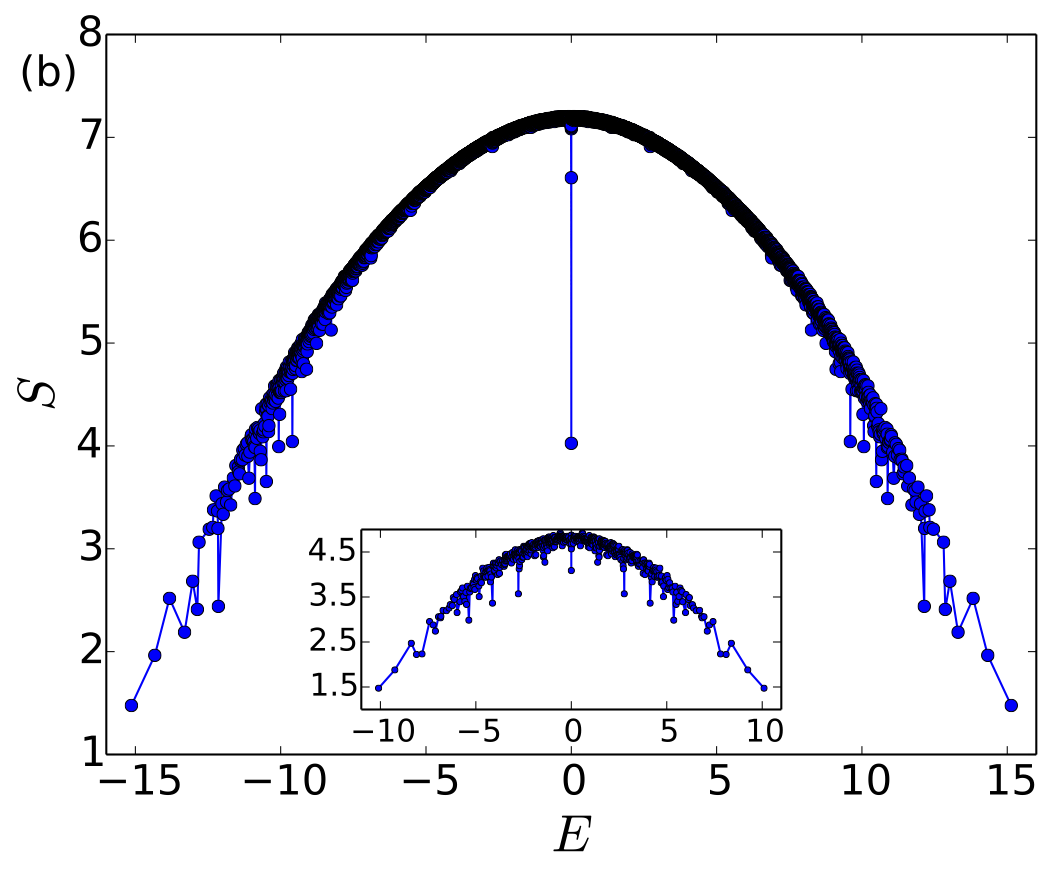}
\end{tabular}
\caption{(a) Overlap of the $\ket{\mathbb{Z}^{(2)}_2\left(k = 0\right)}$ state with the eigenstates $\ket{\psi(E)}$ of the Hamiltonian. The vertical lines represent the energies of the scars as predicted by the FSA (i.e. eigenvalues of $\mH{2}_{\textrm{FSA}}$, see Eq.~(\ref{eq:H1FSA})) (b) Entanglement entropy of the eigenstates of $\mH{2}$ for $N = 12$ and (inset shows the same quantity for $N = 8$). Note the strong hybridization of most of the outlier eigenstates (scars) with the rest of the spectrum with increasing system size. Note that the low entanglement state at $E = 0$ is perhaps a consequence of the fact that there are exponentially many states with $E = 0$ (see Sec.~\ref{sec:zeromodes}). Data shown for PBC in the quantum number sector $(k, X) = (0, +1)$.}
\label{fig:scars}
\end{figure*}

We now demonstrate the FSA for $\Hp$. Note that we directly work with general $p$, and this analysis reduces to that of the PXP model in Refs.~[\onlinecite{turner2017quantum}] and [\onlinecite{turner2018quantum}] by setting $p = 1$ and using the mapping of Eq.~(\ref{eq:duallatticemapping}).
We first construct the Krylov subspace $\mK{p}_+$ defined as 
\begin{equation}
	\mK{p}_+ \equiv \mKdef{\ztwo{p}}{\mH{p}_+}, 
\label{eq:FSAkrylov}
\end{equation}
where $\Hp_+$ is the charge-raising part of the Hamiltonian, shown in Eq.~(\ref{eq:Hpsplit}) and illustrated in App.~\ref{app:symmetries}, and $\Hp_-\ket{\mathbb{Z}_2} = 0$.
The basis vectors of $\mK{p}_+$ are 
\begin{equation}
    \ket{\fsa{p}_j} \equiv \frac{1}{\sqrt{\norm{p}_j}}\left(\mH{p}_+\right)^j\ztwo{p},\;\; j \geq 0, 
\label{eq:fsabasis}
\end{equation}
where $\norm{p}_j$ is a normalization factor. The $\ket{\fsa{p}_j}$'s are all guaranteed to be orthogonal since they have different charges. Indeed, $\ket{\fsa{p}_j}$ has a charge $\Qp = -N p^2/2 + j$ because $\mH{p}_+$ is a charge raising operator.
Furthermore, as discussed in App.~\ref{app:charge} (see Eqs.~(\ref{eq:chargez2pbc}) and (\ref{eq:chargez2obc})), the highest charge configuration in $\mK{p}$ is the configuration of Eq.~(\ref{eq:Z2genp}) translated by one unit cell, and it has a charge $\mQ{p} = Np^2/2$ (resp. $\mQ{p} = Np^2/2 - p$) for PBC (resp. OBC), and thus $\mK{p}_+$ is a Hilbert space of dimension $D^{(p)}_+ = (N p^2 +1)$ (resp. $D^{(p)}_+ = N p^2 + 1 - p$).
The FSA is an approximation that $\mK{p}_+$ is closed under the action of the total Hamiltonian $\mH{p}$.
Since $\mH{p}$ is of the form of Eq.~(\ref{eq:Hpsplit}) (i.e. a sum of charge raising and lowering operators), using Eq.~(\ref{eq:fsabasis}) we obtain
\begin{equation}
    \mH{p}\ket{\fsa{p}_j} = \tridiag{p}_{j+1}\ket{\fsa{p}_{j+1}} + \Hp_-\ket{\fsa{p}_{j}},
\label{eq:Hpactionfsa}
\end{equation}
where $\tridiag{p}_j = \sqrt{\norm{p}_j/\norm{p}_{j-1}}$, where $\norm{p}$ is the normalization factor defined in Eq.~(\ref{eq:fsabasis}). 
The crucial approximation of the FSA is thus
\begin{equation}
    \Hp_-\ket{\fsa{p}_j} \approx \tridiag{p}_j\ket{\fsa{p}_{j-1}}.
\label{eq:fsaapprox}
\end{equation}
While the approximation of Eq.~(\ref{eq:fsaapprox}) is only justified because of matching the charge $\Qp$, we will show that this assumption leads to accurate predictions of the energies of the quantum scars in this model. 
Thus, using Eqs.~(\ref{eq:Hpactionfsa}) and (\ref{eq:fsaapprox}), the Hamiltonian $\mH{p}$ restricted to $\mK{p}_+$ is a $(D^{(p)}_+ - 1)$-dimensional tridiagonal matrix in the FSA approximation that reads
\begin{equation}
    \mH{p}_{\textrm{FSA}} = 
    \begin{bmatrix}
        0 & \tridiag{p}_1 & 0 & \cdots & \cdots & 0 \\
        \tridiag{p}_1 & 0 & \tridiag{p}_2 & 0 & \ddots & \vdots \\
        0 & \tridiag{p}_2 & \ddots & \ddots & \ddots & \vdots \\
        \vdots & \ddots & \ddots & \ddots & \ddots & 0 \\
        \vdots & \ddots & \ddots & \ddots & 0 & \tridiag{p}_{D^{(p)}_+ - 1} \\
        0 & \cdots & \cdots & 0 & \tridiag{p}_{D^{(p)}_+ - 1} & 0 \\
    \end{bmatrix}.
\label{eq:H1FSA}
\end{equation}
Indeed, the eigenstates of FSA Hamiltonian are known to reproduce the energies of the scarred eigenstates of the PXP model to a good approximation.\cite{turner2017quantum, turner2018quantum} These results are equivalent to those for the Hamiltonian $\mH{1}$.
We thus expect that the eigenstates of $\Hp_{\textrm{FSA}}$ to be close to the quantum scars of $\Hp$. 

\begin{figure*}
\centering
\begin{tabular}{cc}
\includegraphics[scale = 0.43]{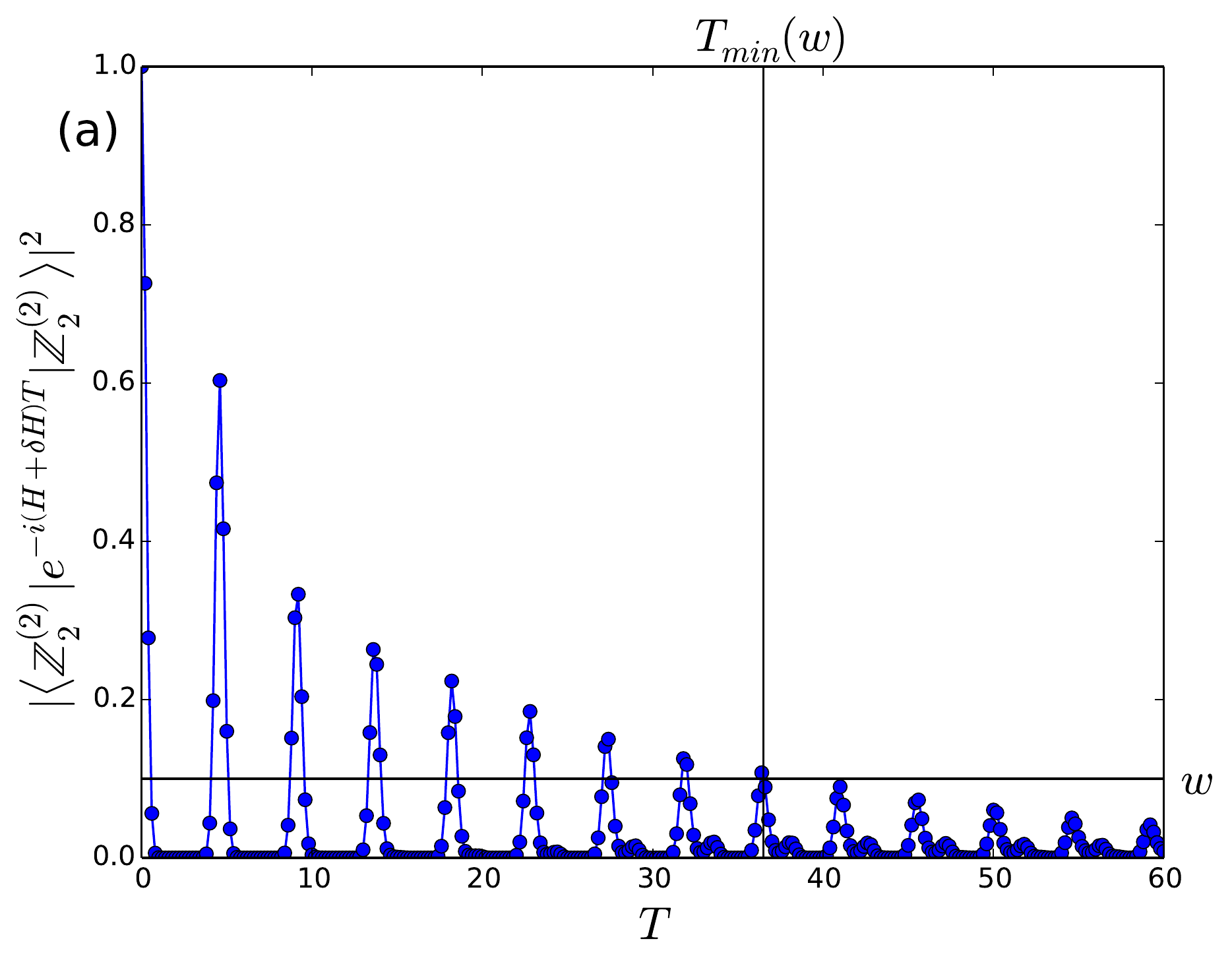}&\includegraphics[scale = 0.43]{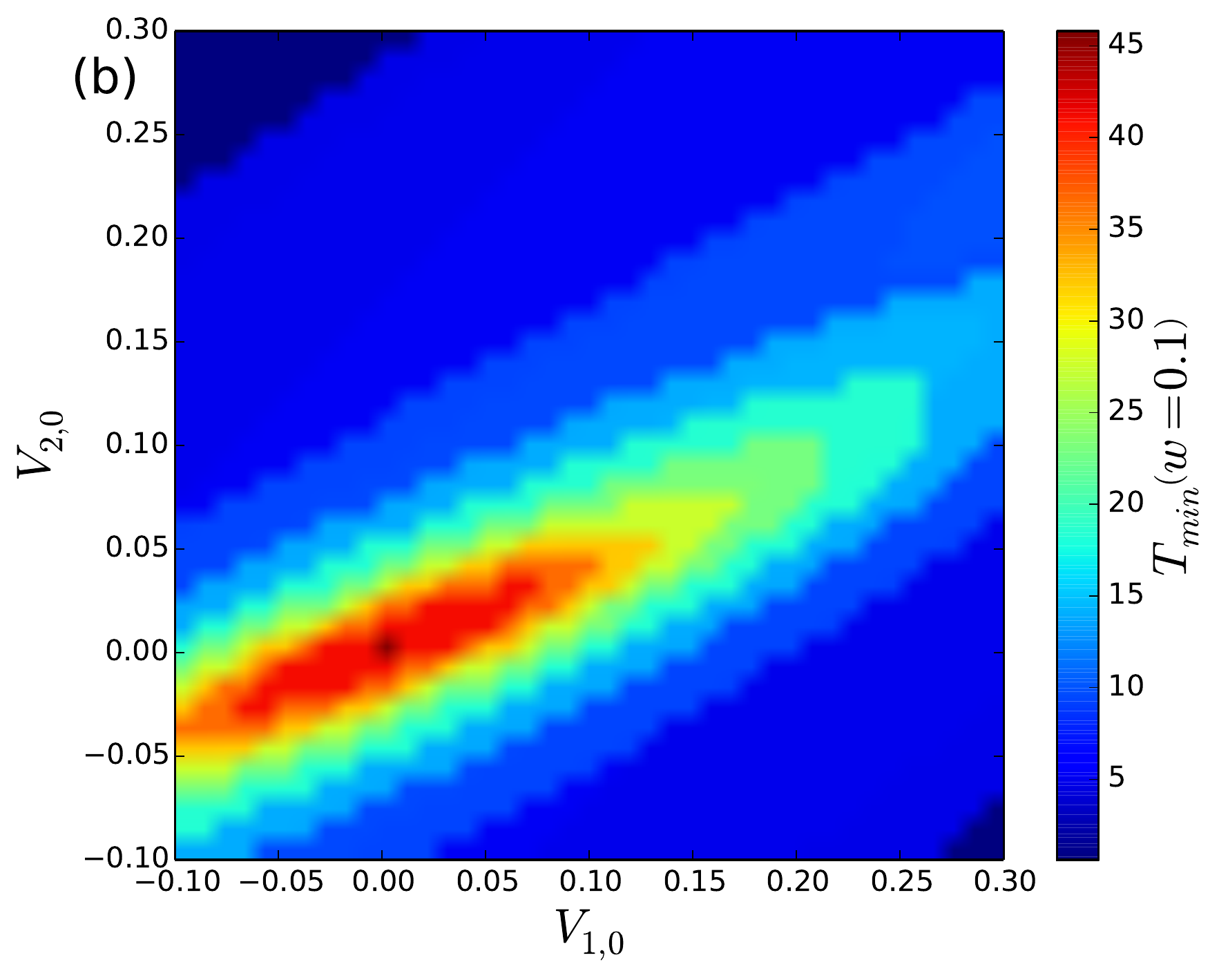}
\end{tabular}
\caption{(Color online) (a) Fidelity of the $\ket{\mathbb{Z}_2\left(k = 0\right)}$ state with $(V_{1,0}, V_{2,0}) = (0.08, 0.04)$. (b) Late time fidelity of the $\ket{\mathbb{Z}_2\left(k = 0\right)}$ state upon the addition of electrostatic terms. Clearly the line $V_{1,0} = 2V_{2,0}$ has stronger revivals than other points in parameter space. Data shown for PBC and $N = 16$ in the quantum number sector $(k, X) = (0, +1)$.}
\label{fig:stability}
\end{figure*}

We numerically test these aspects for $p = 2$. We first define the momentum $k = 0$ eigenstates constructed from the state $\ztwo{2}$ as
\begin{equation}
    \ket{\mathbb{Z}^{(2)}_2\left(k = 0\right)} \equiv \frac{\ztwo{2} + \mathcal{T}\ztwo{2}}{\sqrt{2}},
\end{equation}
where $\mathcal{T}$ is the translation operator by one unit cell. Note that $\mathcal{T}^2\ket{\mathbb{Z}^{(2)}_2} = \ket{\mathbb{Z}^{(2)}_2}$.
In Fig.~\ref{fig:timevolscars}a, we plot the bipartite entanglement entropy $S$ of the states $e^{-i \mH{2}T}\ket{\mathbb{Z}^{(2)}_2\left(k = 0\right)}$ and $e^{-i \mH{2}T}\init{2}$ states with the bipartition being one half of the system containing $N/2$ consecutive unit cells. 
Note that while $\init{2}$ thermalizes quickly, the EE for the $\ket{\mathbb{Z}^{(2)}_2\left(k = 0\right)}$ undergoes small oscillations before saturating close to the expected thermal value. 
Further, the EE growth of $\ket{\mathbb{Z}^{(2)}_2\left(k = 0\right)}$ is subballistic, contrary to the ballistic (linear in $T$) growth observed for $\init{2}$, a characteristic of typical initial states for nonintegrable models.\cite{kim2013ballistic}
Furthermore, as shown in Fig.~\ref{fig:timevolscars}b, the fidelity of $\ket{\mathbb{Z}^{(2)}_2\left(k = 0\right)}$  state, defined as $|\bra{\mathbb{Z}^{(2)}_2\left(k = 0\right)} e^{-i H T} \ket{\mathbb{Z}^{(2)}_2\left(k = 0\right)}|^2$, shows several strong revivals before oscillating and decaying at long times. As shown in Fig.~\ref{fig:timevolscars}c, the revivals survive up to the largest system size ($N = 20$ unit cells) accessible via exact diagonalization. 
In Fig.~\ref{fig:scars}a, we plot the overlap of the eigenstates with the $\ket{\mathbb{Z}^{(2)}_2\left(k = 0\right)}$ state, which clearly show the existence of a ``tower" of states \emph{approximately} equally spaced in energy that have a high overlap.  
Moreover, we compute $\mH{2}_{\textrm{FSA}}$ numerically and show that the FSA (see Eqs.~(\ref{eq:fsaapprox}) and (\ref{eq:H1FSA})) accurately predicts the energies of these ``scarred states", providing evidence that the Krylov subspace $\mK{2}$ is indeed approximately closed under $\mH{2}$.  
However, the low EE ``scarred states" in the spectrum of $\mH{2}$ appear to strongly hybridize with the rest of the spectrum with increasing $N$, as shown in Fig.~\ref{fig:scars}b.
While the hybridization can be attributed to the fact that $\mK{2}_+$ is only \emph{approximately} closed under $\mH{2}$, we do not understand why the hybridization is weaker in the PXP ($p = 1$) model, although it takes place there too\cite{turner2017quantum, turner2018quantum} and is likely to show a similar EE spectrum as Fig.~\ref{fig:scars}b for larger system sizes. 
\section{Stability of Scars}\label{sec:stability}
We now study the effect of perturbations of the scars obtained in Sec.~\ref{sec:scars}.
The full quantum Hall spectrum exhibits level repulsion,\cite{fremling2018dynamics} and presumably does not exhibit scars. It is thus instructive to study how perturbations inspired from the quantum Hall setup affect the scars obtained in Sec.~\ref{sec:scars}.
In this work, we restrict ourselves to center-of-mass preserving electrostatic perturbations that occur in a Landau level (see Eq.~(\ref{eq:qhhamil})).  
We consider the Hamiltonian
\begin{equation}
    H' = H +  \delta H,
\label{eq:totalhamilpert}
\end{equation}
where $H$ is the pair-hopping Hamiltonian of Eq.~(\ref{eq:pairhopping}) and $\delta H$ is a perturbation. 
We consider the effect of electrostatic terms in Eq.~(\ref{eq:uptopairhop}). 
The perturbation thus reads
\begin{equation}
    \delta H = \sumal{j = 1}{L_b}{\left(V_{1,0} \hat{n}_i \hat{n}_{i+1} + V_{2,0} \hat{n}_i \hat{n}_{i+2}\right)}.
\label{eq:pertelectro}
\end{equation}
Since this is a diagonal perturbation, the Krylov subspaces $\Kp$ are still closed under the action of the perturbed Hamiltonian. 
For convenience we define the operators
\begin{eqnarray}
    &Z^+_{j,n} \equiv U^\dagger_{j,n} U_{j,n},\; Z^-_{j,n} \equiv T_{j,n} T^\dagger_{j,n},\nn \\
    &Z^\zero_{j,n} \equiv T^\dagger_{j,n} T_{j,n} = U_{j,n} U^\dagger_{j,n},
\label{eq:spin1ops}
\end{eqnarray}
i.e., $Z^\alpha_j = \ket{\alpha}\bra{\alpha}$ for $\alpha = +, \zero, -$ (see Eq.~(\ref{eq:Zactions})). 
When $p = 1$, using Eq.~(\ref{eq:spin1dof}), we obtain the expression of $\delta H$ of Eq.~(\ref{eq:pertelectro}) within the constrained subspace $\mK{1}$ as 
\begin{equation}
    \delta\mH{1} = V_{1,0} \sumal{j = 1}{N_b}{Z^+_{j,1} Z^-_{j+1,1}}.
\label{eq:deltaHactionp=1}
\end{equation}
Using the mapping of Eq.~(\ref{eq:duallatticemapping}), the operator of Eq.~(\ref{eq:deltaHactionp=1}) maps on to the perturbation $V_{1,0} \sumal{j}{}{(1 + \sigma^z_j)/2}$.
The effect of this perturbation on the scars of the PXP model were briefly studied in Ref.~[\onlinecite{turner2018quantum}], where numerical evidence suggested that the scars are stable upon the addition of small $V_{1,0} < 1$.
For general $p$, the electrostatic terms within the constrained subspace can be written in terms of $Z^\alpha_{j,n}$'s (see Eq.~(\ref{eq:pertelectrogenp})).
To diagnose revivals in the presence of electrostatic terms we use a measure $T_{\textrm{min}}\left(w\right)$, defined as the minimum time after which the fidelity is always less than $w$, as depicted in Fig.~\ref{fig:stability}a. 
We plot $T_{\textrm{min}}\left(w = 0.1\right)$ for several values of $V_{1,0}$ and $V_{2,0}$ in Fig.~\ref{fig:stability}b and observe that revivals are stable for small strengths of electrostatic terms particularly when $V_{1, 0} = 2 V_{2,0}$. 
In contrast to electrostatic terms, certain longer range pair-hopping terms generically do not preserve the Krylov subspaces $\Kp$, invalidating the study of the Hamiltonian $\Hp$.  
For example, consider $p = 1$.
The action of terms of $C_{3, 1}$ of Eq.~(\ref{eq:qhhamil}) $\cd_{j+1} \cd_{j + 3} c_{j + 4} c_{j}$ and $C_{3, 2}$ of Eq.~(\ref{eq:qhhamil}) $\cd_j \cd_{j + 5} c_{j + 3} c_{j + 2}$ for example read
\begin{eqnarray}
    &&\cd_{j+1} \cd_{j + 3} c_{j+4} c_{j}\ket{\ \bminus\ \bzero\ } = \ket{\ \bzero\ \bminus\ } \nn \\
    &&\cd_j \cd_{j + 5} c_{j + 3} c_{j + 2}\ket{\ \bplus\ \bminus\ } = \ket{\ \bminus\ \bplus\ },
\label{eq:P32action}
\end{eqnarray}
configurations that are not allowed in $\mK{1}$. 
We defer the detailed study of the interplay of Krylov subspaces introduced by various long-range hopping terms for future work.
Based on numerical observations for $p \leq 3$ and the similar structure of the problem for higher values of $p$, we make the following {\bf conjecture}: signatures of scars are stable to small perturbations of the pair-hopping Hamiltonian that preserve the Krylov subspaces $\mathcal{K}^{(p)}$.
\section{Conclusions}\label{sec:conclusions}
We have studied a particular pair-hopping Hamiltonian, a one dimensional model that arises within a Landau level in the thin-torus limit of the Quantum Hall effect.
At filling $\nu = 1/3$, the pair-hopping Hamiltonian restricted to a particular constrained Krylov subspace space exactly maps onto the PXP model,\cite{turner2017quantum} which shows the existence of quantum many-body scars\cite{turner2017quantum, turner2018quantum, ho2018periodic} and exact strong ETH-violating eigenstates.\cite{lin2018exact}
We showed that the phenomenology generalizes to filling factors $\nu = 2/5$ and $\nu = 3/7$,  where ``maximally squeezed" charge density wave configurations showed revivals and subballistic growth of entanglement entropy \emph{in spite of} strong hybridization of the scarred eigenstates, contrary to the typical behavior in nonintegrable models.
Due to the similar structure of the problem at all filling factors $\nu = p/(2p + 1)$, we expect similar phenomena to occur for some higher values of $p$, although that is hard to demonstrate numerically for $p \geq 4$. 
Furthermore, we numerically explored the stability of the revivals under electrostatic terms that appear in the quantum Hall Hamiltonian. 
We emphasize that in this work we have only studied one particular Krylov subspace at each of the filling factors.  
Indeed, this model exhibits exponentially many other Krylov subspaces similar to the models studied in Refs.~[\onlinecite{sala2019ergodicity}] and [\onlinecite{khemani2019local}], which we study in detail in Ref.~[\onlinecite{moudgalya2019thermalization}].  
It is likely that similar phenomena occur with longer range pair-hopping terms that arise in the quantum Hall setting (see Eq.~(\ref{eq:qhhamil})) and also multi-body hopping terms that arise in the case of non-abelian quantum Hall states. 
It would be interesting to understand the generic structure of such constrained Hilbert spaces at arbitrary filling factors, whether or not they exhibit quantum scars, and if there is any connection to quantum Hall physics.   
On the mathematical side, it would be interesting to better understand constrained Hilbert spaces $\Kp$ and the Hamiltonians $\Hp$. 
For example, the Hilbert space of the PXP model (and consequently $\mK{1}$) can be related to the configuration space of the Baxter Hard Square Model, which gives rise to Bethe Ansatz integrable models that resemble the PXP model.\cite{fendley2004competing}
Moreover, the same Hilbert space $\mK{1}$ arises in chains of Fibonacci anyons.
An interesting question is to explore any of these connections naturally generalize to the Hilbert spaces $\Kp$. 
On the physical side, an important challenge is to identify physical interactions and regimes that naturally lead to the dominance of pair-hopping terms in the quantum Hall Hamiltonian.
This could provide a new route to the experimental realization of quantum scars.  
\section*{Acknowledgements}
We thank Dan Arovas, Wen Wei Ho, Vedika Khemani, Rahul Nandkishore, Zlatko Papi\'{c}, Abhinav Prem, and Shivaji Sondhi for useful discussions. The research is supported by the Department of Energy Grant No. de- sc0016239,  Simons Investigator Grant No. ONRN00014-14-1-0330,  the Packard Foundation, the Schmidt Fund for Innovative Research and the work is further supported by the National Science Foundation EAGER Grant No. noaawd1004957 and No. NSFMRSEC DMR- 1420541.
\appendix
\onecolumngrid
\section{Obtaining $\boldsymbol{V^{(l)}_{k,m}}$ for general potentials}\label{app:Vkmgeneral}
In this appendix, we outline a general procedure for obtaining $V^{(l)}_{k,m}$ for general potentials in the $l$-th Landau level on the cylinder geometry. Note that similar calculations have been performed in the literature, for example in Refs.~[\onlinecite{lee2015geometric}] and [\onlinecite{rotondo2016devil}].
We start with the Fourier representation of the potential 
\begin{equation}
    V\left(\br\right) = \sumal{\bq}{}{V\left(\bq\right) e^{i \bq\cdot\br}}.
\label{eq:Vfourier}
\end{equation}
If $\{\psi_{l, \alpha}\left(\br\right)\}$ are the single-particle orbitals in the $l$-th Landau level, we obtain $V^{(l)}_{j_1, j_2, j_3, j_4}$ using Eq.~(\ref{eq:matrixel}):
\begin{eqnarray}
    V^{(l)}_{j_1,j_2,j_3,j_4} &\equiv& \frac{1}{2}\iint{\mathrm{d}^2 \br_1\ \mathrm{d}^2 \br_2\ \psi^\ast_{l, j_1}\left(\br_1\right)\psi^\ast_{l, j_2}\left(\br_2\right) V\left(\br_1 - \br_2\right)\psi_{l, j_3}\left(\br_2\right)\psi_{l, j_4}\left(\br_1\right)} \nn \\
    &=& \frac{1}{2}\sumal{\bq}{}{\iint{\mathrm{d}^2 \br_1\ \mathrm{d}^2 \br_2\ \psi^\ast_{l, j_1}\left(\br_1\right)\psi^\ast_{l, j_2}\left(\br_2\right)V\left(\bq\right) e^{i \bq\cdot\left(\br_1 - \br_2\right)}\psi_{l, j_3}\left(\br_2\right)\psi_{l, j_4}\left(\br_1\right)}} \nn \\
    &\equiv& \frac{1}{2}\sumal{\bq}{}{V\left(\bq\right) I^{(l)}_{j_1, j_4}\left(\bq\right) I^{(l)}_{j_2, j_3}\left(-\bq\right)},
\label{eq:Vmatrixel}
\end{eqnarray}
where we have defined
\begin{equation}
    I^{(l)}_{\alpha,\beta}\left(\bq\right) \equiv \int{\mathrm{d}^2 \br\ \psi^\ast_{l, \alpha}\left(\br\right) e^{i \bq\cdot \br}\psi_{l, \beta}\left(\br\right)}.
\label{eq:Ialphabeta}
\end{equation}
Using Eq.~(\ref{eq:Vmn}), the expression for $V^{(l)}_{k,m}$ can then be written as
\begin{eqnarray}
    &V^{(l)}_{k,m} = \frac{1}{2}\sumal{\bq}{}{V\left(\bq\right)\left[I^{(l)}_{j + m, j}\left(\bq\right)I^{(l)}_{j+k, j+k + m}\left(-\bq\right) - I^{(l)}_{j+m, j+k+m}\left(\bq\right)I^{(l)}_{j+k, j}\left(-\bq\right) \right.} \nn \\
    &\left.+ I^{(l)}_{j+k, j+k+m}\left(\bq\right)I^{(l)}_{j+m, j}\left(-\bq\right) - I^{(l)}_{j+k, j}\left(\bq\right)I^{(l)}_{j+m, j+k+m}\left(-\bq\right) \right] \nn \\
    &\equiv \frac{1}{2}\sumal{\bq}{}{V\left(\bq\right)J^{(l)}_{k,m}\left(\bq\right)}.
\label{eq:Vkmsimp}
\end{eqnarray}
We now compute $J^{(l)}_{k,m}\left(\bq\right)$, which is independent of the potential $V\left(\bq\right)$.
As in the main text, we work in the Landau gauge where $\vec{A} = B y \hat{x}$. 
On the cylinder geometry, the single-particle wavefunctions read
\begin{equation}
    \psi_{l, \alpha}\left(\br\right) = \frac{1}{\sqrt{\sqrt{\pi} 2^l l! L_y}}H_l\left(x + \frac{2 \pi \alpha}{L_y}\right) e^{-\frac{1}{2}\left(x + \frac{2 \pi \alpha}{L_y}\right)^2} e^{\frac{2 \pi i \alpha y}{L_y}},
\label{eq:LLcylinder}
\end{equation}
where $H_l(x)$ is the $l$-th Hermite polynomial in $x$.
Using Eqs.~(\ref{eq:LLcylinder}) and (\ref{eq:Ialphabeta}), we obtain the expression for $I^{(l)}_{\alpha,\beta}\left(\bq\right)$ as
\begin{eqnarray}
    I^{(l)}_{\alpha,\beta}\left(\bq\right) &=& \frac{1}{2^p p! \sqrt{\pi} L_y}\int_{-\infty}^{\infty}{\mathrm{d}x\ H_l\left(x + \frac{2 \pi \alpha}{L_y}\right)H_l\left(x + \frac{2 \pi \beta}{L_y}\right) e^{-\frac{1}{2}\left(x + \frac{2 \pi \alpha}{L_y}\right)^2 - \frac{1}{2}\left(x + \frac{2 \pi \beta}{L_y}\right)^2 + i q_x x}\int_{0}^{L_y}{\mathrm{d}y\ e^{\frac{2 \pi i \left(\alpha-\beta\right)y}{L_y} + i q_y y }}} \nn \\
    &=& \frac{1}{2^p p! \sqrt{\pi}}\int_{-\infty}^{\infty}{\mathrm{d}x\ H_l\left(x + \frac{2 \pi \alpha}{L_y}\right)H_l\left(x + \frac{2 \pi \beta}{L_y}\right) e^{-\frac{1}{2}\left(x + \frac{2 \pi \alpha}{L_y}\right)^2 - \frac{1}{2}\left(x + \frac{2 \pi \beta}{L_y}\right)^2 + i q_x x}\delta_{\frac{2 \pi \left(\beta-\alpha\right)}{L_y}, q_y}} \nn \\
    &=& \frac{1}{2^p p! \sqrt{\pi}}\int_{-\infty}^{\infty}{\mathrm{d}x\ H_l\left(x\right)H_l\left(x + \frac{2 \pi \left(\beta - \alpha\right)}{L_y}\right) e^{-\frac{x^2}{2} - \frac{1}{2}\left(x + \frac{2 \pi \left(\beta-\alpha\right)}{L_y}\right)^2 + i q_x \left(x - \frac{2 \pi \alpha}{L_y}\right)}\delta_{\frac{2 \pi \left(\beta-\alpha\right)}{L_y}, q_y}} \nn \\
    &=& \frac{1}{2^p p! \sqrt{\pi}}\int_{-\infty}^{\infty}{\mathrm{d}x\ H_l\left(x\right)H_l\left(x + q_y\right) e^{-\frac{x^2}{2} - \frac{\left(x + q_y\right)^2}{2} + i q_x \left(x - \frac{2 \pi \alpha}{L_y}\right)}\delta_{\frac{2 \pi \left(\beta-\alpha\right)}{L_y}, q_y}} \nn \\
    &=& \frac{1}{2^p p! \sqrt{\pi}}\int_{-\infty}^{\infty}{\mathrm{d}x\ H_l\left(x\right)H_l\left(x + q_y\right) e^{-x^2 + i x \left(q_x + i q_y\right) -\frac{q_y^2}{2}  - \frac{2 \pi i \alpha q_x}{L_y}}\delta_{\frac{2 \pi \left(\beta-\alpha\right)}{L_y}, q_y}} \nn \\
    &=& \frac{1}{2^p p! \sqrt{\pi}}\delta_{\frac{2 \pi \left(\beta-\alpha\right)}{L_y}, q_y} e^{-\frac{|\bq|^2}{4} - \frac{\pi i\left(\alpha + \beta\right) q_x}{L_y}} \int_{-\infty}^{\infty}{\mathrm{d}x\ H_l\left(x\right)H_l\left(x + q_y\right) e^{-\left(x - \frac{i (q_x + i q_y)}{2}\right)^2}} \nn \\
    &=& \frac{1}{2^p p! \sqrt{\pi}}\delta_{\frac{2 \pi \left(\beta-\alpha\right)}{L_y}, q_y} e^{-\frac{|\bq|^2}{4} - \frac{\pi i\left(\alpha + \beta\right) q_x}{L_y}} \int_{-\infty}^{\infty}{\mathrm{d}x\ H_l\left(x + i \frac{q_x + i q_y}{2}\right)H_l\left(x + i \frac{q_x - i q_y}{2}\right) e^{-x^2}} \nn \\
    &=& \delta_{\frac{2 \pi \left(\beta-\alpha\right)}{L_y}, q_y} e^{-\frac{|\bq|^2}{4} - \frac{\pi i\left(\alpha + \beta\right) q_x}{L_y}} L_l\left(\frac{|\bq|^2}{2}\right),
\end{eqnarray}
where we have used the identity\cite{gradshteyn2014table}
\begin{equation}
    \frac{1}{2^l l! \sqrt{\pi}}\int_{-\infty}^{\infty}{\mathrm{d}x\ H_l\left(x + i a\right) H_l\left(x + i b\right) e^{-x^2}} = L_l\left(2 a b\right),
\end{equation}
where $L_l(x)$ is the $l$-th Laguerre polynomial in $x$.
Before computing $J^{(l)}_{km}\left(\bq\right)$, for convenience we compute
\begin{equation}
    K^{(l)}_{\alpha,\beta,\gamma}\left(\bq\right) \equiv I^{(l)}_{\alpha, \alpha + \gamma}\left(\bq\right) I^{(l)}_{\beta + \gamma, \beta}\left(-\bq\right) =  e^{-\frac{|\bq|^2}{2}}\left(L_l\left(\frac{|\bq|^2}{2}\right)\right)^2 \delta_{\frac{2\pi \gamma}{L_y}, q_y} e^{\frac{2\pi i\left(\beta-\alpha\right)q_x}{L_y}}.
\label{eq:Kpdefnapp}
\end{equation}
Using Eqs.~(\ref{eq:Vkmsimp}) and (\ref{eq:Kpdefnapp}), $J^{(l)}_{km}\left(\bq\right)$ can then be written as
\begin{eqnarray}
    J^{(l)}_{km}(\bq) &=& K^{(l)}_{j + m, j + k + m, -m}(\bq) - K^{(l)}_{j + m, j, k}(\bq) + K^{(l)}_{j + k, j, m}\left(\bq\right) - K^{(l)}_{j + k, j + k + m, -k}\left(\bq\right) \nn \\
    &=& e^{-\frac{|\bq|^2}{2}}\left(L_l\left(\frac{|\bq|^2}{2}\right)\right)^2 \left[\delta_{-\frac{2\pi m}{L_y}, q_y}  e^{\frac{2\pi i k q_x}{L_y}} - \delta_{\frac{2\pi k}{L_y}, q_y}  e^{-\frac{2\pi i m q_x}{L_y}} + \delta_{\frac{2\pi m}{L_y}, q_y}  e^{-\frac{2\pi i k q_x}{L_y}} - \delta_{-\frac{2\pi k}{L_y}, q_y}  e^{\frac{2\pi i m q_x}{L_y}}\right]. \nn \\
\label{eq:Jpform}
\end{eqnarray}
$V^{(l)}_{k, m}$ can then be computed using Eq.~(\ref{eq:Vkmsimp}).
In fact, it is convenient to write $V^{(l)}_{k,m}$ in terms of $W^{(l)}_{k,m}$, which is defined as
\begin{eqnarray}
    W^{(l)}_{k,m} = \sumal{\bq}{}{V(\bq) e^{-\frac{|\bq|^2}{2}}\left(L_l\left(\frac{|\bq|^2}{2}\right)\right)^2 \delta_{\frac{2\pi m}{L_y}, q_y} e^{\frac{2 \pi i k q_x}{L_y}}}, \;\;\; V_{k,m} = W_{k,-m} - W_{-m,k} + W_{-k,m} - W_{m,-k}.
\label{eq:Wabdefn}
\end{eqnarray}
$W^{(l)}_{k,m}$ can be simplified to
\begin{eqnarray}
    W^{(l)}_{k,m} &=& \sumal{\bq}{}{V(q_x, q_y) e^{-\frac{|\bq|^2}{2}}\left(L_l\left(\frac{|\bq|^2}{2}\right)\right)^2 \delta_{\frac{2\pi m}{L_y}, q_y} e^{\frac{2 \pi i k q_x}{L_y}}} = \int_{-\infty}^{\infty}{\mathrm{d}q_x\ V\left(q_x, \frac{2\pi b}{L_y}\right)\left(L_l\left(\frac{q_x^2 + \frac{4\pi^2b^2}{L_y^2}}{2}\right)\right)^2 e^{-\frac{q_x^2 + \frac{4\pi^2b^2}{L_y^2}}{2}} e^{\frac{2\pi i a q_x}{L_y}}} \nn \\
    &=& \int_{-\infty}^{\infty}{\mathrm{d}x\ \left(x^2 + B^2\right)\left(L_l\left(\frac{x^2 + B^2}{2}\right)\right)^2 e^{-\frac{x^2 + B^2}{2}} e^{i A x}} = e^{-\frac{A^2 + B^2}{2}} \int_{-\infty}^{\infty}{\mathrm{d}x\ V\left(x, B\right)\left(L_l\left(\frac{x^2 + B^2}{2}\right)\right)^2  e^{-\frac{\left(x - i A\right)^2}{2}}},\nn \\
\label{eq:Wabeasy}
\end{eqnarray}
where in the last line we have changed the integration variable, and we have defined $A = \frac{2 \pi k}{L_y}$, $B = \frac{2\pi m}{L_y}$.

We now illustrate an example using the short-range Haldane-Trugman-Kivelson potential,\cite{haldane1983fractional, trugman1985exact}
\begin{equation}
    V(\br) = \nabla^2\delta^{(2)}(\br) = -\sumal{\bq}{}{|\bq|^2 e^{i\bq\cdot\br}}\;\;\;\implies\;\;\; V(q_x, q_y) = -\left(q_x^2 + q_y^2\right).
\label{eq:TKfourier}
\end{equation}
We compute $W^{(l)}_{k,m}$ using Eq.~(\ref{eq:Wabeasy}), which reduces to
\begin{equation}
    W^{(l)}_{k,m} = -e^{-\frac{2\pi^2\left(k^2 + m^2\right)}{L_y^2}}\int_{-\infty}^{\infty}{\mathrm{d}x\ (x^2 + B^2)\left(L_l\left(\frac{x^2 + B^2}{2}\right)\right)^2  e^{-\frac{\left(x - i A\right)^2}{2}}}. 
\label{eq:WabTK}
\end{equation}
We have not been able to obtain a useful closed form expression for $W^{(l)}_{k,m}$ of Eq.~(\ref{eq:WabTK}) for general $l$.
Here we list $V^{(l)}_{k,m}$ for the lowest two Landau levels: 
\begin{eqnarray}
    &V^{(0)}_{k,m} = \frac{16 \pi^2}{L_y^2} (k-m) (k+m) e^{-\frac{2\pi^2\left(k^2 + m^2\right)}{L_y^2}}\nn \\
    &V^{(1)}_{k,m} = \frac{16 \pi^2}{L_y^2}(k - m) (k + m) \left(15 - \frac{24 \pi^2}{L_y^2}(k^2 + m^2) + \frac{16 \pi^4}{L_y^4} \left(k^2 - m^2\right)^2\right)e^{-\frac{2\pi^2\left(k^2 + m^2\right)}{L_y^2}}.
\label{eq:TKlist}
\end{eqnarray}
\twocolumngrid
\section{Properties of operators $T$ and $U$}\label{app:TUproperties}
Here we list the properties of single spin operators $T$ and $U$ defined in Eq.~(\ref{eq:TUdefn}) that are used extensively in the main text. 
The action of $T$ and $U$ on spin-1 basis states read
\begin{eqnarray}
    &&T \ket{+} = 0,\; T \ket{\zero} = -\ket{-},\; T\ket{-} = 0,\nn \\
    &&T^\dagger\ket{+} = 0,\; T^\dagger\ket{\zero} = 0,\; T^\dagger\ket{-} = -\ket{\zero} \nn \\
    &&U \ket{+} = \ket{\zero},\; U\ket{\zero} = 0,\; U \ket{-} = 0 \nn \\
    &&U^\dagger \ket{+} = 0,\; U^\dagger \ket{\zero} = \ket{+},\; U^\dagger\ket{-} = 0.
\label{eq:TUactions}
\end{eqnarray}
From Eq.~(\ref{eq:TUactions}), we deduce the following properties
\begin{eqnarray}
    &&T^\dagger T\ket{+} = 0,\; T^\dagger T \ket{\zero} = \ket{\zero},\; T^\dagger T \ket{-} = 0,\nn \\
    &&T T^\dagger\ket{+} = 0,\; T T^\dagger\ket{\zero} = 0,\; T T^\dagger\ket{-} = \ket{-},\nn \\
    &&U^\dagger U\ket{+} = \ket{+},\; U^\dagger U \ket{\zero} = 0,\; U^\dagger U \ket{-} = 0,\nn \\
    &&U U^\dagger\ket{+} = 0,\; U U^\dagger\ket{\zero} = \ket{\zero},\; U U^\dagger\ket{-} = 0.
\label{eq:TUdaggeractions}
\end{eqnarray}
\section{Constrained Hilbert Space Dimension}\label{app:dimension}
In this appendix, we derive the Hilbert space dimension $D^{(p)}_N$ for the Hilbert spaces $\Kp$, and we show that --for large $N$-- $D^{(p)}_N$ scales as
\begin{equation}
    D^{(p)}_N \sim \left(\lambda^{(p)}\right)^N.
\label{eq:quantdef}
\end{equation}
We refer to $\lambda^{(p)}$ as the \emph{quantum dimension} of the constrained Hilbert space $\Kp$.
For the sake of simplicity, we restrict ourselves to OBC throughout this section. 
In the following subsections, we show the steps to count $D^{(p)}_N$ and $\lambda^{(p)}$ for general $p$.
Explicit examples of the counting can be found in App.~\ref{app:explicitcount}.
\subsection{Preliminaries}
Before deriving $D^{(p)}$, we present an outline of the procedure we use to count the Hilbert space dimension, and introduce the notations and concepts we will require throughout this appendix.
As an illustrative example, we use $N = 3$ (three unit cells), and $p = 2$ (two spin-1's per unit cell). 
According to constraints (c1), (c2), and (c3) discussed in Sec.~\ref{sec:pb2p+1}, the Hilbert space is spanned by several configurations, for example 
\begin{eqnarray}
    &\textrm{(i)}\; \fbox{\zero\ \zero}\ \fbox{\zero\ \zero}\ \fbox{\zero\ \zero}\ ,\ \textrm{(ii)}\; \fbox{\plus\ \zero}\ \fbox{\zero\ \minus}\ \fbox{\zero\ \zero}\ \nn \\
    &\textrm{(iii)}\; \fbox{\zero\ \zero}\ \fbox{\plus\ \plus}\ \fbox{\minus\ \minus}\ ,\ \textrm{(iv)}\; \fbox{\zero\ \plus}\ \fbox{\minus\ \plus}\ \fbox{\minus\ \zero}\ .
\label{eq:exampleconfigs}
\end{eqnarray}
We refer to the configurations allowed in the Hilbert space as \emph{valid} configurations.
Thus, the Hilbert space dimension is defined as
\begin{center}
\begin{tabular}{cc}
    $D^{(p)}_N$: & Number of valid configurations of $N$ unit cells.\\
\end{tabular}
\end{center}
To define a valid configuration, we introduce the following notation. 
Configurations of $N$ unit cells are represented by $\ket{\vec{\sigma}} = \ket{\sigma_1\sigma_2 \cdots \sigma_N}$, where $\sigma_j$ is the configuration of the $j$-th unit cell (i.e. a configuration of $p$ spin-1's obtained after adding $(p-1)$ pseudozeroes as illustrated in Sec.~\ref{sec:pb2p+1}). 
We define the following quantities associated with the unit cell configuration $\sigma_j$:
\begin{center}
\begin{tabular}{cc}
    $P_{\sigma_j}$: & Number of $+$'s in $\sigma_j$ \\
    $M_{\sigma_j}$: & Number of $-$'s in $\sigma_j$ \\
\end{tabular}
\end{center}
Using this language, a valid configuration is defined as follows:
\begin{definition}
A \emph{valid} configuration of $N$ unit cells is a configuration $\ket{\vec{\sigma}}$ that is in $\Kp$ with OBC, i.e. it satisfies the following properties (c1), (c2), and (c3) listed in Sec.~\ref{sec:pb2p+1}:
\begin{enumerate}
    \item[(c1)] $P_{\sigma_j} = M_{\sigma_{j+1}}$ $\forall j$, $1 \leq j \leq N - 1$.
    \item[(c2)] In any unit cell $\sigma_j$, the $-$ always appear to the left of a $+$. 
    \item[(c3)] $P_{\sigma_N} = 0$, $M_{\sigma_1} = 0$. This constraint only appears for OBC, as explained in Sec.~\ref{sec:pb2p+1}.
\end{enumerate}
For example, when $p = 2$, consider the configurations such as $\twoket{\zero +}{- -}$, $\ket{\ \fbox{\zero +}\ \fbox{+ \minus}\ \fbox{\zero \minus}\ }$, and $\ket{\ \fbox{\zero +}\ \fbox{\minus \zero}\ \fbox{\zero +}\ }$. The first configuration violates (c1), the second violates (c2), and the third violates (c3). Thus, these configurations are not valid configurations.
On the other hand, configurations of Eq.~(\ref{eq:exampleconfigs}) are valid since they satisfy all three constraints. 
We henceforth suppress the index $p$ in the Hilbert space dimension $D^{(p)}_N$ since we will always be working with a fixed $p$.
\end{definition}
Each of the valid configurations can be thought to be composed of \emph{connected} configurations of $n \leq N$ unit cells, which are valid configurations of $n$ unit cells that cannot be divided into valid configurations of $m < n$ unit cells.   
For example, consider the configurations (i)-(iv) in Eq.~(\ref{eq:exampleconfigs}).
They are composed of the following connected configurations
\begin{eqnarray}
    &&\textrm{(i)}:\ \ket{\ \fbox{\ \zero\ \zero\ }\ }, \ket{\ \fbox{\ \zero\ \zero\ }\ }, \ket{\ \fbox{\ \zero\ \zero\ }\ } \nn \\
    &&\textrm{(ii)}:\ \ket{\ \fbox{\ \plus\ \zero\ }\ \fbox{\ \zero\ \minus\ }\ }, \ket{\ \fbox{\ \zero\ \zero\ }\ } \nn \\
    &&\textrm{(iii)}:\ \ket{\ \fbox{\ \zero\ \zero\ }\ }, \ket{\ \fbox{\ \plus\ \plus\ }, \fbox{\ \minus\ \minus\ }\ } \nn \\
    &&\textrm{(iv)}:\ \ket{\ \fbox{\ \zero\ \plus\ }\ \fbox{\ \minus\ \plus\ }\ \fbox{\ \minus\ \zero\ }\ }.
\label{eq:exampleconnectedconfig}
\end{eqnarray}
None of the configurations in Eq.~(\ref{eq:exampleconnectedconfig}) can be further divided into valid configurations. 
Since each valid configuration is composed of several connected configurations, we can count the number of valid configurations by counting the number of connected configurations and placing them adjacent to each other. 
We thus focus on counting the number of connected configurations of $N$ unit cells. 
Formally, connected configurations are defined as follows:
\begin{definition}
A \emph{connected} configuration of $N$ unit cells is a valid configuration of $N$ unit cells for which \emph{no} subconfiguration of $n < N$ consecutive unit cells is a valid configuration. That is, in addition to being a valid configuration, a connected configuration $\ket{\vec{\sigma}}$ satisfies the following properties:
\begin{enumerate}
    \item[(d1)] $P_{\sigma_1} \geq 1 $, $M_{\sigma_N} \geq 1$. That is, the first or the last unit cell should not have the configuration $\fbox{$\ \zero\ \cdots\ \zero$}$. If either of these conditions are violated, the subconfiguration consisting of the first unit cell or the subconfiguration consisting of last unit cell forms a valid configuration. 
    \item[(d2)] $P_{\sigma_j} \geq 1$ and $M_{\sigma_j} \geq 1$\; $\forall j$, $2 \leq j \leq N - 1$. If this condition is violated, i.e. $P_{\sigma_j} = 0$ (resp. $M_{\sigma_j} = 0$) for some $j$, $2 \leq j \leq N - 1$, then the configuration $\ket{\sigma_1\cdots\sigma_{j-1}}$ (resp. $\ket{\sigma_{j+1}\cdots\sigma_N}$) is a valid configuration.
\end{enumerate}
\end{definition}
For example, when $p = 2$ and $N = 3$, the configuration (iv) of Eq.~(\ref{eq:exampleconfigs}) $\ket{\ \fbox{\zero\ \plus}\ \fbox{\minus\ \plus}\ \fbox{\minus\ \zero}\ }$ is connected because all of its subconfigurations violate the constraint (c3). However, the configuration (ii) of Eq.~(\ref{eq:exampleconfigs}) $\ket{\ \fbox{\plus\ \zero}\ \fbox{\zero\  \minus}\ \fbox{\zero\ \zero}\ }$ is not connected, because the subconfiguration $\ket{\ \fbox{\zero \zero}\ }$ of the latter is a valid configuration (see Eq.~(\ref{eq:exampleconnectedconfig})).
Further, we define the quantity
\begin{center}
    \begin{tabular}{p{1cm}p{7cm}}
        $C_N$: & Number of valid connected configurations of $N$ unit cells\\
    \end{tabular}
\end{center}
\subsection{Counting $C_N$}
We now focus on counting $C_N$.
In order to count the number of connected configuration via recursion relation, we first establish a mapping between each connected configuration $\ket{\vec{\tau}[N-1]}$ of $(N-1)$ unit cells and  a set of connected configurations $\{\ket{\vec{\tau}[N]}\}$ of $N$ unit cells: 
\begin{equation}
	\ket{\vec{\tau}[N-1]} \rightarrow \{\ket{\vec{\tau}[N]}\}.
\label{eq:uniquemapping}
\end{equation}
We choose the mapping such that two different configurations $\ket{\vec{\tau}[N-1]}$ and $\ket{\vec{\sigma}[N-1]}$ of $(N-1)$ unit cells respectively map onto sets $\{\vec{\tau}[N]\}$ and $\{\vec{\sigma}[N]\}$ without any common elements. 
Then the number of connected configurations of $N$ unit cells is the sum of the cardinalities of all sets $\{\ket{\vec{\tau}[N]}\}$ generated by all connected configurations $\ket{\vec{\tau}[N-1]}$ of $(N-1)$ unit cells. 
We now illustrate one such mapping.
Consider a configuration $\ket{\vec{\tau}[N-1]} = \ket{\tau_2 \tau_3 \cdots \tau_N}$ of $(N-1)$ unit cells.
Since $\ket{\vec{\tau}[N-1]}$ is a connected configuration, according to (c3) and (d1), $P_{\tau_2} \geq 1$ and $M_{\tau_2} = 0$. 
One way of constructing the mapping of Eq.~(\ref{eq:uniquemapping}) is to consider the set $\{\ket{\vec{\tau}[N]}\}$, where $\ket{\vec{\tau}[N]}$ is a connected configuration of the form  $\ket{\vec{\tau}[N]} = \ket{\tau_1\widetilde{\tau}_2\tau_3\cdots\tau_N}$, where $\tau_3, \tau_4,\cdots \tau_N$ are the unit cell configurations of $\ket{\vec{\tau}[N-1]}$, and $\widetilde{\tau}_2$ has the same position of the $+$'s as $\tau_2$, and $\tau_1$ is any configuration such that $\ket{\vec{\tau}[N]}$ is a valid connected configuration (in particular, $P_{\tau_1} \geq 1$ according to (d1)).
Note that in order for $\ket{\vec{\tau}[N]}$ to be connected, the unit cell configuration $\tau_2$ of $\ket{\vec{\tau}[N-1]}$ necessarily has to be modified to some $\widetilde{\tau}_2$, since $M_{\tau_2} = 0$ (using the constraint (c3) for $\ket{\vec{\tau}[N-1]}$) and $M_{\widetilde{\tau}_2} \geq 1$ (using the constraint (d2) for $\ket{\vec{\tau}[N]}$).
For two distinct configurations $\ket{\vec{\tau}[N-1]} = \ket{\tau_2 \tau_3 \cdots \tau_N}$ and $\ket{\vec{\sigma}[N-1]} = \ket{\sigma_2 \sigma_3 \cdots \sigma_N}$, which differ by the configuration of at least one unit cell, the mapping generates sets $\{\ket{\vec{\tau}[N]}\}$ and $\{\ket{\vec{\sigma}[N]}\}$ with no elements in common.
This is trivial if $\sigma_j \neq \tau_j$ for any $j \geq 3$.
If $\sigma_2 \neq \tau_2$, which consist of only $+$'s and $\zero$'s (as a consequence of (c3)), $\sigma_2$ and $\tau_2$ differ by the position of the $+$'s. 
Since by construction $\widetilde{\sigma}_2$ and $\widetilde{\tau}_2$ have the same positions of the $+$'s as $\sigma_2$ and $\tau_2$ respectively, $\widetilde{\sigma}_2 \neq \widetilde{\tau}_2$ if $\sigma_2 \neq \tau_2$.
Thus the sets $\{\ket{\vec{\tau}[N]}\}$ and $\{\ket{\vec{\sigma}[N]}\}$ do not have any elements in common if the configurations $\ket{\vec{\tau}[N-1]}$ and $\ket{\vec{\sigma}[N-1]}$ differ by at least one unit cell.    
For example, consider the following configurations $\ket{\vec{\tau}[N-1]}$ with $N -1 = 2$ unit cells and $p = 3$ spin-1's per unit cell: 
\begin{eqnarray}
    &&\textrm{(i)}:\ \ket{\ \fbox{\zero\ \zero\ \plus\ }\ \fbox{\zero\ \minus\ \zero\ }\ } \nn \\
    &&\textrm{(ii)}:\ \ket{\ \fbox{\zero\ \plus\ \zero\ }\ \fbox{\zero\ \zero\ \minus}\ } \nn \\
    &&\textrm{(iii)}:\ \ket{\ \fbox{\plus\ \plus\ \zero\ }\ \fbox{\zero\ \minus\ \minus}\ } \nn \\
    &&\textrm{(iv)}:\ \ket{\ \fbox{\plus\ \plus\ \plus\ }\ \fbox{\minus\ \minus\ \minus}\ }. 
\label{eq:recursionexample}
\end{eqnarray}
The configurations $\{\ket{\vec{\tau}[N]}\}$ of $N = 3$ unit cells that can be constructed from each of the configurations (i)-(iv) in Eq.~(\ref{eq:recursionexample}) are
\begin{eqnarray}
    &\hspace{-5mm}\textrm{(i)}: \{\ket{\ \fbox{\plus\ \zero\ \zero\ }\ \fbox{\minus\ \zero\ \plus\ }\ \fbox{\zero\ \minus\ \zero\ }\ } ,\nn \\
    &\ket{\ \fbox{\plus\ \zero\ \zero\ }\ \fbox{\zero\ \minus\ \plus\ }\ \fbox{\zero\ \minus\ \zero\ }\ }, \nn \\
    &\ket{\ \fbox{\zero\ \plus\ \zero\ }\ \fbox{\minus\ \zero\ \plus\ }\ \fbox{\zero\ \minus\ \zero\ }\ }, \nn \\
    &\ket{\ \fbox{\zero\ \plus\ \zero\ }\ \fbox{\zero\ \minus\ \plus\ }\ \fbox{\zero\ \minus\ \zero\ }\ }, \nn \\
    &\ket{\ \fbox{\zero\ \zero\ \plus\ }\ \fbox{\minus\ \zero\ \plus\ }\ \fbox{\zero\ \minus\ \zero\ }\ }, \nn \\
    &\ket{\ \fbox{\zero\ \zero\ \plus\ }\ \fbox{\zero\ \minus\ \plus\ }\ \fbox{\zero\ \minus\ \zero\ }\ }, \nn \\
    &\ket{\ \fbox{\plus\ \plus\ \zero\ }\ \fbox{\minus\ \minus\ \plus\ }\ \fbox{\zero\ \minus\ \zero\ }\ }, \nn \\
    &\ket{\ \fbox{\plus\ \zero\ \plus\ }\ \fbox{\minus\ \minus\ \plus\ }\ \fbox{\zero\ \minus\ \zero\ }\ }, \nn \\
    &\ket{\ \fbox{\zero\ \plus\ \plus\ }\ \fbox{\minus\ \minus\ \plus\ }\ \fbox{\zero\ \minus\ \zero\ }\ } \}\nn \\
    &\hspace{-5mm}\textrm{(ii)}: \{\ket{\ \fbox{\zero\ \zero\ \plus}\ \fbox{\minus\ \plus\ \zero\ }\ \fbox{\zero\ \zero\ \minus}\ }, \nn \\
    &\ket{\ \fbox{\zero\ \plus\ \zero}\ \fbox{\minus\ \plus\ \zero\ }\ \fbox{\zero\ \zero\ \minus}\ }, \nn \\
    &\ket{\ \fbox{\plus\ \zero\ \zero}\ \fbox{\minus\ \plus\ \zero\ }\ \fbox{\zero\ \zero\ \minus}\ }\} \nn \\
    &\hspace{-5mm}\textrm{(iii)}: \{ \}\ \textrm{(No configuration possible)} \nn \\
    &\hspace{-5mm}\textrm{(iv)}: \{ \}\ \textrm{(No configuration possible)}. 
\label{eq:recursionexamplenew}
\end{eqnarray}
Note that none of the sets (i)-(iv) in Eq.~(\ref{eq:recursionexamplenew}) have common elements.
Moreover, since the constraints (d1) and (d2) are constraints only between nearest neighbor unit cells, the cardinality of the set $\{\ket{\vec{\tau}[N]}\}$ \emph{only} depends on the configuration of the \emph{leftmost} unit cell $\tau_2$ of $\ket{\vec{\tau}[N-1]}$.
Hence, for the purposes of counting the number of connected configurations, it is sufficient to keep track of the number of connected configurations with a fixed configuration of the leftmost unit cell. Thus, we introduce the quantity
\begin{center}
    \begin{tabular}{p{1cm}p{7cm}}
        $C_{N, \tau}:$ & Number of connected configurations $\ket{\vec{\sigma}}$ of $N$ unit cells that have the leftmost unit cell configuration $\sigma_1 = \tau$. Note that as a consequence of (c3) $\tau$ only consists of $+$'s and $\zero$'s with $P_{\tau} \geq 1$, and is not a valid configuration because by definition the configuration $\ket{\vec{\sigma}}$ is connected. \\
    \end{tabular}
\end{center}
We express $C_{N}$ in terms of $C_{N,\tau}$ as
\begin{equation}
    C_N = \sumal{\tau \in \mathcal{L}}{}{C_{N, \tau}},
\label{eq:CLCLsig}
\end{equation}
where $\mathcal{L}$, the set of possible configurations of the leftmost unit cell, defined as:
\begin{center}
    \begin{tabular}{p{1cm}p{7cm}}
        $\mathcal{L}$: & Set of single unit cell configurations that reside on the leftmost unit cell of a connected configuration of $N \geq 2$ unit cells. That is, they are the set of configurations of the unit cell $\tau$ that satisfy $P_{\tau} \geq 1$ and $M_{\tau} = 0$ (since these are satisfied by (d1) and (c3) respectively).  \\
    \end{tabular}
\end{center}
For example, when $p = 3$, the configurations of the leftmost unit cell shown in the examples in Eqs.~(\ref{eq:recursionexample}) and (\ref{eq:recursionexamplenew}) all belong to the set $\mathcal{L}$.
Before proceeding with the counting of $C_N$, we set the notation for the elements in the set $\mathcal{L}$. 
Since each spin-1 in the leftmost unit cell can be in either the states $\zero$ or $+$, and we exclude the state $\fbox{$\zero\zero\cdots\zero$}$, the number of configurations in $\mathcal{L}$ is $2^p - 1$. Note that it is convenient to think of configurations in $\mathcal{L}$ as numbers between $1$ and $2^{p} - 1$ by viewing the configuration as a binary number with $1$ and $0$ representing $+$ and $\zero$.
For example, when $p = 3$, the set $\mathcal{L}$ has 7 configurations that are shown in the first column in Table~\ref{tab:tauprops}. There we label each of the configurations by binary numbers from $1$ to $7$ by replacing $\textrm{\zero}$ with $0$ and $\textrm{\plus}$ with $1$.
For example, the binary representation of $\tau = \fbox{+ \zero \zero}$ is $100$, i.e. $4$ in decimal notation. 
This binary/decimal notation provides a natural ordering of the configurations in $\mathcal{L}$ that we will rely on to write vectors in the basis labelled by configurations of in $\mathcal{L}$. 
In the following, we will abuse notation and use $\tau$ to denote the decimal number as well as the corresponding configuration in $\mathcal{L}$. For example, when $\tau_j = 4$ and $p = 3$, we mean $\tau_j = \fbox{+ \zero \zero}$ since $100$ is the binary representation of $4$. 
In order to obtain the cardinality of $\{\ket{\vec{\tau}[N]}\}$ given a configuration $\ket{\vec{\tau}[N-1]}$, we first determine the number $\mt_{\tau_1, \tau_2}$, which is the number of $N$ unit cell connected configurations of the form $\ket{\vec{\tau}[N]} = \ket{\tau_1\widetilde{\tau}_2\tau_3\cdots\tau_N}$, where $\tau_3, \tau_4,\cdots \tau_N$ are the unit cell configurations of $\ket{\vec{\tau}[N-1]}$, and $\widetilde{\tau}_2$ and $\tau_2$ are unit cell configurations which have the same positions of the $+$'s, as shown in Eqs.~(\ref{eq:recursionexample}) and (\ref{eq:recursionexamplenew}).
Given a configuration $\tau_1$ (and hence $P_{\tau_1}$, the number of $+$'s in $\tau_1$ is specified), the number of choices of $\widetilde{\tau}_2$ depends uniquely on $\tau_2$.  
In particular, according to (c2), the $-$'s can be inserted into $\tau_2$ only in the $Z_{\tau_2}$ leading $\zero$'s of $\tau_2$, where we define
\begin{center}
    \begin{tabular}{p{1cm}p{7cm}}
        $Z_{\tau}$: & Number of leading $\zero$'s in the single unit cell configuration $\tau \in \mathcal{L}$, i.e. the number of consecutive $\zero$'s in the left end of the configuration.  \\
    \end{tabular}
\end{center}
For example, when $p = 7$ and $\tau = \fbox{\zero\zero+\zero\zero\zero+}$ ($\tau = 17$ in the decimal representation), $Z_{\tau} = 2$. 
Since $P_{\tau_1} = M_{\widetilde{\tau}_2} \geq 1$ (according to (c1) and (d2)) and $M_{\tau_1} = 0$ (according to (c3) and (d1)), $\mt_{\tau_1, \tau_2}$ is the number of ways to insert $P_{\tau_1}$ $-$'s into the $Z_{\tau_2}$ leading $\zero$'s of the unit cell configuration $\tau_2$, and it reads
\begin{equation}
    \mt_{\tau_1, \tau_2} = \binom{Z_{\tau_2}}{P_{\tau_1}},\;\;\;\textrm{for}\;\;\; 1\leq \tau_1, \tau_2 \leq 2^p - 1, 
\label{eq:mtdefn}
\end{equation}
where we have abused notation for $\tau_1$ and $\tau_2$ to represent both the configuration as well as the corresponding decimal number. 
For example, if $N = 3$, $p = 3$, as seen in Eqs.~(\ref{eq:recursionexample}) and (\ref{eq:recursionexamplenew}), for $\ket{\vec{\tau}[2]} = \ket{\ \fbox{\zero \zero \plus}\ \fbox{\zero \minus \zero}\ }$ and $\tau_1 = \fbox{\zero + \zero}$, there are two allowed configurations for $\ket{\vec{\tau}[3]}$: $\ket{\ \fbox{\zero + \zero}\ \fbox{\minus \zero \plus}\ \fbox{\zero \minus \zero}\ }$ and $\ket{\ \fbox{\zero + \zero}\ \fbox{\zero \minus \plus}\ \fbox{\zero \minus \zero}\ }$. In this example, $P_{\tau_1} = 1$ and $\tau_2 = \fbox{\zero \zero +}$, and hence $Z_{\tau_2} = 1$. Thus, according to Eq.~(\ref{eq:mtdefn}), $\mt_{\tau_1, \tau_2} = 2$. 
$\mathcal{T}$ can also be expressed as a $\left(2^p - 1\right) \times \left(2^p - 1\right)$ matrix with bases as the configurations in $\mathcal{L}$.
For example, when $p = 3$, the configurations in $\mathcal{L}$ are given in the first column of Table~\ref{tab:tauprops}, along with the corresponding values of $Z_{\tau}$ and $P_{\tau}$. 
Consequently, using Eq.~(\ref{eq:mtdefn}), the matrix $\mathcal{T}$ reads (using the ordering specified by the binary representations in Table~\ref{tab:tauprops})
\begin{equation}
    \mathcal{T} = 
    \begin{bmatrix}
        2 & 2 & 1 & 2 & 1 & 1 & 0 \\
        1 & 1 & 0 & 1 & 0 & 0 & 0 \\
        1 & 1 & 0 & 1 & 0 & 0 & 0 \\
        0 & 0 & 0 & 0 & 0 & 0 & 0 \\
        0 & 0 & 0 & 0 & 0 & 0 & 0 \\
        0 & 0 & 0 & 0 & 0 & 0 & 0 \\
        0 & 0 & 0 & 0 & 0 & 0 & 0 \\
    \end{bmatrix}.
\end{equation}
\begin{table}
\begin{tabular}{|c|c|c|c|c|}
\hline
\boldsymbol{$\tau$} & {\bf Binary} & {\bf Decimal} & \boldsymbol{$Z_{\tau}$} & \boldsymbol{$P_\tau$}\\
\hline
\zero\zero+ & 001 & 1 & 2 & 1\\
\hline 
\zero+\zero & 010 & 2 & 1 & 1\\
\hline 
\zero++ & 011 & 3 & 1 & 2\\
\hline
+\zero\zero & 100 & 4 & 0 & 1\\
\hline 
+\zero+ & 101 & 5 & 0 & 2\\
\hline 
++\zero & 110 & 6 & 0 & 2\\
\hline 
+++ & 111 & 7 & 0 & 3\\
\hline
\end{tabular}
\caption{Table of properties of all the 7 ($2^p - 1$)configurations $\tau \in \mathcal{L}$ for $p = 3$. The binary representations of the configurations are obtained by replacing $\zero$ by 0 and $+$ by 1.}
\label{tab:tauprops}
\end{table}
Given $\mt_{\tau_1, \tau_2}$, the cardinality of the set $\{\ket{\vec{\tau}[N]}\}$ given $\ket{\vec{\tau}[N-1]} = \ket{\tau_2\tau_3\cdots \tau_N}$ is given by $\sumal{\tau_1 \in \mathcal{L}}{}{\mt_{\tau_1, \tau_2}}$.
Thus, we obtain
\begin{equation}
	C_N = \sumal{\tau_1, \tau_2 \in \ml}{}{\mt_{\tau_1, \tau_2} C_{N-1, \tau_2}}. 
\label{eq:cardinality}
\end{equation}
Using Eqs.~(\ref{eq:cardinality}) and (\ref{eq:CLCLsig}), we deduce that $C_{N,\tau_1}$ is related to $C_{N-1, \tau_2}$ via the relation
\begin{equation}
    C_{N,\tau_1} = \sumal{\tau_2 \in \ml}{}{\mt_{\tau_1, \tau_2} C_{N-1,\tau_2}}.
\label{eq:CLsigrecursion}
\end{equation}
In Eq.~(\ref{eq:CLsigrecursion}), $\mt$ can be viewed as a $(2^p-1) \times (2^p-1)$ ``transfer matrix". 
Applying Eq.~(\ref{eq:CLsigrecursion}) repeatedly and using Eq.~(\ref{eq:CLCLsig}), we obtain
\begin{equation}
    C_{N} = \sumal{\{\tau_j \in \ml\}}{}{\mt_{\tau_1, \tau_2} \mt_{\tau_2, \tau_3} \cdots \mt_{\tau_j, \tau_{j+1}} \cdots \mt_{\tau_{N-2}, \tau_{N-1}} C_{2, \tau_{N-1}}},
\label{eq:CLexpand}
\end{equation}
We now focus on computing $C_{2, \tau}$, the number of connected two unit cell configurations of the form $\ket{\tau \sigma}$ for some $\sigma$, which is required to evaluate $C_N$ using Eq.~(\ref{eq:CLexpand}). 
Such a configuration satisfies $P_{\tau} = M_{\sigma} \geq 1$ (according to (c1) and (d1)).
Moreover, $M_{\tau} = P_{\sigma} = 0$ according to (c3). 
For example, the only connected 2 unit cell configurations when $p = 2$ are
\begin{eqnarray}
    &\twoket{\zero+}{-\zero}, \twoket{+\zero}{-\zero}, \twoket{++}{--}\nn \\
    &\twoket{\zero+}{\zero-}, \twoket{\zero+}{-\zero}.\nn \\
\end{eqnarray} 
For a fixed unit cell configuration $\tau$, the number of connected 2 unit cell configurations of the form $\ket{\tau \sigma}$ is the number of distinct configurations $\sigma$ with $M_{\sigma} = P_{\tau}$ $-$'s (since the configuration $\sigma$ does not have any $+$'s). 
Thus, we obtain
\begin{equation}
    C_{2, \tau} = \binom{p}{P_\tau}. 
\label{eq:C2tau}
\end{equation}
Thus, using Eqs.~(\ref{eq:CLexpand}) and (\ref{eq:C2tau}), and noting that 
$C_1 = 1$ (because $\ket{\ \fbox{\zero$\cdots$\zero}\ }$ is the only connected configuration when $N = 1$) we can express $C_N$ as 
\begin{eqnarray}
    C_N &=& \threepartdef{1}{N = 1}{\sumal{\tau \in \mathcal{L}}{}{\binom{p}{P_\tau}}}{N = 2}{\sumal{\{\tau_j \in \ml\}}{}{\mt_{\tau_1, \tau_2} \cdots \mt_{\tau_{N-2}, \tau_{N-1}} \binom{p}{P_{\tau_{N-1}}}}}{N \geq 3} \nn \\
    &=& \threepartdef{1}{N = 1}{
    \begin{bmatrix}
    1 & 1 & \cdots & 1
    \end{bmatrix} 
    \begin{bmatrix}
        \binom{p}{P_1} \\
        \binom{p}{P_2} \\
        \vdots \\
        \binom{p}{P_{2^p - 1}}
    \end{bmatrix}}{N = 2}{\begin{bmatrix}
    1 & 1 & \cdots & 1
    \end{bmatrix}  \mt^{N-2} \begin{bmatrix}
        \binom{p}{P_1} \\
        \binom{p}{P_2}\\
        \vdots \\
        \binom{p}{P_{2^p - 1}}
    \end{bmatrix}}{N \geq 3}.
\label{eq:CLexprint}
\end{eqnarray}
In Eq.~(\ref{eq:CLexprint}), we have abused notation for $\tau$ to represent both the configuration in $\mathcal{L}$ as well as the index of the configuration (the correspondence being one to one using the binary representation). 
For example, $P_2$ should be understood as $P_{\tau = \fbox{$\zero\zero\cdots\zero+\zero$}}$. 
The expression for $C_N$ in Eq.~(\ref{eq:CLexprint}) can thus be rewritten as
\begin{equation}
    C_N = \threepartdef{1}{N = 1}{h^T v}{N = 2}{h^T \mt^{N-2} v}{N \geq 3},
\label{eq:CLexpr}
\end{equation}
where $h$ and $v$ are $\left(2^p - 1\right)$-dimensional vectors whose bases are labelled by the configurations in the set $\mathcal{L}$. 
In Eq.~(\ref{eq:CLexpr}), the components of $h$ and $v$ thus read
\begin{equation}
    h_{\tau} = 1,\;\;\;
    v_{\tau} = \binom{p}{P_\tau},\;\;\;1 \leq \tau \leq 2^p - 1.
\label{eq:lrdefn}
\end{equation}
Thus, using Eq.~(\ref{eq:lrdefn}) and the values of $P_\tau$ given in Table~\ref{tab:tauprops}, the vectors $h$ and $v$ read (using the ordering specified by the binary representations in Table~\ref{tab:tauprops})
\begin{equation}
    h =
    \begin{bmatrix}
        1 \\
        1 \\
        1 \\
        1 \\
        1 \\
        1 \\
        1
    \end{bmatrix},\;\;\;
    v = 
    \begin{bmatrix}
        3 \\
        3 \\
        3 \\
        3 \\
        3 \\
        3 \\
        1
    \end{bmatrix}.
\end{equation}
\subsection{Counting $D_N$}
Given the number of connected configurations $C_n$ for every size $n < N$, we describe two ways to obtain the total Hilbert space dimension. 
First, a recursion relation for the Hilbert space dimension for a chain of $N$ unit cells can be obtained by noting that a connected configuration of size $n$ ($1 \leq n \leq N - 1$) can be appended to any configuration of size $N - n$ to obtain a valid configuration with $N$ unit cells.
Thus, the recursion relation reads
\begin{equation}
    D_N = C_N + \sumal{n = 1}{N-1}{D_{N-n} C_n}.
\label{eq:DLrecursion}
\end{equation}
Alternately, $D_N$ can be computed directly without the use of recursion. This method will enable an estimation of the quantum dimension of the constrained Hilbert space in App.~\ref{app:quantdim}. 
Given the number of connected configurations $C_n$ for every size $n < N$, the total Hilbert space dimension $D_N$ can be obtained by viewing connected configurations as the fundamental building blocks of any valid configuration, as illustrated in Eqs.~(\ref{eq:exampleconfigs}) and (\ref{eq:exampleconnectedconfig}).
Thus, any valid configuration of $N$ unit cells can be constructed by placing one or more connected configurations adjacent to each other.
Since we know the number of connected configurations of $n$ unit cells to be $C_n$, we will now count the number of valid configurations by counting the number of ways connected configurations can be placed adjacent to each other to construct valid configurations of $N$ unit cells.  
Hence we define the quantity
\begin{center}
    \begin{tabular}{p{2cm}p{6cm}}
        $C_{\left(j_1, j_2, \cdots, j_N\right)}$: & Number of valid configurations that are composed of $j_1$ connected configurations of 1 unit cell, $j_2$ connected configurations of $2$ unit cells, and so on up to $j_N$ connected configurations of $N$ unit cells. Here we do not impose any restrictions on the total number of unit cells in the valid configuration, which is given by $\sumal{l = 1}{N}{l j_l}$.   \\
    \end{tabular}
\end{center}
Given that the number of connected configurations of $n$ unit cells is $C_n$, the number $C_{\left(j_1, j_2, \cdots, j_N\right)}$ is given by the standard combinatorics result
\begin{equation}
    C_{\left(j_1, j_2, \cdots, j_N\right)} \equiv \binom{j_1 + j_2 + \dots + j_N}{j_1,\  j_2,\  \dots,\  j_N} \left(C_1\right)^{j_1} \left(C_2\right)^{j_2} \dots \left(C_N\right)^{j_N},
\label{eq:numconnectedvarious}
\end{equation}
where 
\begin{equation}
    \binom{j_1 + j_2 + \dots + j_N}{j_1,\  j_2,\  \dots,\  j_N} = \frac{\left(j_1 + j_2 + \dots + j_N\right)!}{j_1!j_2! \dots j_N!}
\label{eq:multinomial}
\end{equation}
is a multinomial coefficient.
Thus, using Eq.~(\ref{eq:numconnectedvarious}), we obtain the expression for total number of valid configurations, i.e. the Hilbert space dimension as 
\begin{eqnarray}
    &D_N =  \sumal{\{j_k\} = 0}{N}{C_{\left(j_1, j_2, \cdots, j_N\right)} \delta\left({\sumal{l=1}{N}{l j_l}, N}\right)} \nn \\ 
    &= \sumal{\{j_k\} = 0}{N}{\binom{j_1 + j_2 + \dots + j_N}{j_1,\  j_2,\  \dots,\  j_N} \left(C_1\right)^{j_1} \left(C_2\right)^{j_2} \dots \left(C_N\right)^{j_N} \delta\left({\sumal{l = 1}{N}{l j_l}, N}\right)}, \nn \\
\label{eq:DLalternate}
\end{eqnarray}
where we have imposed the constraint that the chain has $N$ unit cells using the Kronecker delta function $\delta\left(\sumal{l=1}{N}{lj_l}, N\right)$. 
\subsection{Examples of Counting}\label{app:explicitcount}
We now provide explicit examples of the counting of $D_N$ when $p = 1$ and $p = 2$. 
\subsubsection{$p = 1$}
When $p = 1$, the only configuration in the set $\mathcal{L}$ is $\tau = \fbox{+}$, which has $Z_\tau = 0$ and $P_\tau = 1$.
Thus, using Eq.~(\ref{eq:mtdefn}), $\mt$ is a $1 \times 1$ matrix 
\begin{equation}
    \mt = 
    \begin{bmatrix}
        0
    \end{bmatrix}.
\label{eq:mtp=1}
\end{equation}
Furthermore, using Eq.~(\ref{eq:lrdefn}), we obtain
\begin{equation}
    h =
    \begin{bmatrix}
        1
    \end{bmatrix}\;\;\;
    v = 
    \begin{bmatrix}
        1
    \end{bmatrix}.
\label{eq:lrp=1}
\end{equation}
Thus, when $p = 1$, using Eqs.~(\ref{eq:CLexpr}), (\ref{eq:mtp=1}) and (\ref{eq:lrp=1}), we obtain
\begin{equation}
    C_N = \twopartdef{1}{N = 1, 2}{0}{N \geq 3},
\label{eq:CLp=1}
\end{equation}
which means that one cannot obtain a connected configuration of more than two unit cells. Indeed, the longest connected configuration with $p = 1$ is $\ket{\ \fbox{+}\ \fbox{\minus}\ }$.
Substituting Eq.~(\ref{eq:CLp=1}) into Eq.~(\ref{eq:DLrecursion}), we obtain
\begin{equation}
    D_N = D_{N-1} + D_{N-2}, 
\end{equation}
which is the usual Fibonacci recursion relation.
Since $D_1 = 1$ (the only valid configuration is $\oneket{\zero}$) and $D_2 = 2$ (the valid configurations are $\twoket{+}{-}$ and $\twoket{\zero}{\zero}$), we obtain 
\begin{equation}
    D_N = F_{N + 1},
\end{equation}
where $F_n$ is the $n$-th Fibonacci number. 
The same result can be obtained using Eqs.~(\ref{eq:DLalternate}) and (\ref{eq:CLp=1}):
\begin{eqnarray}
    D_N &=& \sumal{j_1 + 2j_2 = N}{}{\binom{j_1 + j_2}{j_1}} \nn \\
    &=& \sumal{j = 0}{\lfloor{\frac{N}{2}}\rfloor}{\binom{N - j}{j}} \nn \\
    &=& F_{N+1}.
\end{eqnarray}
\subsubsection{$p = 2$}
When $p = 2$, the set $\mathcal{L}$, i.e. the set of all possible configurations of the leftmost unit cell in a connected configuration, has three configurations 
\begin{equation}
    \fbox{\zero+},\ \fbox{+\zero},\ \fbox{++}, 
\label{eq:vboundp=2}
\end{equation}
which have $Z_\tau = 1, 0, 0$ and $P_\tau = 1, 1, 2$ respectively. 
Thus, using Eq.~(\ref{eq:mtdefn}), $\mt$ is a $3 \times 3$ matrix that reads
\begin{equation}
    \mt = 
    \begin{bmatrix}
        1 & 1 & 0 \\
        0 & 0 & 0 \\
        0 & 0 & 0 \\
    \end{bmatrix}.
\label{eq:mtp=2}
\end{equation}
Similarly, using Eq.~(\ref{eq:lrdefn}) and the values of $Z_\tau$ and $P_\tau$ for the configurations in Eq.~(\ref{eq:vboundp=2}), $h$ and $v$ are 3-dimensional vectors that read
\begin{equation}
    h = 
    \begin{bmatrix}
        1 \\
        1 \\
        1
    \end{bmatrix}\;\;\;
    v = 
    \begin{bmatrix}
        2 \\
        2 \\
        1
    \end{bmatrix}.
\label{eq:lrp=2}
\end{equation}
Thus, using Eqs.~(\ref{eq:mtp=2}), (\ref{eq:lrp=2}) and (\ref{eq:CLexpr}), we obtain
\begin{equation}
    C_N = \threepartdef{1}{N = 1}{5}{N = 2}{4}{N \geq 3},
\label{eq:CLp=2}
\end{equation}
and the Hilbert space dimension $D_N$ can be computed numerically using Eqs.~(\ref{eq:CLp=2}) and (\ref{eq:DLalternate}).
\subsection{Quantum Dimension}\label{app:quantdim}
We now estimate the quantum dimension for the constrained Hilbert space for general $p$ using Eqs.~(\ref{eq:CLexpr}) and (\ref{eq:DLalternate}) by performing a saddle point approximation with $N$ as the large parameter.
We first apply Stirling's approximation
\begin{equation}
    n! \approx \sqrt{2 \pi n} \left(\frac{n}{e}\right)^n
\label{eq:stirling}
\end{equation}
to the multinomial coefficient of Eq.~(\ref{eq:multinomial}) and obtain 
\begin{equation}
    \binom{j_1 + j_2 + \dots + j_N}{j_1,\  j_2,\  \dots,\  j_N} \approx \frac{1}{\left(2\pi\right)^{\frac{N-1}{2}}}\sqrt{\frac{\sumal{l = 1}{N}{j_l}}{\prodal{l=1}{N}{j_l}}} \frac{\left(\sumal{l = 1}{N}{j_l}\right)^{\left(\sumal{l = 1}{N}{j_l}\right)}}{\prodal{l=1}{N}{j_l^{j_l}}}.
\label{eq:multinomialstirling1}
\end{equation}
Introducing the notations
\begin{equation}
    x_k = \frac{j_k}{N},\;\;1 \leq k \leq N, \nn \\
\end{equation}
and
\begin{eqnarray}
    &H\left(\{x_k\}\right) \equiv \left(\sumal{l=1}{N}{x_l}\right) \log\left(\sumal{l=1}{N}{x_l}\right) - \sumal{l=1}{N}{\left(x_l \log x_l\right)} \nn \\
    &= \sumal{l=1}{N}{\left(x_l \log\left(\frac{\sumal{k=1}{N}{x_k}}{x_l}\right)\right)},
\label{eq:Hdefn}
\end{eqnarray}
Eq.~(\ref{eq:multinomialstirling1}) reads
\begin{widetext}
\begin{equation}
    \binom{j_1 + j_2 + \dots + j_N}{j_1,\  j_2,\  \dots,\  j_N} \approx  \frac{1}{\left(2\pi N\right)^{\frac{N-1}{2}}}\sqrt{\frac{\sumal{l = 1}{N}{x_l}}{\prodal{l=1}{N}{x_l}}} \frac{\left(\sumal{l = 1}{N}{x_l}\right)^{N\left(\sumal{l = 1}{N}{x_l}\right)}}{\prodal{l=1}{N}{x_l^{N x_l}}} = \frac{1}{\left(2\pi N\right)^{\frac{N-1}{2}}}\sqrt{\frac{\sumal{l = 1}{N}{x_l}}{\prodal{l=1}{N}{x_l}}} e^{N H\left(\{x_k\}\right)}
\label{eq:multinomialstirling}
\end{equation}
Consequently, for large $N$ Eq.~(\ref{eq:DLalternate}) can be written as
\begin{eqnarray}
    D_N &=& \sumal{\{j_k\} = 0}{N}{\frac{1}{\left(2\pi N\right)^{\frac{N-1}{2}}}\sqrt{\frac{\sumal{l = 1}{N}{x_l}}{\prodal{l=1}{N}{x_l}}} e^{N \left(H\left(\{x_k\}\right) + \sumal{l = 1}{N}{x_l \log C_l}\right)} \delta\left({\sumal{l = 1}{N}{l x_l}, 1}\right)}, \nn \\
    &=& \bigint_{0}^{1}{\left[\prodal{l=1}{N}{\mathrm{d}x_l}\frac{N^{\frac{N+1}{2}}}{\left(2\pi\right)^{\frac{N-1}{2}}}\sqrt{\frac{\sumal{l = 1}{N}{x_l}}{\prodal{l=1}{N}{x_l}}} \delta\left(\sumal{l =1}{N}{l x_l} - 1\right)  \exp\left(N \left(H\left(\{x_k\}\right) + \sumal{l = 1}{N}{x_l \log C_l}\right) \right)\right]}
\label{eq:DLintegral}
\end{eqnarray}
We now want to obtain an approximation of Eq.~(\ref{eq:DLintegral}) for large $N$. 
A standard method is to apply the saddle point approximation, which, for a single variable reads
\begin{equation}
    \int_{a}^{b}{\mathrm{d}x\ g(x) e^{N f(x)}} \approx \int_{a}^{b}{\mathrm{d}x\ g(x) e^{N f(x_0) + \frac{N}{2}f''(x_0) (x - x_0)^2}} = g(x_0) e^{N f(x_0)} \sqrt{\frac{2 \pi}{N |f''(x_0)|}},
\label{eq:saddlepoint}
\end{equation}
where $f'(x_0) = 0$ such that $a < x_0 < b$ and $f''(x_0) < 0$.
While a rigorous saddle point approximation with multiple variables in the presence of a constraint ($\delta$ function) requires careful treatment, we follow Eq.~(\ref{eq:saddlepoint}) and approximate the exponential-in-$N$ dependence of $D_N$ in Eq.~(\ref{eq:DLintegral}) as
\begin{equation}
    D_N \sim \exp\left(N \left(H\left(\{y_k\}\right) + \sumal{l=1}{N}{y_l\log C_l}\right)\right),
\label{eq:DLsaddlepoint}
\end{equation}
where $\{y_k\}$ are the parameters $\{x_k\}$ at which the function $H\left(\{x_k\}\right) + \sumal{l = 1}{N}{x_l \log C_l}$ has a ``saddle point" in the presence of the constraint $\sumal{l = 1}{N}{l y_l} = 1$.  
That is, introducing a Lagrange multiplier $\log\lambda$ that enforces $\sumal{l=1}{N}{l y_l} = 1$ we obtain
\begin{eqnarray}
    &&\left.\frac{\partial}{\partial x_k}\left(H\left(\{x_k\}\right) + \sumal{l =1}{N}{x_l \log C_l} -  \log\lambda\left(\sumal{l=1}{N}{l x_l} - 1\right)\right)\right|_{\{x_k\} = \{y_k\}}\hspace{-13mm}  = 0, \label{eq:saddlepointcondition1} \nn \\
    && \\
    &&\left.\frac{\partial}{\partial \log\lambda}\left(H\left(\{x_k\}\right) + \sumal{l =1}{N}{x_l \log C_l} - \log\lambda\left(\sumal{l=1}{N}{l x_l} - 1\right)\right)\right|_{\{x_k\} = \{y_k\}}\hspace{-13mm} = 0, \;\;\forall k,\;\; 1 \leq k \leq N.
\label{eq:saddlepointcondition2}
\end{eqnarray}
When evaluated, Eq.~(\ref{eq:saddlepointcondition1}) reads
\begin{equation}
    \log\left(\sumal{l = 1}{N}{y_l}\right) - \log y_k + \log C_k  - k \log\lambda = 0 \implies \lambda^k y_k = C_k \sumal{l = 1}{N}{y_l},\;\;\forall k,\; 1 \leq k \leq N.
\label{eq:saddlepointequation}
\end{equation}
Evaluating Eq.~(\ref{eq:DLsaddlepoint}) using Eq.~(\ref{eq:saddlepointequation}), we obtain 
\begin{eqnarray}
    D_N &\sim& \exp\left(N\left(-\sumal{l = 1}{N}{x_l \log\left(\lambda^{-l} C_l\right)} + \sumal{l = 1}{N}{x_l \log C_l} \right)\right) \nn \\
    &\sim& \exp\left(N\left(\log \lambda\sumal{l = 1}{N}{l x_l}\right)\right) \nn \\
    &\sim& \lambda^N,
\end{eqnarray}
where we have used Eq.~(\ref{eq:saddlepointcondition2}) or
\end{widetext}
\begin{equation}
    \sumal{l = 1}{N}{l x_l} = 1. 
\end{equation} 
Thus, $\lambda$ is the quantum dimension defined in Eq.~(\ref{eq:quantdef}).
$\lambda$ can be obtained using Eq.~(\ref{eq:saddlepointequation}), which can be written as
\begin{eqnarray}
        &C_k \lambda^{-k} = \frac{y_k}{\sumal{l = 1}{N}{y_l}} \nn \\
    &\implies \sumal{k = 1}{N}{C_k \lambda^{-k}} = 1 \nn \\
    &\implies \lambda^N - \sumal{k = 1}{N}{C_k \lambda^{N-k}} = 0. 
\label{eq:lambdaeqn1}
\end{eqnarray}
Using Eq.~(\ref{eq:CLexpr}), Eq.~(\ref{eq:lambdaeqn1}) can be rewritten as
\begin{eqnarray}
    \lambda^N &=& \lambda^{N-1} + \lambda^{N-2} h^T v + \sumal{k=3}{N}{h^T \mt^{k-2} v \lambda^{N-k}}.
\label{eq:lambdaeqn2}
\end{eqnarray}
To compute the quantum dimension, we want to obtain an equation independent of $N$.
We thus simplify Eq.~(\ref{eq:lambdaeqn2}) further. 
We numerically observe that $\mt$ is diagonalizable for $p \leq 8$. Assuming $\mt$ is always diagonalizable, if written in terms of its eigenstates as
\begin{equation}
    \mt = \sumal{m = 1}{2^p-1}{\theta_m r_m l_m^T},
\label{eq:Teigstates}
\end{equation}
Eq.~(\ref{eq:lambdaeqn2}) simplifies to
\begin{eqnarray}
    \lambda^N - \lambda^{N-1} - \lambda^{N-2}h^T v &=&  \sum_{k=3}^{N}{\sum_{m=1}^{2^p-1}{\left(h^T r_m\right)\left(l_m^T v\right) \theta_m^{k-2}\lambda^{N-k}}} \nn \\
    &=& \lambda^{N-2} \sum_{j = 1}^{N-2}{\sum_{m = 1}^{2^p-1}{e_m \left(\frac{\theta_m}{\lambda}\right)^j}} \nn \\
    &=& \lambda^{N-2}\sum_{m=1}^{2^p-1}{e_m \theta_m\frac{1 - \left(\frac{\theta_m}{\lambda}\right)^{N-2}}{\lambda - \theta_m}} \nn \\
\label{eq:lambdaeqn3}
\end{eqnarray}
where we have defined
\begin{equation}
    e_m \equiv \left(h^T r_m\right)\left(l_m^T v\right).
\label{eq:emdefn}
\end{equation}
For large $N$ and small $p$, Eq.~(\ref{eq:lambdaeqn3}) simplifies to
\begin{equation}
    \lambda^2 - \lambda - h^T v = \sumal{m=1}{2^p-1}{\frac{e_m \theta_m}{\lambda - \theta_m}},
\label{eq:lambdaeqnfin}
\end{equation}
which is a $(p + 1)$-th degree equation for $p \geq 1$. Note that Eq.~(\ref{eq:lambdaeqnfin}) is only valid if $\left(\frac{\theta_1}{\lambda}\right)^N \rightarrow 0$: that is, if the largest eigenvalue of $\mathcal{T}$, $\theta_1$ satisfies $\theta_1 < \lambda$, which we have self-consistently verified for $p \leq 8$ in Table~\ref{tab:qdims}. 
We now provide examples of the computation of the quantum dimension. 
When $p = 1$, using Eqs.~(\ref{eq:mtp=1}), (\ref{eq:lrp=1}), and (\ref{eq:emdefn}), (\ref{eq:lambdaeqnfin}) reduces to 
\begin{equation}
    \lambda^2 - \lambda - 1 = 0, 
\end{equation}
and thus the quantum dimension $\lambda^{(1)}$ is the Golden ratio
\begin{equation}
    \lambda^{(1)} = \varphi = \frac{1 + \sqrt{5}}{2}.
\label{eq:lambdap1}
\end{equation}
When $p = 2$, using Eqs.~(\ref{eq:mtp=2}), (\ref{eq:lrp=2}) and (\ref{eq:emdefn}), one directly obtains
\begin{equation}
    e_1 = 4,\;\;\;\theta_1 = 1,\;\;\; h^T v = 5.
\end{equation}
Thus, Eq.~(\ref{eq:lambdaeqnfin}) simplifies to
\begin{eqnarray}
    \lambda^3 - 2\lambda^2 - 4\lambda + 1  &=& 0 \nn \\
    \implies \lambda \approx 3.16425.
\end{eqnarray}
For other values of $p$, Eq.~(\ref{eq:lambdaeqnfin}) can be solved numerically. The quantum dimensions for a few values of $p$ are tabulated in Table~\ref{tab:qdims}.
We empirically find that the quantum dimension $\lambda^{(p)}$  roughly scales as
\begin{equation}
    \lambda^{(p)} \sim  2^{p-1} \varphi. 
\label{eq:qdimroughscaling}
\end{equation}
However, we observe that the true Hilbert space dimension converges to its asymptotic scaling extremely slowly in $N$.
\begin{table}
\begin{tabular}{|c|c|c|}
\hline
\boldsymbol{$p$} & \boldsymbol{$\lambda^{(p)}$} & \boldsymbol{$\theta^{(p)}_1$} \\
\hline
1 & 1.61803 & 0\\
\hline 
2 & 3.16425 & 1\\
\hline 
3 & 6.35065 & 3.303\\
\hline
4 & 12.9044 & 8.309\\
\hline 
5 & 26.3557 & 18.9515\\
\hline 
6 & 53.7857 & 41.2559\\
\hline 
7 & 109.464 & 87.5446\\
\hline 
8 & 222.135 & 182.926\\
\hline
\end{tabular}
\caption{Table of quantum dimensions $\lambda^{(p)}$ and $\theta^{(p)}_1$, the eigenvalue of largest magnitude of the transfer matrix $\mathcal{T}$ for the constrained Hilbert space at filling $\nu = \frac{p}{2p + 1}$, $1 \leq p \leq 8$.}
\label{tab:qdims}
\end{table}
\section{Transformation of charge under symmetries}\label{sec:chargesym}
In this appendix, we show the transformation of the charge $\Qp_{\vec{\sigma}}$ of Eq.~(\ref{eq:chargegenp}) under the symmetry $\mI\mP$. 
The quantities in Eq.~(\ref{eq:Xdefn}) transform under $\mI\mP$ as
\begin{eqnarray}
    &&P_{\sigma_j} \rightarrow M_{\sigma_{N + 1 - j}},\;\; M_{\sigma_j} \rightarrow P_{\sigma_{N + 1 - j}}, \nn \\
    &&X^{(P)}_{\sigma_j} \rightarrow X^{(M)}_{\sigma_{N + 1 - j}},\;\; X^{(M)}_{\sigma_j} \rightarrow X^{(P)}_{\sigma_{N + 1 - j}},
\label{eq:eletransform}
\end{eqnarray}
where the inversion center is the center of the chain, on a site (resp. bond) if $N$ is odd (resp. even).
Thus, using Eqs.~(\ref{eq:chargegenp}) and (\ref{eq:eletransform}), the charge $\Qp_{\vec{\sigma}}$ transforms as
\begin{eqnarray}
    &\Qp_{\vec{\sigma}} \rightarrow \sumal{j = 1}{N_b}{(-1)^{j+1} \left(\frac{M_{\sigma_{N + 1 -j}} + P_{\sigma_{N - j}}}{2} - \left(X^{(M)}_{\sigma_{N + 1 - j}} + X^{(P)}_{\sigma_{N - j}}\right)\right)} \nn \\
    &= \sumal{k = N - N_b}{N - 1}{(-1)^{N - k - 1} \left(\frac{M_{\sigma_{k + 1}} + P_{\sigma_{k}}}{2} - \left(X^{(M)}_{\sigma_{k + 1}} + X^{(P)}_{\sigma_{k}}\right)\right)} \nn \\
    &= (-1)^N \sumal{j = N - N_b}{N - 1}{(-1)^{j + 1} \left(\frac{P_{\sigma_{j}} + M_{\sigma_{j + 1}}}{2} - \left(X^{(P)}_{\sigma_{j}} + X^{(M)}_{\sigma_{j + 1}}\right)\right)} \nn \\
\label{eq:Qsymtransform}
\end{eqnarray}
Using Eq.~(\ref{eq:Qsymtransform}) and that $N_b = N - 1$ (resp. $N_b = N$) for OBC (resp. PBC), we obtain
\begin{eqnarray}
    &&\Qp_{\vec{\sigma}}  \rightarrow (-1)^N \Qp_{\vec{\sigma}} + \nn \\
    &&\twopartdef{0}{\textrm{OBC}}{2 \delta_{N,\textrm{odd}}\left(\frac{P_{\sigma_{N}} + M_{\sigma_{1}}}{2} - \left(X^{(P)}_{\sigma_{N}} + X^{(M)}_{\sigma_{1}}\right)\right)}{\textrm{PBC}}.\nn \\
\label{eq:Qfullsymtransform}
\end{eqnarray}
Thus, the \emph{parity} of $\Qp_{\vec{\sigma}}$ is invariant under $\mI\mP$, and the $\mathbb{Z}_2$ index $\Cp_{\vec{\sigma}}$ defined in Eq.~(\ref{eq:Z2index}) transforms under $\mI\mP$ as
\begin{equation}
    \Cp_{\vec{\sigma}} \rightarrow \Cp_{\vec{\sigma}}. 
\label{eq:Z2symtransform}
\end{equation}
\section{Structure of the effective Hamiltonians $\boldsymbol{\mH{p}}$} \label{app:symmetries}
In this appendix we discuss the action of the Hamiltonians $\Hp$ in the constrained Hilbert space $\Kp$.
We also show that it can be written as a sum of charge-raising and charge-lowering operators, where the charge $\mQ{p}_{\vec{\sigma}}$ is defined in Eq.~(\ref{eq:chargegenp}). 
\subsection{\boldsymbol{$p = 1$}}
The action of the Hamiltonian $\mH{1}$ is shown in Eq.~(\ref{eq:1b3scattering}). 
We split the Hamiltonian as
\begin{equation}
    \mH{1} = \mH{1}_+ + \mH{1}_-
\label{eq:p1split}
\end{equation}
where
\begin{equation}
    \mH{1}_\pm = \sumal{j = 1}{N_b}{\left(\mH{1}_\pm\right)_{j,j+1}},    
\end{equation}
and the non-vanishing actions of $\mH{1}_+$ and $\mH{1}_-$ are (written compactly)
\begin{eqnarray}
    &&\hspace{0mm}\left(\mH{1}_+ \right)_{j, j+1} \overset{\;\;j\;\;j+1}{\ket{\cdots \bzero \spa \bzero \cdots}} = \overset{\;\;j\;\;j+1}{\ket{\cdots\spa \bplus \spa \bminus \cdots}} \; \textrm{if $j$ is even}\nn \\
    &&\hspace{0mm}\left(\mH{1}_+ \right)_{j, j+1} \overset{\;\;j\;\;j+1}{\ket{\cdots \bplus \spa \bminus\cdots}} = \overset{\;\;j\;\;j+1}{\ket{\cdots\bzero \spa \bzero\cdots}}\; \textrm{if $j$ is odd}\nn \\
    &&\hspace{0mm}\left(\mH{1}_- \right)_{j, j+1} \overset{\;\;j\;\;j+1}{\ket{\cdots \bzero \spa \bzero \cdots}} = \overset{\;\;j\;\;j+1}{\ket{\cdots\spa \bplus \spa \bminus \cdots}} \; \textrm{if $j$ is even}\nn \\
    &&\hspace{0mm}\left(\mH{1}_-\right)_{j, j+1} \overset{\;\;j\;\;j+1}{\ket{\cdots \bplus \spa \bminus\cdots}} = \overset{\;\;j\;\;j+1}{\ket{\cdots\bzero \spa \bzero\cdots}}\; \textrm{if $j$ is odd}.\nn \\
\label{eq:Hpmaction}
\end{eqnarray}
The actions of Eq.~(\ref{eq:Hpmaction}) are chosen such that $\mH{1}_+$ and $\mH{1}_-$ are charge-raising and charge-lowering parts of the action of the Hamiltonian $\mH{1}$ (see Eq.~(\ref{eq:1b3scattering})), where the charge $\mQ{1}$ is defined in Eq.~(\ref{eq:chargep1}).
Note that $[\Hp_+, \Hp_-] \neq 0$.
Using Eq.~(\ref{eq:Hpmaction}), if
\begin{equation}
    \left(\mH{1}_\pm\right)_{j,j+1}\ket{\vec{\sigma}} = \ket{\vec{\tau}}
\end{equation}
for product configurations $\ket{\vec{\sigma}}$ and $\ket{\vec{\tau}}$, then exactly one of the following holds:
\begin{enumerate}
    \item[(i)] $P_{\tau_j} = P_{\sigma_j} \pm 1$ and $M_{\tau_{j+1}} = M_{\sigma_{j+1}} \pm 1$ if $(-1)^j = \pm 1$
    \item[(ii)] $P_{\tau_j} = P_{\sigma_j} \mp 1$ and $M_{\tau_{j+1}} = M_{\sigma_{j+1}} \mp 1$ if $(-1)^j = \mp 1$
\end{enumerate}
Thus, using the definition of charge in Eq.~(\ref{eq:chargep1}), in either of the cases (i) or (ii) we obtain
\begin{eqnarray}
    \mQ{1}_{\vec{\tau}} = \mQ{1}_{\vec{\sigma}} \pm  1.
\end{eqnarray}
Thus, $\mH{1}_+$ and $\mH{1}_-$ are the charge raising and lowering operators respectively.
In the operator language, $\mH{1}_+$ and $\mH{1}_-$ read
\begin{eqnarray}
    &&\mH{1}_+ = -\sumal{j\ \textrm{even}}{}{U^\dagger_{j,1} T_{j+1,1}} - \sumal{j\  \textrm{odd}}{}{T^\dagger_{j+1,1} U_{j,1}} \nn \\
    &&\mH{1}_- = -\sumal{j\ \textrm{even}}{}{T^\dagger_{j+1,1} U_{j,1}} - \sumal{j\  \textrm{odd}}{}{U^\dagger_{j,1} T_{j+1,1}}. 
\label{eq:FSAhamil1/3}
\end{eqnarray}
Thus, using Eqs.~(\ref{eq:effectivehamil1b3}) and (\ref{eq:FSAhamil1/3}), Eq.~(\ref{eq:p1split}) is verified. 
\subsection{General $p$}
A similar property holds for the Hamiltonian $\Hp$ of Eq.~(\ref{eq:genericfillinghamil}) as well. 
That is, it can be split into two parts as
\begin{equation}
    \Hp = \Hp_+ + \Hp_-,
\label{eq:hamilsplit}
\end{equation}
where
\begin{equation}
    \Hp_\pm = \sumal{j = 1}{N}{\left(\Hp_\pm\right)_{j}} + \sumal{j = 1}{N_b}{\left(\Hp_\pm\right)_{j, j+1}} 
\end{equation}
where $N_b = N - 1$ for OBC and $N_b = N$ for PBC. 
Further, 
\begin{equation}
    \left(\Hp_\pm\right)_j = \sumal{n = 1}{p-1}{\left(\left(\Hp_\pm\right)_{j}\right)_{n,n+1}}.
\end{equation}
The non-vanishing actions of $\Hp_+$ and $\Hp_-$ read 
\begin{widetext}
\begin{eqnarray}
    &&\left(\left(\Hp_+\right)_j\right)_{n,n+1}\overset{n\ n+1}{\oneket{\cdots \zero + \cdots}} = \overset{n\ n+1}{\oneket{\cdots + \zero \cdots}}\;\; \textrm{if $j$ is even}\nn \\
    &&\left(\left(\Hp_-\right)_j\right)_{n,n+1}\overset{n\ n+1}{\oneket{\cdots \zero + \cdots}} = \overset{n\ n+1}{\oneket{\cdots + \zero \cdots}}\;\; \textrm{if $j$ is odd}\label{eq:ic1}
\end{eqnarray}%
\begin{eqnarray}
    &&\left(\left(\Hp_+\right)_j\right)_{n,n+1}\overset{n\ n+1}{\oneket{\cdots + \zero \cdots}} = \overset{n\ n+1}{\oneket{\cdots \zero + \cdots}}\;\; \textrm{if $j$ is odd} \nn \\
    &&\left(\left(\Hp_-\right)_j\right)_{n,n+1}\overset{n\ n+1}{\oneket{\cdots + \zero \cdots}} = \overset{n\ n+1}{\oneket{\cdots \zero + \cdots}}\;\; \textrm{if $j$ is even} \label{eq:ic2}
\end{eqnarray}%
\begin{eqnarray}
    &&\left(\left(\Hp_+\right)_j\right)_{n,n+1}\overset{n\ n+1}{\oneket{\cdots \zero - \cdots}} = \overset{n\ n+1}{\oneket{\cdots - \zero \cdots}}\;\; \textrm{if $j$ is odd}\nn \\
    &&\left(\left(\Hp_-\right)_j\right)_{n,n+1}\overset{n\ n+1}{\oneket{\cdots \zero - \cdots}} = \overset{n\ n+1}{\oneket{\cdots - \zero \cdots}}\;\; \textrm{if $j$ is even}\label{eq:ic3}
\end{eqnarray}
\begin{eqnarray}
    &&\left(\left(\Hp_+\right)_j\right)_{n,n+1}\overset{n\ n+1}{\oneket{\cdots - \zero \cdots}} = \overset{n\ n+1}{\oneket{\cdots \zero - \cdots}}\;\; \textrm{if $j$ is even} \nn \\
    &&\left(\left(\Hp_-\right)_j\right)_{n,n+1}\overset{n\ n+1}{\oneket{\cdots - \zero \cdots}} = \overset{n\ n+1}{\oneket{\cdots \zero - \cdots}}\;\; \textrm{if $j$ is odd} \label{eq:ic4}
\end{eqnarray}
\begin{eqnarray}
    &&\left(\Hp_+\right)_{j, j + 1}\overset{j\;\;j+1}{\twoket{\cdots \zero}{\zero \cdots}} = \overset{j\;\;j+1}{\twoket{\cdots +}{- \cdots}}\;\; \textrm{if $j$ is odd} \nn \\
    &&\left(\Hp_-\right)_{j, j + 1}\overset{j\;\;j+1}{\twoket{\cdots \zero}{\zero \cdots}} = \overset{j\;\;j+1}{\twoket{\cdots +}{- \cdots}}\;\; \textrm{if $j$ is even} \label{eq:ic5}
\end{eqnarray}
\begin{eqnarray}
    &&\left(\Hp_+ \right)_{j, j+1}\overset{j\;\;j+1}{\twoket{\cdots +}{- \cdots}} = \overset{j\;\;j+1}{\twoket{\cdots \zero}{\zero \cdots}} \;\; \textrm{if $j$ is even}.\nn \\
    &&\left(\Hp_- \right)_{j, j+1}\overset{j\;\;j+1}{\twoket{\cdots +}{- \cdots}} = \overset{j\;\;j+1}{\twoket{\cdots \zero}{\zero \cdots}} \;\; \textrm{if $j$ is odd}.\label{eq:ic6}
\end{eqnarray}
\end{widetext}
As we will shortly show, the actions in Eqs.~(\ref{eq:ic1}) to (\ref{eq:ic6}) have been chosen so that $\Hp_+$ and $\Hp_-$ are the charge-raising and charge-lowering parts of the actions of $\Hp$ (see Eqs.~(\ref{eq:borderunitaction}) and (\ref{eq:withinunitaction})), where the charge $\Qp$ is defined in Eq.~(\ref{eq:chargegenp}).
Using Eqs.~(\ref{eq:ic1}) to (\ref{eq:ic6}), if
\begin{equation}
    \left(\left(\mH{p}_\pm\right)_j\right)_{n,n+1}\ket{\vec{\sigma}} = \ket{\vec{\tau}} 
\end{equation}
or
\begin{equation}
    \left(\mH{p}_\pm\right)_{j,j+1}\ket{\vec{\sigma}} = \ket{\vec{\tau}} 
\end{equation}
for product configurations $\ket{\vec{\sigma}}$ and $\ket{\vec{\tau}}$, then exactly one of the following holds:
\begin{enumerate}
    \item[(i)] $X^{(P)}_{\tau_j} = X^{(P)}_{\sigma_j} \pm 1$ if $(-1)^j = \mp 1$ (Eq.(\ref{eq:ic1}) or Eq.~(\ref{eq:ic2}))
    \item[(ii)] $X^{(M)}_{\tau_j} = X^{(M)}_{\sigma_j} \pm 1$ if $(-1)^j = \pm 1$ (Eq.(\ref{eq:ic3}) or Eq.~(\ref{eq:ic4}))
    \item[(iii)] $P_{\tau_j} = P_{\sigma_j} \pm 1$, $M_{\tau_{j+1}} = M_{\sigma_{j+1}} \pm 1$, $X^{(P)}_{\tau_j} = X^{(P)}_{\sigma_j} \pm 1$, $X^{(M)}_{\tau_{j+1}} = X^{(M)}_{\sigma_{j+1}} \pm 1$ if $(-1)^j = \pm 1$ (Eq.(\ref{eq:ic5}) or Eq.~(\ref{eq:ic6}))
\end{enumerate}
Thus, using Eq.~(\ref{eq:chargegenp}), in all the cases (i), (ii) and (iii) we obtain
\begin{equation}
    \mathcal{Q}_{\vec{\tau}} = \mathcal{Q}_{\vec{\sigma}} \pm 1.
\label{eq:chargechange}
\end{equation}
In the operator language, $\Hp_+$ and $\Hp_-$ respectively read
\begin{eqnarray}
    &\Hp_+ = -\sumal{j\ \textrm{even}}{}{U^\dagger_{j, p} T_{j+1, 1}} -\sumal{j\ \textrm{odd}}{}{T^\dagger_{j+1,1} U_{j,p}} \nn \\
    &+\sumal{j\ \textrm{odd}}{}{\ \sumal{n = 1}{p-1}{\left(U^\dagger_{j,n+1} U_{j,n} + T^\dagger_{j,n+1} T_{j,n}\right)}} \nn \\
    &+ \sumal{j\ \textrm{even}}{}{\ \sumal{n = 1}{p-1}{\left(U^\dagger_{j, n} U_{j,n+1} + T^\dagger_{j,n} T_{j,n+1}\right)}},
\label{eq:FSAhamilgeneral+}
\end{eqnarray}
and
\begin{eqnarray}
    &\Hp_- = -\sumal{j\ \textrm{even}}{}{T^\dagger_{j+1,1} U_{j,p}} - \sumal{j\ \textrm{odd}}{}{U^\dagger_{j, p} T_{j+1, 1}}\nn \\
    &+\sumal{j\ \textrm{odd}}{}{\ \sumal{n = 1}{p-1}{\left(U^\dagger_{j, n} U_{j,n+1} + T^\dagger_{j,n} T_{j,n+1}\right)}}\nn \\
    &+ \sumal{j\ \textrm{even}}{}{\ \sumal{n = 1}{p-1}{\left(U^\dagger_{j,n+1} U_{j,n} + T^\dagger_{j,n+1} T_{j,n}\right)}}.
\label{eq:FSAhamilgeneral-}
\end{eqnarray}
Using Eqs.~(\ref{eq:FSAhamilgeneral+}), (\ref{eq:FSAhamilgeneral-}), and (\ref{eq:genericfillinghamil}), Eq.~(\ref{eq:hamilsplit}) is verified. 
\section{Symmetry-Protected Zero-modes}\label{app:zeromodes}
To obtain a lower-bound on the number of zero-modes, we need to obtain $N^p_e$ and $N^p_o$ (see Eq.~(\ref{eq:zerobounduseful})). That is, we need to study symmetric product configurations (i.e. those that satisfy $\ket{\vec{\sigma}} = \mI\mP\ket{\vec{\sigma}}$, see Eq.~(\ref{eq:productinvariant})) and their $\mathbb{Z}_2$ indices $\Cp_{\vec{\sigma}} = \pm 1$ as defined in Eq.~(\ref{eq:Z2index}). 
As shown in Eqs.~(\ref{eq:chargep1}) and (\ref{eq:chargegenp}), the charge for a product configuration $\Qp_{\vec{\sigma}}$ is of the form
\begin{equation}
    \Qp_{\vec{\sigma}} = \sumal{j = 1}{N_b}{\Qp_{\sigma_j, \sigma_{j+1}}}, 
\label{eq:Qpstruct}
\end{equation}
where
\begin{equation}
    \Qp_{\sigma_j,\sigma_{j+1}} \equiv (-1)^{j + 1}\left(\frac{P_{\sigma_j} + M_{\sigma_{j+1}}}{2} - \left(X^{(P)}_{\sigma_j} + X^{(M)}_{\sigma_{j+1}}\right)\right).
\label{eq:Qpjjdefn}
\end{equation} 
If the configuration $\vec{\sigma}$ is an $\mI\mP$-symmetric product state, according to Eq.~(\ref{eq:eletransform}), the following identities hold
\begin{eqnarray}
    &&\Qp_{\sigma_j, \sigma_{j+1}} = (-1)^N \Qp_{\sigma_{N - j}, \sigma_{N - j + 1}},\;\;\forall j,\ 1 \leq j < \frac{N}{2}\nn \\
    &&\Qp_{\sigma_{\frac{N}{2}}, \sigma_{\frac{N}{2} + 1}} = P_{\sigma_{\frac{N}{2}}} - 2 X^{(P)}_{\sigma_{\frac{N}{2}}}\;\;\;\textrm{if $N$ is even},\nn \\
    &&\Qp_{\sigma_{N}, \sigma_{1}} = P_{\sigma_{N}} - 2 X^{(P)}_{\sigma_{N}}\;\;\;\textrm{for PBC}.
\label{eq:sigmajcondition}
\end{eqnarray}
For such a configuration $\vec{\sigma}$, its $\mathbb{Z}_2$ index $\Cp_{\vec{\sigma}} = (-1)^{\Qp_{\vec{\sigma}}}$ can be written as 
\begin{equation}
    \Cp_{\vec{\sigma}} = \fourpartdef{(-1)^{P_{\sigma_{\frac{N}{2}}}}}{\textrm{OBC}, N\ \textrm{even}}{1}{\textrm{OBC}, N\ \textrm{odd}}{(-1)^{P_{\sigma_{\frac{N}{2}}} + P_{\sigma_{{N}}}}}{\textrm{PBC}, N\  \textrm{even}}{(-1)^{P_{\sigma_{{N}}}}}{\textrm{PBC}, N\ \textrm{odd}}. 
\label{eq:Cpsimp}
\end{equation}
Since $N^p_o = 0$ when $N$ is odd for OBC, according to Eq.~(\ref{eq:zerobounduseful}) the number of zero modes is lower bounded by the total number of (even) symmetric product states. Since symmetric product configurations can be uniquely determined by the configuration of half of the chain, we expect the number of zero modes to scale as $D^{(p)}_{N/2} \sim \sqrt{D^{(p)}_N}$. 
We believe that for a large $N$ the same scaling holds for systems with PBC and those with OBC. 
The lower bound for the number of zero models has been calculated exactly for the PXP model,\cite{turner2018quantum} which is equivalent to the $p = 1$ case.
\section{Lowest and Highest Charged States}\label{app:charge}
In this appendix, we obtain the lowest charge and highest charge configurations for $N$ unit cells with OBC and PBC.
We focus on the case where $N$ is even, since this is the relevant one for the discussion of quantum many-body scars in Sec.~\ref{sec:scars}.
\subsection{\boldsymbol{$p = 1$}}
\subsubsection{PBC}
As shown in Eq.~(\ref{eq:chargep1}), the charge for a configuration when $p = 1$ is defined as (for even $N$)
\begin{eqnarray}
    &\mQ{1}_{\vec{\sigma}} = \sumal{j = 1}{N}{(-1)^{j} \left(\frac{P_{\sigma_j} + M_{\sigma_{j+1}}}{2}\right)} \nn \\
    &= \sumal{j = 1}{N}{(-1)^{j} \left(\frac{P_{\sigma_j} - M_{\sigma_{j}}}{2}\right)}, 
\label{eq:chargep1app}
\end{eqnarray}
where
\begin{equation}
    P_{\sigma_j} = \twopartdefoth{1}{\sigma_j = +}{0}, \;\;\; M_{\sigma_j} = \twopartdefoth{1}{\sigma_j = -}{0}.
\label{eq:PMdefn}
\end{equation}
Using Eq.~(\ref{eq:chargep1app}), the lowest (resp. highest) charge configuration can be obtained by setting
\begin{equation}
    P_{\sigma_j} - M_{\sigma_j} = \twopartdef{+1}{\textrm{$j$ is odd (resp. even)}}{-1}{\textrm{$j$ is even (resp. odd)}}.
\label{eq:lowestchargecond}
\end{equation}
Thus the lowest and highest charge states, which we respectively call $\ztwo{1}$ and $\ztwop{1}$, read (for even $N$)
\begin{eqnarray}
    &&\ztwo{1} = \ket{\ \bplus\ \bminus\ \bplus\cdots\bminus\ \bplus\ \bminus\ } \nn \\
    &&\ztwop{1} = \ket{\ \bminus\ \bplus\ \bminus\cdots\bplus\ \bminus\ \bplus\ },
\label{eq:z2z2ppbc}
\end{eqnarray}
which have charges
\begin{equation}
    \mQ{1}_{\mathbb{Z}_2} = - N/2,\;\;\; \mQ{1}_{\mathbb{Z}_2'} = + N/2.
\label{eq:chargez2pbc}
\end{equation}
\subsubsection{OBC}
For open boundary conditions, the definition $\Qp_{\vec{\sigma}}$ for even $N$ reads
\begin{eqnarray}
    \mQ{1}_{\vec{\sigma}} &=& \sumal{j = 1}{N-1}{(-1)^{j} \left(\frac{P_{\sigma_j} + M_{\sigma_{j+1}}}{2}\right)}\nn \\
    &=& \sumal{j = 2}{N-1}{(-1)^{j} \left(\frac{P_{\sigma_j} - M_{\sigma_{j}}}{2}\right)} - \frac{P_{\sigma_1} - M_{\sigma_{N}}}{2}.\nn \\
\label{eq:chargep1appobc}
\end{eqnarray}
The lowest (resp. highest) configuration can then be obtained by satisfying Eq.~(\ref{eq:lowestchargecond}) for all $j$, $2 \leq j \leq N-1$.
Moreover, the lowest (resp. highest) charge configuration should satisfy $P_{\sigma_1} = 1$ (resp. $P_{\sigma_1} = 0$) and $M_{\sigma_N} = 1$ (resp. $M_{\sigma_N} = 1$). 
Thus the lowest and highest charge states for OBC (for even $N$) read
\begin{eqnarray}
    &&\ztwo{1} = \ket{\ \bplus\ \bminus\ \bplus\cdots\bminus\ \bplus\ \bminus\ } \nn \\
    &&\ztwop{1} = \ket{\ \bzero\ \bplus\ \bminus\cdots\bplus\ \bminus\ \bzero\ },
\label{eq:z2z2pobc}
\end{eqnarray}
which have charges (according to Eq.~(\ref{eq:chargep1}))
\begin{equation}
    \mQ{1}_{\mathbb{Z}_2} = - N/2,\;\;\; \mQ{1}_{\mathbb{Z}_2'} = + N/2 - 1.
\label{eq:chargez2obc}
\end{equation}
\subsection{General \boldsymbol{$p$}}\label{app:chargegenp}
\subsubsection{PBC}
As shown in Eq.~(\ref{eq:chargegenp}), the charge $\Qp_{\vec{\sigma}}$ for PBC is defined as (for even $N$) 
\begin{eqnarray}
    \Qp_{\vec{\sigma}} &=& \sumal{j = 1}{N}{(-1)^{j+1} \left(\frac{P_{\sigma_j} + M_{\sigma_{j+1}}}{2} - \left(X^{(P)}_{\sigma_j} + X^{(M)}_{\sigma_{j+1}}\right)\right)},\nn \\
    &=& \sumal{j = 1}{N}{(-1)^{j+1} \left(\frac{P_{\sigma_j} - M_{\sigma_{j}}}{2} - \left(X^{(P)}_{\sigma_j} - X^{(M)}_{\sigma_{j}}\right)\right)}.
\label{eq:chargegenpapp}
\end{eqnarray}
where 
\begin{eqnarray}
    &P_{\sigma_j} = \sumal{l = 1}{p}{\delta_{\sigma_{jl}, +}}, \;\;\; M_{\sigma_j} = \sumal{l = 1}{p}{\delta_{\sigma_{jl}, -}} \nn \\
	&X^{(P)}_{\sigma_j} = \sumal{l = 1}{p}{(p + 1 - l)\  \delta_{\sigma_{jl}, +}} \nn \\
	&X^{(M)}_{\sigma_j} = \sumal{l = 1}{p}{l\  \delta_{\sigma_{jl}, -}},
\label{eq:Xdefnapp}
\end{eqnarray}
where $\sigma_{jl}$ is the configuration of the $l$-th spin in the $j$-th unit cell.
The lowest charge possible for a system of $N$ unit cells are configurations that maximize $\left(\left(P_{\sigma_j} - M_{\sigma_j}\right)/2 - X^{(P)}_{\sigma_j} + X^{(M)}_{\sigma_j}\right)$ when $j$ is odd and minimize it when $j$ is even. 
Thus, the lowest charge configuration can be obtained by setting
\begin{eqnarray}
    &X^{(P)}_{\sigma_{j}} = \frac{p(p+1)}{2} \delta_{j, \textrm{odd}}\;\;\; X^{(M)}_{\sigma_j} = \frac{p(p+1)}{2} \delta_{j, \textrm{even}} \nn \\
    &P_{\sigma_j} = p\ \delta_{j, \textrm{odd}}, \;\;\; M_{\sigma_j} = p\ \delta_{j, \textrm{even}}.
\label{eq:charges}
\end{eqnarray}
Similarly, the highest charge for a system of $N$ unit cells are configurations that minimize and maximize $\left(\left(P_{\sigma_j} - M_{\sigma_j}\right)/2 - X^{(P)}_{\sigma_j} + X^{(M)}_{\sigma_j}\right)$ when $j$ is odd and even respectively.
Thus, the highest charge configuration satisfies
\begin{eqnarray}
    &X^{(P)}_{\sigma_{j}} = \frac{p(p+1)}{2} \delta_{j, \textrm{even}}\;\;\; X^{(M)}_{\sigma_j} = \frac{p(p+1)}{2} \delta_{j, \textrm{odd}} \nn \\
    &P_{\sigma_j} = p\ \delta_{j, \textrm{even}}, \;\;\; M_{\sigma_j} = p\ \delta_{j, \textrm{odd}}.
\label{eq:charges2}
\end{eqnarray}
Thus, the lowest and highest charged configurations, which we refer to as $\ztwo{p}$ and $\ztwop{p}$ respectively, read (for even $N$)
\begin{eqnarray}
    &&\ztwo{p} = \ket{\ \fbox{$+\cdots+$}\ \fbox{$- \cdots-$}\ \cdots\ \fbox{$+\cdots+$}\ \fbox{$- \cdots-$}\ } \nn \\
    &&\ztwop{p} = \ket{\ \fbox{$-\cdots-$}\ \fbox{$+ \cdots+$}\ \cdots\ \fbox{$-\cdots-$}\ \fbox{$+ \cdots+$}\ }. \nn \\
\label{eq:Z2genpapppbc}
\end{eqnarray}
They have charges
\begin{equation}
    \Qp_{\mathbb{Z}_2} = - \frac{N p^2}{2},\;\;\; \Qp_{\mathbb{Z}_2'} = + \frac{N p^2}{2}. 
\label{eq:chargez2z2pgenppbc}
\end{equation}
\subsubsection{OBC}
For open boundary conditions, the definition of $\Qp_{\vec{\sigma}}$ for even $N$ reads
\begin{eqnarray}
    &\Qp_{\vec{\sigma}} = \sumal{j = 1}{N-1}{(-1)^{j+1} \left(\frac{P_{\sigma_j} + M_{\sigma_{j+1}}}{2} - \left(X^{(P)}_{\sigma_j} + X^{(M)}_{\sigma_{j+1}}\right)\right)},\nn \\
    &= \sumal{j = 2}{N-1}{(-1)^{j+1} \left(\frac{P_{\sigma_j} - M_{\sigma_{j}}}{2} - X^{(P)}_{\sigma_j} + X^{(M)}_{\sigma_{j}}\right)} \nn \\
    &+ \frac{P_{\sigma_1} - M_{\sigma_N}}{2} - X^{(P)}_{\sigma_1} + X^{(M)}_{\sigma_N}.
\label{eq:chargegenpobcapp}
\end{eqnarray}
Thus, the lowest (resp. highest) charge configurations can be obtained by satisfying Eq.~(\ref{eq:charges}) for all $j$, $2 \leq j \leq N - 1$. 
Furthermore, the we set $P_{\sigma_1} = p$ (resp. $P_{\sigma_1} = 0$), $X^{(P)}_{\sigma_1} = \frac{p(p+1)}{2}$ (resp. $X^{(P)}_{\sigma_1} = 0$), $M_{\sigma_N} = p$ (resp. $M_{\sigma_N} = 0$) and $X^{(M)}_{\sigma_N} = \frac{p(p + 1)}{2}$ (resp. $X^{(M)}_{\sigma_N} = 0$) for the lowest (resp. highest) charge configurations. 
Thus, these states for OBC and even $N$ read
\begin{eqnarray}
    &&\ztwo{p} = \ket{\ \fbox{$+\cdots+$}\ \fbox{$- \cdots-$}\ \cdots\ \fbox{$+\cdots+$}\ \fbox{$- \cdots-$}} \nn \\
    &&\ztwop{p} = \ket{\ \fbox{$\zero\cdots\zero$}\ \fbox{$+ \cdots+$}\ \cdots\ \fbox{$-\cdots-$}\ \fbox{$\zero\cdots\zero$}},\nn \\
\label{eq:Z2genpappobc}
\end{eqnarray}
which respectively have charges
\begin{equation}
    \Qp_{\mathbb{Z}_2} = - \frac{N p^2}{2},\;\;\; \Qp_{\mathbb{Z}_2'} = + \frac{N p^2}{2} - p^2.
\label{eq:chargez2z2pgenpobc}
\end{equation}
\section{Action of the electrostatic terms}
Here we consider the action of the electrostatic terms in the spin-1 basis. 
The electrostatic part of the Hamiltonian within a Landau level on a thin-torus reads (see Eq.~(\ref{eq:pertelectro}))
\begin{equation}
    \delta H = \sumal{j = 1}{L_b}{\left(V_{1,0} \hat{n}_j \hat{n}_{j+1} + V_{2,0} \hat{n}_j \hat{n}_{j+2}\right)}.
\label{eq:pertelectroapp}
\end{equation}
Note that it is diagonal in the spin-1 basis.
The only non-vanishing action of $\delta H$ on configurations of orbitals are
\begin{equation}
    \delta H\ket{1\ 1} = V_{1,0}\ket{1\ 1},\;\; \delta H\ket{1\ \ast\ 1} = V_{2,0}\ket{1\ \ast\ 1}, 
\label{eq:deltaHactionorbital}
\end{equation}
where $\ast$ is either $0$ or $1$. 
For convenience, we define the spin-1 operators (see Eq.~(\ref{eq:spin1ops}))
\begin{equation}
    Z^+ \equiv U^\dagger U,\; Z^- \equiv T T^\dagger,\; Z^\zero \equiv T^\dagger T = U U^\dagger.
\label{eq:spin1opsapp}
\end{equation}
Using Eq.~(\ref{eq:TUdaggeractions}) in App.~\ref{app:TUproperties}, the non-vanishing actions of these operators on the spin-1 basis states read
\begin{equation}
    Z^+\ket{+} = \ket{+},\;\; Z^-\ket{-} = \ket{-},\;\; Z^\zero\ket{\zero} = \ket{\zero}.
\label{eq:Zactions}
\end{equation}
We then obtain the action of $H$ in terms of the spin-1 basis states, i.e. $\delta \mH{p}$.
Similar to Eq.~(\ref{eq:Hpgeneral}), we split $\delta\mH{p}$ as
\begin{equation}
    \delta\mH{p} = \sumal{j = 1}{N}{\left(\delta\mH{p}\right)_j} + \sumal{j = 1}{N_b}{\left(\delta\mH{p}\right)_{j,j+1}}, 
\end{equation} 
where $N_b = N$ with PBC and $N_b = N - 1$ with OBC. 
\subsection{$\boldsymbol{p = 1}$}
When $p = 1$, using Eq.~(\ref{eq:deltaHactionorbital}), the only non-vanishing action of $\delta H$ on the states in $\mK{1}$ is %
\begin{eqnarray}
    &V_{1,0} \hat{n}_j \hat{n}_{j+1} \overset{j\;j+1}{\ket{\cdots\ \fbox{0\ 0\ 1}\ \fbox{1\ 0\ 0}\ \cdots}} \nn \\
    &= V_{1,0}\overset{j\;j+1}{\ket{\cdots\ \fbox{0\ 0\ 1}\ \fbox{1\ 0\ 0}\ \cdots}},
\end{eqnarray}
which, in terms of the spin-1 degrees of freedom reads (see Eq.~(\ref{eq:spin1dof}))
\begin{equation}
    \left(\delta \mH{1}\right)_{j, j + 1}\overset{j\;\;\;\;j+1}{\ket{\ \cdots\ \bplus\ \bminus\ \cdots\ }} = V_{1,0}\overset{j\;\;\;\;j+1}{\ket{\ \cdots\ \bplus\ \bminus\ \cdots\ }}.
\label{eq:deltaH1action}
\end{equation}
Indeed, nearest neighbor unit cells can only have the configurations shown in Eq.~(\ref{eq:nnallowed})).
Thus, the perturbation reads
\begin{equation}
    \delta\mH{1} = V_{1,0} \sumal{j = 1}{N_b}{Z^+_{j,1} Z^-_{j+1,1}}.   
\end{equation}
Using Eq.~(\ref{eq:deltaH1action}) and the mapping of Eq.~(\ref{eq:duallatticemapping}) to the spin-1/2 degrees of freedom, the non-vanishing action of the perturbation $\delta H_d$ on the dual lattice corresponding to $\delta\mH{1}$ reads
\begin{equation}
    \left(\delta H_d\right)_{j-\frac{1}{2}, j + \frac{1}{2}, j + \frac{3}{2}} \overset{j + \frac{1}{2}}{\ket{\ \downarrow\ \uparrow\ \downarrow\ }} = V_{1,0}\overset{j + \frac{1}{2}}{\ket{\ \downarrow\ \uparrow\ \downarrow\ }}.
\end{equation}
Thus, the operator form of $\delta H_d$ reads (for PBC)
\begin{equation}
    \delta H_d = \sumal{l = \frac{3}{2}}{N + \frac{1}{2}}{\mathcal{P}_{l - 1} \left(\frac{1 + \sigma^z_l}{2}\right) \mathcal{P}_{l + 1}},
\end{equation}
where $\mathcal{P}_{l}$ is defined in Eq.~(\ref{eq:njdefn}).
This is one of the perturbations to the PXP model studied in Ref.~[\onlinecite{turner2018quantum}], and is different from the class of perturbations studied in Refs.~[\onlinecite{khemani2018signatures}] and [\onlinecite{choi2018emergent}].
\subsection{General $\boldsymbol{p}$}
For general $p$, we observe that the non-vanishing actions of the terms of $\delta H$ read (after the addition of pseudozeroes)
\begin{widetext}
\begin{eqnarray}
    &V_{1,0} \hat{n}_j \hat{n}_{j+1}\overset{\;\;\;j\;\;\;\;\;j+1}{\ket{\ \fbox{$\cdots\ 0\ 0\ 1\ [0]\ 1\ 0\ \cdots$}\ }} = V_{1,0}\overset{\;\;\;j\;\;\;\;\;j+1}{\ket{\ \fbox{$\cdots\ 0\ 0\ 1\ [0]\ 1\ 0\ \cdots$}\ }} \nn \\
    &V_{1,0} \hat{n}_j \hat{n}_{j+1}\overset{j\;\;\;\;\;j+1\;\;\;}{\ket{\ \fbox{$\cdots\ 0\ 1\ [0]\ 1\ 0\ 0\ \cdots$}\ }} = V_{1,0}\overset{j\;\;\;\;\;j+1\;\;\;}{\ket{\ \fbox{$\cdots\ 0\ 1\ [0]\ 1\ 0\ 0\ \cdots$}\ }} \nn \\
    &V_{2,0} \hat{n}_j \hat{n}_{j+2}\overset{\;\;\;\;\;\;j\;\;\;\;\;\;\;\;\;j+2}{\ket{\ \fbox{$\cdots\ 0\ 0\ 1\ [0]\ 0\ 1\ \cdots$}\ }} = V_{2,0}\overset{\;\;\;\;\;\;j\;\;\;\;\;\;\;\;\;j+2}{\ket{\ \fbox{$\cdots\ 0\ 0\ 1\ [0]\ 0\ 1\ \cdots$}\ }} \nn \\
    &V_{2,0} \hat{n}_j \hat{n}_{j+2}\overset{j\;\;\;\;\;\;\;j+2}{\ket{\ \fbox{$\cdots\ 0\ 1\ 0\ [0]\ 1\ 0\ \cdots$}\ }} = V_{2,0}\overset{j\;\;\;\;\;\;\;j+2}{\ket{\ \fbox{$\cdots\ 0\ 1\ 0\ [0]\ 1\ 0\ \cdots$}\ }} \nn \\
    &V_{2,0} \hat{n}_j \hat{n}_{j+2}\overset{j\;\;\;\;\;\;\;\;j+2\;\;\;\;\;\;\;}{\ket{\ \fbox{$\cdots\ 1\ 0\ [0]\ 1\ 0\ 0\ \cdots$}\ }} = V_{2,0}\overset{j\;\;\;\;\;\;\;\;j+2\;\;\;\;\;\;\;}{\ket{\ \fbox{$\cdots\ 1\ 0\ [0]\ 1\ 0\ 0\ \cdots$}\ }} \nn \\
    &V_{1,0} \hat{n}_j \hat{n}_{j+1}\overset{j\;j+1}{\ket{\ \fbox{$\cdots\ 0\ 0\ 1$}\ \fbox{$1\ 0\ 0\ \cdots$}\ }} = V_{1,0}\overset{j\;j+1}{\ket{\ \fbox{$\cdots\ 0\ 0\ 1$}\ \fbox{$1\ 0\ 0\ \cdots$}\ }}, \nn \\
\end{eqnarray}
\end{widetext}
where $[0]$ depicts a pseudozero, and $0$ may be a pseudozero.
In terms of spin-1 degrees of freedom defined in Eq.~(\ref{eq:spin1dof}), these actions read
\begin{eqnarray}
    &\left(\delta\mH{p}\right)_j\overset{j}{\ket{\ \fbox{$\cdots\ \plus\ \zero\ \cdots$}\ }} = V_{1,0}\overset{j}{\ket{\ \fbox{$\cdots\ \plus\ \zero\ \cdots$}\ }} \nn \\
    &\left(\delta\mH{p}\right)_j\overset{j}{\ket{\ \fbox{$\cdots\ \zero\ -\ \cdots$}\ }} = V_{1,0}\overset{j}{\ket{\ \fbox{$\cdots\ \zero\ -\ \cdots$}\ }} \nn \\
\end{eqnarray}
\begin{eqnarray}
    &\left(\delta\mH{p}\right)_j\overset{j}{\ket{\ \fbox{$\cdots\ \plus\ \plus\ \cdots$}\ }} = V_{2,0}\overset{j}{\ket{\ \fbox{$\cdots\ \plus\ \plus\ \cdots$}\ }} \nn \\
    &\left(\delta\mH{p}\right)_j\overset{j}{\ket{\ \fbox{$\cdots\ \zero\ \zero\ \cdots$}\ }} = V_{2,0}\overset{j}{\ket{\ \fbox{$\cdots\ \zero\ \zero\ \cdots$}\ }} \nn \\
    &\left(\delta\mH{p}\right)_j\overset{j}{\ket{\ \fbox{$\cdots\ -\ -\ \cdots$}\ }} = V_{2,0}\overset{j}{\ket{\ \fbox{$\cdots\ -\ -\ \cdots$}\ }} \nn \\
    &\left(\delta\mH{p}\right)_{j,j+1}\overset{j\;\;j+1}{\ket{\ \fbox{$\cdots\ \plus$}\ \fbox{$-\ \cdots$}\ }} = V_{1,0}\overset{j\;\;j+1}{\ket{\ \fbox{$\cdots\ \plus$}\ \fbox{$-\ \cdots$}\ }}. \nn \\
\end{eqnarray}
Thus, using Eq.~(\ref{eq:Zactions}), the expression for the perturbation reads
\begin{eqnarray}
    &\delta\mH{p} = \sumal{j = 1}{N_b}{V_{1,0} Z^+_{j, p} Z^-_{j+1,1}} \nn \\
    &+ \sumal{j = 1}{N}{\ \sumal{n = 1}{p-1}{\left(V_{1,0}\left(Z^+_{j,n} Z^\zero_{j,n+1} + Z^\zero_{j,n} Z^-_{j,n+1}\right)\right.}} \nn \\
    &\left.+ V_{2,0}\left(Z^+_{j,n} Z^+_{j,n+1} + Z^\zero_{j,n} Z^\zero_{j,n+1} + Z^-_{j,n} Z^-_{j,n+1}\right)\right).  
\label{eq:pertelectrogenp}
\end{eqnarray}
\bibliography{pair_hopping}
\end{document}